\documentclass[a4paper,12pt]{article}
\pdfoutput=1
\usepackage{epsfig}
\usepackage{amssymb}
\usepackage{amsfonts}
\usepackage{amsmath}
\usepackage{euscript}
\usepackage{verbatim}
\usepackage{latexsym}
\usepackage{graphicx}
\usepackage{caption}
\usepackage{float}
\usepackage{mathtools}
\usepackage{hyperref}
\usepackage{subcaption}
\usepackage{tikz}

\usepackage{listings}

\usepackage{booktabs}

\newif\ifdtup

\catcode`\@=11

\@addtoreset{equation}{section}

\def\@normalsize{\@setsize\normalsize{15pt}\xiipt\@xiipt
\abovedisplayskip 14pt plus3pt minus3pt%
\belowdisplayskip \abovedisplayskip
\abovedisplayshortskip \z@ plus3pt%
\belowdisplayshortskip 7pt plus3.5pt minus0pt}

\def\small{\@setsize\small{13.6pt}\xipt\@xipt
\abovedisplayskip 13pt plus3pt minus3pt%
\belowdisplayskip \abovedisplayskip
\abovedisplayshortskip \z@ plus3pt%
\belowdisplayshortskip 7pt plus3.5pt minus0pt
\def\@listi{\parsep 4.5pt plus 2pt minus 1pt
     \itemsep \parsep
     \topsep 9pt plus 3pt minus 3pt}}

\relax

\catcode`@=12

\catcode`\@=11

\def\section{\@startsection{section}{1}{\z@}{3.5ex plus 1ex minus
   .2ex}{2.3ex plus .2ex}{\large\bf}}

\def\SymBoxes#1#2#3#4{\newdimen\un@t \un@t#3%
\raisebox{#1}{\rule{#2\un@t}{#4}\hskip-#2\un@t
\@tempdimb\un@t \advance\@tempdimb by-#4\@tempcntb#2\relax%
\@whilenum{\@tempcntb>0}\do{
\rule{#4}{\un@t}\hskip\@tempdimb \advance\@tempcntb by\m@ne}%
\hskip-#2\un@t \rule[\un@t]{#2\un@t}{#4}%
\rule[\un@t]{#4}{#4}\hskip-#4
\rule{#4}{\un@t}}\hskip-#4}                

\begin{document}

\newcommand{\beq}{\begin{equation}}
\newcommand{\eeq}{\end{equation}}
\newcommand{\bea}{\begin{eqnarray}}
\newcommand{\eea}{\end{eqnarray}}
\newcommand{\beas}{\begin{eqnarray*}}
\newcommand{\eeas}{\end{eqnarray*}}
\newcommand{\defi}{\stackrel{\rm def}{=}}
\newcommand{\non}{\nonumber}
\newcommand{\bquo}{\begin{quote}}
\newcommand{\enqu}{\end{quote}}
\renewcommand{\(}{\begin{equation}}
\renewcommand{\)}{\end{equation}}
\def \eqn#1#2{\begin{equation}#2\label{#1}\end{equation}}

\def\e{\epsilon}
\def\IZ{{\mathbb Z}}
\def\IR{{\mathbb R}}
\def\IC{{\mathbb C}}
\def\IQ{{\mathbb Q}}
\def\de{\partial}
\def\Tr{ \hbox{\rm Tr}}
\def\H{ \hbox{\rm H}}
\def\HE{ \hbox{$\rm H^{even}$}}
\def\HO{ \hbox{$\rm H^{odd}$}}
\def\K{ \hbox{\rm K}}
\def\Im{ \hbox{\rm Im}}
\def\Ker{ \hbox{\rm Ker}}
\def\const{\hbox {\rm const.}}
\def\o{\over}
\def\im{\hbox{\rm Im}}
\def\re{\hbox{\rm Re}}
\def\bra{\langle}\def\ket{\rangle}
\def\Arg{\hbox {\rm Arg}}
\def\Re{\hbox {\rm Re}}
\def\Im{\hbox {\rm Im}}
\def\exo{\hbox {\rm exp}}
\def\diag{\hbox{\rm diag}}
\def\longvert{{\rule[-2mm]{0.1mm}{7mm}}\,}
\def\a{\alpha}
\def\dag{{}^{\dagger}}
\def\tq{{\widetilde q}}
\def\p{{}^{\prime}}
\def\W{W}
\def\N{{\cal N}}
\def\hsp{,\hspace{.7cm}}

\def\br{\nonumber}
\def\IZ{{\mathbb Z}}
\def\IR{{\mathbb R}}
\def\IC{{\mathbb C}}
\def\IQ{{\mathbb Q}}
\def\IP{{\mathbb P}}
\def \eqn#1#2{\begin{equation}#2\label{#1}\end{equation}}

\newcommand{\C}{\ensuremath{\mathbb C}}
\newcommand{\Z}{\ensuremath{\mathbb Z}}
\newcommand{\R}{\ensuremath{\mathbb R}}
\newcommand{\rp}{\ensuremath{\mathbb {RP}}}
\newcommand{\cp}{\ensuremath{\mathbb {CP}}}
\newcommand{\vac}{\ensuremath{|0\rangle}}
\newcommand{\vact}{\ensuremath{|00\rangle}                    }
\newcommand{\oc}{\ensuremath{\overline{c}}}
\newcommand{\psizero}{\psi_{0}}
\newcommand{\phizero}{\phi_{0}}
\newcommand{\hzero}{h_{0}}
\newcommand{\psiin}{\psi_{\rh}}
\newcommand{\phiin}{\phi_{\rh}}
\newcommand{\hin}{h_{\rh}}
\newcommand{\rh}{r_{h}}
\newcommand{\rb}{r_{b}}
\newcommand{\psibnd}{\psi_{0}^{b}}
\newcommand{\psibndp}{\psi_{1}^{b}}
\newcommand{\phibnd}{\phi_{0}^{b}}
\newcommand{\phibndp}{\phi_{1}^{b}}
\newcommand{\gbnd}{g_{0}^{b}}
\newcommand{\hbnd}{h_{0}^{b}}
\newcommand{\zh}{z_{h}}
\newcommand{\zb}{z_{b}}
\newcommand{\man}{\mathcal{M}}
\newcommand{\hbr}{\bar{h}}
\newcommand{\tbr}{\bar{t}}

\begin{titlepage}

\def\thefootnote{\fnsymbol{footnote}}

\begin{center}
{\large
{\bf A Bottom-Up Approach to Black Hole Microstates
}
}
\end{center}

\begin{center}
Vaibhav Burman$^a$\footnote{\texttt{vaibhav2021@iisc.ac.in  }}, 
Chethan Krishnan$^a$\footnote{\texttt{chethan.krishnan@gmail.com}} \  

\end{center}

\renewcommand{\thefootnote}{\arabic{footnote}}

\begin{center}

$^a$ {Center for High Energy Physics,\\
Indian Institute of Science, Bangalore 560012, India}\\

\end{center}
\vspace{-0.15in}
\begin{center} {\bf Abstract} \end{center} 
In \cite{Burman}, we argued that a sliver at the black hole mass in the Hilbert space of a quantum field theory on an AdS black hole with a stretched horizon, has many desirable features of black hole microstates. A key observation is that the stretched horizon requires a finite Planck length, i.e., finite-$N$. Therefore it is best viewed as a ``quantum horizon" -- a proxy for the UV-complete bulk description, and not directly an element in the bulk EFT. It was shown in \cite{Burman} that despite the manifest absence of the interior, the 2-point function in the sliver is indistinguishable from the smooth horizon correlator, up to the Page time. In this paper, instead of boundary correlators, we work directly in the bulk, and demonstrate the appearance of the bulk Hartle-Hawking correlator in the large-$N$ limit. This is instructive because it shows that the analytic continuation across the horizon is emergent. It is a $bulk$ transition to a Type III algebra and provides a structural distinction between black holes and weakly coupled thermal systems. We also identify a mechanism for universal code subspaces and interior tensor factors to appear via a quantum horizon version of thermal factorization. Our claims apply directly only within a “Page window”, so they are not in immediate tension with firewall arguments. During a Page window, typical heavy microstates probed by light single trace operators respond with effectively smooth horizons in low-point correlators. We work in 2+1 dimensions to be concrete, but expect our results to hold in all higher dimensions. We discuss some important differences between our approach and the conventional fuzzball program, and also argue that probe-notions (like infalling boundary conditions) must be distinguished from microstate-notions (like size of the Einstein-Rosen bridge) to make  meaningful statements about post-Page smoothness. 




\vspace{1.6 cm}
\vfill

\end{titlepage}

\tableofcontents
\setcounter{footnote}{0}

\section{The Quantum Horizon}

Black hole horizons necessarily have entropy $S= A/4G$. This means that if the conventional horizon is part of the description of a state, it is impossible to avoid having unaccounted entropy. The bulk description of a black hole {\em micro}state therefore cannot (by definition) contain the familiar notion of a horizon. Something will have to {\em resolve} the metric-based description of a horizon if we want the state to not carry any entropy.

Information paradox is fundamentally a Lorentzian problem \cite{Hawking, Page, Mathur, AMPS}. This makes it natural that a Hilbert space perspective and an understanding of the states in the Hilbert space will likely be unavoidable for a satisfying resolution of the information paradox. Much of the recent progress on replica wormholes, instead relies on Euclidean path integrals to compute certain combinations of matrix elements related to entanglement entropy. It is unclear from these calculations whether there is a sense in which the horizon is smooth after the Page time. In fact, a prima facie reading of some aspects of the calculation suggests that the horizon is in fact {\em not} smooth after the Page time, while other aspects of it suggest that it is\footnote{These statements are implicit in previous works, but we pose it as a sharp puzzle about post-Page smoothness in Appendix \ref{State-flip}.}. We will have more to say about the status of the post-Page horizon later, but here it suffices to note that it is highly desirable to have some understanding of the nature of black hole microstates in quantum gravity. 

In string theory, we have some understanding of BPS black hole microstates which are extremal and supersymmetric \cite{Sen, Strominger-Vafa}. Most of this understanding is in the weak coupling limit. In the strongly coupled gravitational description, things are much less satisfactory. The conventional fuzzball program\footnote{We will refer to the effort to construct black hole microstates in supergravity as the ``conventional fuzzball program". We will distinguish that from the ``fuzzball paradigm". The latter we will take to be the broad idea that black hole microstates do not have a conventional interior. Our work lies within the broadly defined fuzzball paradigm, but not the conventional fuzzball program.} attempts to construct microstates of BPS black holes in the bulk in supergravity \cite{Mathur1, BenaReview}, but there are reasons to suspect that the currently known supergravity microstates are analogues of BPS multi-graviton states in the bulk rather than true black hole microstates \cite{ChiMing1, ChiMing2}\footnote{These constructions are of course still of interest, because they manage to evade the no-hair theorems of classical general relativity using the compact extra dimensions of string theory.}. Whatever the eventual status of the conventional fuzzball program, it comes with a crucial drawback -- the microstates that we are most interested in are those at finite temperature and far from extremality. For black holes which are at finite temperature, we have no real understanding of typical microstates from a top-down perspective in string theory.

Given this state of affairs, we can ask the following question -- is it possible to {\em engineer} an instructive picture of bulk microstates using {\em bottom-up} arguments? The difficulty with this is of course that we will have to rely on some indirect reasoning, since we do not have a non-perturbative definition of string theory. So where do we even begin? 

We will start by incorporating some of our expectations into a minimal set of demands.
\begin{itemize}
    \item As mentioned above, we would like our construction to be in Lorentzian signature.
    \item Again as mentioned above, we would like to stay away from supersymmetry and extremality. Even though we lose some forms of analytic control by doing this, all is not lost -- this is because we can plausibly expect to draw some general conclusions because black holes are a generic prediction of general relativity and are expected to be thermal (ie., suitably generic) states.
    \item We can work in the context of AdS/CFT where we have some intuition about the states in the theory \cite{WittenAdSCFT, WittenAdSBH}.
    \item It seems reasonable as a first task to consider the black hole as an isolated object, and not treat it as an open system. This means that we should avoid the complications of the small AdS black hole, which can completely evaporate away because it comes with its own sink. We will instead look for models of microstates for the large AdS black hole.
\end{itemize}
With these demands, we are therefore immediately working in a pretty concrete setting. We can look to model the microstates of a large AdS black hole, which we know is dual to a small sliver of heavy microstates in the CFT. What can be a bottom-up way to engineer such a heavy CFT microstate from the bulk?

Should we completely give up the metric? Ultimately, this is likely what we will have to do in a full quantum theory of gravity\footnote{See \cite{Gautam} for an interesting recent paper, where this expectation is realized in a toy model.}. After all, we are looking for finite-$N$ physics, which corresponds to a finite Planck-length in the bulk. If we adopt this stand, it will also mean that we are back to square one -- because this is tantamount to demanding a finite-$N$ description of string theory, which we presently do not have.

But perhaps one can make some progress by noting that it is {\em not} the metric description itself that is responsible for the unaccounted entropy. The culprit here is the horizon. There are well-known metric descriptions of the bulk geometry, which we believe do {\em not} carry any entropy -- empty AdS or flat space being the simplest examples. Therefore perhaps one can make some progress by giving up the horizon in a meaningful way in an otherwise metric-based description. 

Clearly we are looking at a brick-wall\cite{tHooft}/stretched horizon\cite{STU}/fuzzball\cite{MathurFuzz}/firewall\cite{AMPS}-type idea\footnote{We will use the phrases ``brick-wall" and ``stretched horizon", interchangeably. We will view them as bulk UV regulators near the horizon. We have already mentioned fuzzballs. ``Firewall" arguments are arguments that suggest that old black holes have no interiors, (usually) if one assumes smoothness for young black holes.}. These ideas have been around for many decades now in various guises, and refer to the drastic possibility that a black hole has ``no interior". Even though ending the spacetime right before the horizon does (apparent) violence to the principle of equivalence, the idea clearly has a history of refusing to go away. Is there a reasonable and precise way to implement this?

In our setting of the large AdS black hole, there are two immediate problems with the stretched horizon/brick-wall idea. 
\begin{itemize}
    \item Firstly, a heavy microstate in AdS is in equilibrium with its own radiation. So we can {\em not} simply replace the horizon with (say) a Planckian stretched horizon, and then call it a day. There is no natural temperature to this object, there is no radiation.
    \item Secondly, and equally importantly, there is a large {\em negative} stress tensor expectation value at the stretched horizon. The natural time coordinate for us is the CFT time, which is also the same as the exterior Schwarzschild time. Quantizing fields in this set up leads to the Boulware vacuum, and it is known that the stress tensor expectation value diverges to negative infinity at the horizon in the Boulware vacuum. One can view this as an Einstein penalty -- the price to pay for installing the stretched horizon and violating the principle of equivalence. 
\end{itemize} 

In a flat space context, it has been suggested by Israel and Mukohyama \cite{Israel-Mukohyama} that the solution to the second question is to dress the system with a radiation fluid at the Hawking temperature. They did some simple estimates to show that the two divergent contributions cancel. One thing we will note here, is that the ingredient suggested by Israel and Mukohyama is precisely what is needed to fix the problem suggested in the first bullet point as well -- what was missing there, was also thermal radiation. This is a simple qualitative observation, but we feel it is highly suggestive. 

\subsection{Summary}

In a recent paper \cite{Burman}, we tried to provide a ``bottom-up" model for the microstates of a large-AdS black hole in the precise setting of AdS/CFT. A key perspective we adopted was that the stretched horizon is a finite-$N$ effect. In other words, it should really be viewed as a {\em quantum horizon}. This is natural from two perspectives  -- (1) The lore that the stretched horizon is a UV regulator, and (2) the observation that the stretched horizon has to be a {\em finite}\footnote{There are some $\mathcal{O}(1)$ factors in the stretched horizon calculation of black hole thermodynamics that are difficult to fix without a direct understanding of the UV-complete description. This manifests as a one-parameter freedom in the calculations of \cite{Burman,Pradipta,PradiptaNew, tHooft}. But these do not affect the distinction we are making here, which is about finite vs. perturbative.} distance away from the horizon for it to be able to reproduce black hole entropy and thermodynamics \cite{Pradipta}. A UV regulator for a gravitating theory naturally suggests a finite Planck length, so these two perspectives are complementary. The finiteness of Planck scale also implies that we are working with finite-$N$ in the dual holographic description. 


Since the stretched horizon is a bulk object and it captures a finite-$N$ effect, it is natural to view it as an ingredient in the UV complete bulk description. But it is not immediately clear how one can put this perspective to use. As we already mentioned, a ``UV complete bulk description" is nothing but a euphemism for a non-perturbative definition of string theory, which we do not yet have\footnote{We will assume that such a bulk (and not just holographic) description of non-perturbative string theory exists. Note again that the recent results of \cite{Gautam} provide strong evidence for this.}. The observation of \cite{Burman} was that {\em if} one assumes that the resolutions of black hole puzzles are {\em not} reliant on bulk UV-effects far outside the horizon, then one can simply use bulk EFT itself as a proxy for the UV complete description in those regions. This is the sense in which we are not giving up the metric description completely.  A stretched horizon placed about a Planck length outside the horizon can then be a model for a UV-complete calculation, and one expects that computing correlators with this boundary condition can be useful in studying aspects of the UV-complete description. The precise boundary conditions at the stretched horizon turn out not to be important for statements before the Page time, as long as they are not dissipative.

The proposal of \cite{Burman} was that one should quantize field theory with a stretched horizon, but instead of working with the stretched horizon ``vacuum", one should consider a sliver of states at energy equal to the black hole mass. This is a bulk Hilbert space implementation of the Israel-Mukohyama idea in AdS/CFT. It has the advantage that specifying the mass specifies the ensemble, and therefore the rest of the thermodynamic quantities one can compute explicitly. Suitably fixing the one-parameter freedom mentioned previously (arising from the ambiguity in the stretched horizon location), this allowed a complete determination of {\em all} the thermodynamic quantities including the precise pre-factors. In the old 't Hooft calculation, one had to specify both the temperature and the mass in order to get the entropy -- in \cite{Burman} the situation is better because the normal modes were explicitly computed.

Because the Hilbert space set up was completely explicit and one could reproduce the thermodynamics exactly, it was possible to take a further step in \cite{Burman}. One could compute the correlators in typical pure states in the sliver. The remarkable fact about these correlators was that in the large-$N$ (ie., vanishing stretched horizon) limit they reduced to the Hartle-Hawking correlator. The HH thermal correlator is a starting point for bulk reconstruction of the interior and therefore this could be viewed as a hint of the emergence of the smooth interior in the large-$N$ limit. 

Because we could work with a finite stretched horizon, one could also identify the limiting sense in which the smooth horizon is attained. The conclusion of \cite{Burman} was that the smooth horizon in the sense of infalling boundary conditions is a meaningful idea all the way to the Page time despite the presence of the stretched horizon\footnote{Note that the status of bulk EFT after the Page time is not well-understood \cite{AMPS, Marolf-Polchinski}. This is fundamentally because Page time is a finite-$N$ effect, and is not immediately accessible to perturbative bulk EFT arguments.}. It was found that the stretched horizon 2-point functions that stand in for the single-sided correlators on a heavy black hole microstate are indistinguishable from those on a smooth horizon two-sided geometry, up to the Page time. Here, ``indistinguishable" means that the two correlators differ only by exponentially suppressed corrections in the black hole entropy. In the large-$N$ limit, the two correlators are identical\footnote{Note that the Page time is infinite in the large-$N$ limit.}, showing that the smooth horizon is a natural consequence of the classical limit. It should be emphasized that this is an emergent limiting feature of the correlators -- at no point is the boundary condition infalling. The emergence of the smooth horizon in the large-$N$ limit here is to be compared to the emergence of magnetism in the large volume limit of thermodynamics. At finite $N$ or finite volume both phenomena are approximate. A corollary of these facts is that Maldacena's information paradox \cite{MaldacenaEternal}, which kicks in precisely at the Page time, is naturally avoided because of the exponential variance corrections that become important at that timescale.

The correlators that were discussed in \cite{Burman} were extrapolate boundary correlators in AdS/CFT. The final result after the large-$N$ limit was the boundary Hartle-Hawking 2-point function. Since bulk reconstruction can be done in terms of them, this suggested the claims made in \cite{Burman}. But it leaves us with an air of mystery -- what exactly in the bulk is responsible for the emergence of the interior?\footnote{We thank Debajyoti Sarkar for asking this question.} A related but distinct question also arises -- How is it that a black hole is different from other thermal systems in having an interior?

The goal of this paper is to extend the results of \cite{Burman} slightly, make the discussion a bit more explicit from the bulk side, and take steps towards answering some of the above questions. Firstly, we will consider bulk correlators in order to partially de-mystify the origin of the interior. We will demonstrate the emergence of the bulk Hartle-Hawking correlator in the large-$N$ limit. Since we start with a pure one-sided microstate, the bulk points we start with are (by definition) single-sided. But the limiting object (the bulk HH correlator) has the property that it allows a new and natural analytic continuation that was not present at any finite value of the stretched horizon. This is the Kruskal analytic continuation into the interior. The emergence of such a natural analytic continuation seems\footnote{It is conceivable that generic strongly coupled thermal systems have emergent ``interiors" in suitable classical limits. We will comment briefly about this later.} absent in more familiar (weakly coupled) thermal systems. 

Secondly, we discuss $n$-point correlation functions, with $n$ small enough so that we can ignore backreaction. Our explicit calculations are for $n=4$, but our conclusions are more general. This generalizes the 2-point correlator discussion of \cite{Burman}. One of the technical goals behind considering $n$-point correlators is to clarify the nature and significance of factorization in these systems. Our free scalar field which is to be viewed as a perturbation around the black hole geometry, is a candidate for a light single trace operator in the dual gauge theory. In our set-up, since we are working with manifestly free theories, perturbative corrections are absent\footnote{Equivalently, we are only concerned with the leading behavior at large-$N$ and non-perturbative corrections (arising as finite-$N$ effects due to the finiteness of the stretched horizon). The sub-leading perturbative corrections vanish because the theory is free. Note that subleading perturbative corrections do not carry the information we are interested in, so this is an acceptable set up to address the problems we are after. But to discuss perturbative backreaction effects systematically, we will have to include them.}, and the correlators around the vacuum, must manifestly factorize. But as we have emphasized previously, we are not concerned with the ``vacuum" correlators of the stretched horizon -- the microstates of the black hole are the sliver states. We will be interested in factorization in typical pure states on the stretched horizon, up to corrections that are suppressed exponentially in the entropy. In the large-$N$ (vanishing stretched horizon) limit, this reduces to the thermal factorization of Hartle-Hawking correlators in AdS/CFT. The calculation demonstrating these facts is not without some elegance and maybe of broader interest in AdS/CFT. But our primary motivation in considering thermal factorization is that it is a crucial ingredient in our argument that there is an emergent code subspace, which can play a natural role as the interior tensor factor. This is an explicit realization of some observations due to Balasubramanian and collaborators \cite{Bala1, Bala2, Onkar}, we will in particular be concerned with \cite{Onkar}. They argued that a generic mechanism exists for emergent tensor factors to arise as code subspaces in thermal systems thanks to ETH-like physics \cite{Deutsch, Srednicki}. Factorization will play an important role in how this mechanism is realized in our set up. 

One question that is left unaddressed by the discussions in \cite{Onkar} is the question of {\em what} makes the black hole thermal system distinct from more familiar weakly coupled thermal systems -- why do we talk about an interior, only for the black hole? We believe the existence of the large-$N$ analytic continuation across the horizon provides a partial answer to this question. It will be interesting to see if such analytic continuations exist in more familiar (but perhaps strongly coupled) systems in a suitable thermodynamic limit, and whether they allow an interpretation as an ``interior". We do not attempt a complete resolution of this question in this paper, but believe our results suggest directions for future exploration.

The operational notion of smoothness that we use is that the correlators that we find, are consistent with the demand of infalling boundary conditions at the horizon. For a typical state in the sliver, the $t=0$ is defined by the insertion of the first field operator in the 2-point correlator. The thermal result that the 2-pt correlator captures is what the second operator sees, a thermal time\footnote{Thermal time $t_c$ can be viewed as a generalization of ``collision" time that is general enough for strongly coupled systems. Sometimes it is also called the diffusion time. In strongly coupled systems we expect it to be $t_c \sim \beta$. We will not be interested in distinguishing between diffusion time and the somewhat larger scrambling time $t_s \sim \beta \log S$ in this paper. But we will be interested in the time range between (approximately) a thermal/scrambling time and the Page time $t_p \sim \beta S$ at which the bulk smooth horizon correlator starts losing information.} after the initial insertion. But this thermal result involves an exponential temporal decay, which is never-ending if we demand a smooth horizon at all times, and results in information loss. Our stretched horizon sliver correlators match the thermal correlator in this window, but the exponentially suppressed corrections are important after the Page time (this is what prevents information loss in our setting). When there are multi-point correlators, intuitively, thermal factorization applies for pairs of insertions that are within such a ``Page window". We discuss this in some detail to emphasize that infalling boundary conditions as an operational definition of smoothness makes sense only within a Page window -- it is a statement about probes/perturbations. In the concluding section, we discuss {\em what} it can mean for the horizon to be smooth after the Page time, because this needs a satisfying {\em definition} before a clear statement can be made.

Our explicit work-horse in this paper is the (large) BTZ black hole in AdS$_3$/CFT$_2$. This is primarily because the wave equation in BTZ is explicitly solvable in terms of hypergeometric functions. But the structures we see in our discussion are {\em not} crucially reliant on dimension, and we expect them to generalize. We have made an effort to make the discussions in this paper somewhat self-contained so that the reader does not have to consult \cite{Burman}, at least not too much. We have also tried to be fairly explicit in our equations. This is for three reasons -- (1) The normalizations\footnote{We thank Suchetan Das for help with some closely related calculations in \cite{Burman}.} and coefficients of the stretched and smooth horizon calculations play an important role in reproducing precisely the correct result, in the vanishing stretched horizon limit. (2) We wanted to be clear that the claims we make do not contain any hidden caveats beyond the ones we are explicit about. (3) Our discussion of factorization maybe of broader interest and does not seem to have been explicitly worked out in the AdS/CFT context. Finally, we have included a concluding section where we present a critical discussion of the various (often conflicting) viewpoints in the literature regarding microstates and smoothness, and where our own work sits. We have done this for two reasons. We feel that it is crucial to ``steel-man" and not just ignore (as is often done) the various distinct points of view, if we are to get anywhere. We also feel that a critical discussion is useful for leveling the paying field a bit for scientists who are working somewhat in isolation, who do not have access to the narratives of the mainstream groups. In any event, we have tried to share our (necessarily imperfect) overall perspective in the Conclusions.

\section{A Bulk Hilbert Space for Black Hole Microstates}

In this section, we will briefly discuss the key ideas and results of \cite{Burman, Pradipta} that we will use. The discussion will not entirely be a review however -- we have understood the physics of the low-lying normal mode spectrum $\omega_{n,J}$ somewhat better now, and therefore we believe the presentation below incorporates some simplifying insights. It relies on some new things we have learnt \cite{PradiptaNew} by considerations of more general black holes (Kerr-Newman and Cvetic-Youm) and some old observations on wave equations on them \cite{CveticLarsen1, CveticLarsen2, CMS, CK-KerrCFT}. The discussion here is tailored to the BTZ black hole \cite{BTZ, BTZH}.

\subsection{Black Hole Thermodynamics from Normal Modes} 

In our previous paper \cite{Burman}, we worked with the non-rotating BTZ metric 
\bea\label{BTZmetric}
ds^{2}=-\frac{r^2-r_{h}^{2}}{L^{2}}dt^{2}+\frac{L^{2}}{r^2-r_{h}^{2}}dr^{2}+r^{2}d\phi^{2}
\eea
where $r_{h}$ is the horizon radius and $L$ is the AdS radius. In this background, we solved the equation of motion of the massless scalar field $\Phi(r,t,\psi).$ The mode solutions of the scalar field equation can be written in the form 
\bea
\mathcal{U}_{n,J}(r,t\psi)  =\frac{1}{\sqrt{r}}e^{-i\omega_{n,J}t}e^{iJ\psi}\phi_{n,J}(r)
\eea
Because of the boundary condition\footnote{For concreteness, we will take it to be Dirichlet. But there are strong reasons to think that the results we find are largely unaffected by the b.c. choice, as long as they are non-dissipative \cite{Das}. The $J$-dependence of the spectrum will undergo slight variation, but the fact that the spectrum is degenerate in $J$ to a first approximation (which is what goes into the success of our calculations) is unaffected. The key physics underlying our results here is the redshift at the horizon.} at the stretched horizon, 
\bea
\phi_{n,J}(r_h+\epsilon) =0,
\eea
and normalizability demand at the boundary, the spectrum of $\omega$'s becomes quantized. The explicit calculation of these quantized modes is given in \cite{Burman, Pradipta}. Here, we will discuss the results in the low-lying part of the spectrum, which is what turns out to affect the physics we are interested in. While a general determination of the normal modes can quickly turn quite complicated \cite{Pradipta}, the lowest-lying part of the spectrum can be written in the simple form \cite{Pradipta, Burman} 
\bea
\omega_{n,J}\approx\frac{2n\pi r_{h}}{L^{2}\log(\frac{r_h}{2\epsilon})} \equiv n \omega_0. \label{omega0}
\eea
The second equality defines $\omega_0$. There are two observations about the above formula that we will make here. The first is that the $J$-dependence of the low-lying spectrum is (essentially) degenerate. More precisely, the $J$-dependence is exponentially denser than the $n$-dependence \cite{Burman, Pradipta}. Therefore in the expression above, we have ignored the $J$-dependence. The entropy of the black hole is carried by that region of the spectrum where this assumption is valid \cite{Burman,Pradipta}. In \cite{Pradipta}, it was noted that this quasi-degeneracy was ultimately what was behind the success of 't Hooft's brick-wall calculation \cite{tHooft}, which managed to reproduce the area scaling of entropy. The second observation we will make will not play a major role in the present paper, but we believe it is conceptually significant --  $\omega_0$ is set by the scrambling timescale, if one chooses the stretched horizon to be at about a Planck length. Note that scrambling time here is precisely the boundary time that it takes for a free-falling observer to reach a Planckian stretched horizon from the boundary.  

In effect, an observation of \cite{Burman, Pradipta} was  that the thermodynamics of a gas with the spectrum \eqref{omega0}, leads precisely to the detailed thermodynamics of the BTZ black hole, once one fixes the ensemble by choosing the energy to be the black hole mass
\bea
E=M \Big(=\frac{r_h^2}{8 G_N L^2}\Big), \label{Level}
\eea
{\em provided} that $\epsilon$ is small enough and one chooses a cut-off in $J$ defined by
\bea
c=\frac{J_{cut}}{\omega_0 L}, \ \ {\rm where} \ \ c=\frac{3L}{2 G_N}  \label{Jcut}
\eea
is the central charge of AdS$_3$ gravity \cite{BrownHenneaux, StromingerNear}. Note that the condition \eqref{Level} is simply the definition of the ensemble, and necessary in any statistical system. In the presentation of \cite{Burman}, there was an additional condition on $\epsilon$. It was stipulated in \cite{Burman} that the geodesic distance from the horizon to the stretched horizon is precisely a Planck length, this leads to our eqn. \eqref{Planck}. It was appreciated in \cite{Burman} that this was not strictly necessary -- one could relax \eqref{Planck} to allow a free parameter $\alpha$ in \eqref{Planck} while adjusting the $J_{cut}$ accordingly to ensure that \eqref{Jcut} holds, and both the thermodynamics and the correlators will come out correctly\footnote{This allows us to naturally incorporate multiple fields, by scaling the location of the $\epsilon$ appropriately, but we will work with a single scalar field in this paper for concreteness.}. 
Here we are strengthening that observation by phrasing the calculation in terms of the central charge \cite{Burman}. In fact, the observation that the one parameter freedom is naturally related to a central charge is more widely true \cite{PradiptaNew}. It can be used to precisely fix the thermodynamics of very general black holes involving many more charges \cite{PradiptaNew} by considering wave equations on those geometries with stretched horizons. The general message is this -- the thermodynamics of quantum gases associated to fluctuations of fields living on black holes with stretched horizons have precisely the correct structure to reproduce black hole thermodynamics. This is true for very general black holes \cite{PradiptaNew}.

In fact, we suspect that the statement about the existence of $J_{cut}$ can be made more plausible. It is possible that the $J_{cut}$ may {\em not} need to be an extra input, if we stop making the crude approximation that $J$-dependence is degenerate. This makes the calculation significantly more involved, but some efforts in this direction will be reported elsewhere. Note that once one incorporates the true $J$-dependence of the spectrum, even though the calculation gets more complicated, it has the virtue that specifying the ensemble may naturally remove the very high lying $J$-modes. This does {\em not} of course mean that now the answer is fixed without specifying the one-parameter freedom -- what we are doing is after all an IR calculation and the UV regulator {\em must} be specified.  The point however is that now $\epsilon$-dependence can show up in more complicated ways in determining $J_{cut}$. Therefore the $\epsilon$-dependence becomes the natural one-parameter freedom. Said another way, with the approximation of $J$-degeneracy,  the $\epsilon$-dependence drops completely from parts of the calculation and transmutes into the $J_{cut}$-dependence. When the degeneracy approximation is dropped however, it is more natural to work in terms of the $\epsilon$-dependence directly.   

What we will do in the rest of the paper is to fix $\epsilon$ explicitly via the demand that it should be at a geodesic distance of a Planck length above the horizon\footnote{This idea goes back to 't Hooft \cite{tHooft}. But since we are working within an AdS/CFT context, perhaps another natural prescription is to fix the stretched horizon from a boundary anchored distance (instead of a horizon anchored distance). We thank Samir Mathur for a discussion on this.}. This fixes \cite{Burman}
\bea
\frac{r_h}{2 \epsilon} = \frac{L^2}{G_N^2}. \label{Planck}
\eea
This is a convenient choice, but we emphasize that the overall freedom in the problem is best viewed as a single real parameter that is a combination of the location of $\epsilon$ in Planck units, together with $J_{cut}$. In fact, as hinted by the expression \eqref{Jcut} and mentioned above, the thermodynamics of very general black holes  can be fixed by such a choice -- we have been able to show \cite{PradiptaNew} that this amounts to the choice of the central charge in the Kerr-CFT correspondence \cite{CMS, CK-KerrCFT}.

For the BTZ case, let us write down some explicit formulas which will be useful for us in this paper. The ensemble definition \eqref{Level} can be written explicitly as
\bea
\sum_{n,J} \omega_{n,J}N_{nJ} = \frac{r_h^2}{8G_N L^2}.
\eea
The $J$-degeneracy approximation translates this expression into 
\bea
\sum_n n N_n =\frac{\log\Big(\frac{r_h}{2\epsilon}\Big)^2}{96 \pi^2} \label{partition}
\eea
upon using \eqref{Jcut} and the normal modes \eqref{omega0}. Note that this expression is independent of \eqref{Planck} in the degenerate-$J$ approximation. Now \eqref{partition} is a standard problem in integer partitions, and it can be shown (again using the above formulas) that typical partitions have \cite{Burman}
\bea
\langle N_n\rangle = \frac{1}{e^{\beta_H \omega_{n,J}}-1}
\eea
where $\beta_H = \frac{2 \pi L^2}{r_h}$ is the inverse Hawking temperature. Note that apart from choosing the stretched horizon to be close enough to the horizon (which guarantees the normal mode \eqref{omega0} spectrum) the only input we had to make was the choice of $J_{cut}$ as given in \eqref{Jcut}. The latter choice was made to make the precise coefficient of Bekenstein-Hawking entropy to come out right\footnote{Note that the correct scaling with energy of the entropy is a prediction of the explicit calculation using normal modes.}. This automatically guarantees the correct Hawking temperature above, ensuring that the normal modes have correctly reproduced the precise thermodynamics of BTZ. A similar calculation can be done for rotating BTZ \cite{Pradipta}, as well as more general black holes \cite{PradiptaNew}. In all these cases, once the ensemble is specified, there is only a single real number worth of freedom in the calculation -- but that is sufficient to fully fix the thermodynamics.

This calculation \cite{Pradipta,Burman} improves upon the old calculation of 't Hooft \cite{tHooft} in a few ways.
\begin{itemize}
   \item It drops 't Hooft's assumption that the modes need to be semi-classical. We are instead working with honest-to-God normal modes that are aware of tunneling. Crudely, one can think of the semi-classicality demand of 't Hooft, as providing an alternate trick that justifies a cut-off in $J$: Semi-classicality forces one to consider only the modes trapped behind the angular momentum barrier. 
   \item It gives a direct physical understanding of the origin of area-scaling of entropy in terms of the quasi-degeneracy of the modes in $J$. Without this degeneracy, we would get the Planckian black body volume-scaling.
   \item Because we computed the modes explicitly instead of relying on indirect arguments, our calculation is a fully legitimate statistical mechanics calculation once the energy of the ensemble is specified via the mass of the black hole. In 't Hooft's calculation, the energy and temperature both had to be specified because the normal modes were not known.
   \item The energy chosen is precisely the one that is almost certain to cancel the Boulware divergence. The Israel-Mukohyama estimate \cite{Israel-Mukohyama} makes this plausible, but to completely establish this, we need to do a stress tensor expectation value calculation in the sliver state \cite{StressNew}. We do not do that in the present paper, but we provide overwhelming evidence for this by computing correlation functions and showing that they become the Hartle-Hawking correlator in the small $\epsilon$ limit. 
\end{itemize}

\subsection{Highly Excited QFT with a Quantum Horizon}

We can write the general field solution $\Phi$ in the form of a mode expansion
\bea\label{modeexpansionofthefield}
\Phi(r,t,\psi) = \sum_{n,J}\frac{1}{\sqrt{4\pi\omega_{n,J}}}\Bigg(b_{n,J}\mathcal{U}_{n,J}(r,t,\psi)+b^{\dagger}_{n,J}\mathcal{U}_{n,J}^{*}(r,t,\psi)\Bigg)
\eea
The modes $\mathcal{U}_{n, J}$'s are normalized using the Klein-Gordan inner product, and $b_{n,J},b_{n,J}^{\dagger}$ have the usual meaning of the annihilation and creation operators for the modes respectively. The stretched horizon vacuum is defined as 
\bea
b_{n,J}|0\rangle_{SH}=0 \hspace{0.5cm}  \forall\hspace{0.5cm} n\in \mathbb{N},\hspace{0.3cm} J\in \pm\mathbb{Z}^{+}
\eea
This vacuum $|0\rangle_{SH}$ has similarities to the Boulware vacuum. One key technical difference is that the modes are constrained to vanish on the stretched horizon (so for example, the $\omega$ is quantized here). 

We will calculate the two-point correlator in a typical eigenstate built on the stretched horizon vacuum $|0\rangle_{SH}$ with energy equal to that of the mass of the black hole. Such states can be constructed as 
\bea\label{excitedstate}
|N\rangle\equiv |\{N_{m,K}\}\rangle = \prod_{m,K}\frac{1}{\sqrt{N_{m,K}!}}(b^{\dagger})^{N_{m,K}}|0\rangle_{SH}
\eea
We consider states that lie in an energy sliver at the mass of the black hole:
\bea
E = M = \sum_{n,J}\omega_{n,J}N_{n,J}
\eea
In this state, the bulk Wightman function is defined as
\bea
\mathcal{G}_{c}^{+}\equiv\langle\{N_{m,K}\}|\Phi(r,t,\psi)\Phi'(r,t,\psi)|\{N_{m,K}\}\rangle
\eea
In \cite{Burman}, we showed that when we take $\epsilon\rightarrow 0$, the extrapolate boundary limit of the above two-point correlator, which we call $G_{c}^{+}$, became the boundary Hartle-Hawking correlator $G_{HH}$. 

In getting the exact expression for the Hartle-Hawking correlator, the normalization of the field $\phi_{n,J}(r)$ played a role. The detailed calculation of this normalization is done in Appendix C of \cite{Burman}. Here, we will state the result for the normalized field.
\bea
\phi_{n,J}(r) = \frac{T}{L}\sqrt{K}r^{-\frac{1}{2}-\nu}\frac{r_{h}^{1+\nu}}{\Gamma(\Delta)}\Bigg(1-\frac{r_{h}^{2}}{r^{2}}\Bigg)^{-\frac{i\omega_{n,J}L^{2}}{2r_{h}}}{}_2F_{1}(\alpha,\beta;\Delta;y) \label{normfield}
\eea
where
\bea
T =\left[\frac{\Gamma(\beta^{*})\Gamma(\alpha)\Gamma(\beta)\Gamma(\alpha^{*})}{\Gamma(\frac{i\omega L^{2}}{r_{h}})\Gamma(-\frac{i\omega L^{2}}{r_{h}})}\right]^{\frac{1}{2}} \hspace{0.2cm},\hspace{0.2cm} K = \frac{1}{\log(\frac{r_{h}}{2\epsilon})}\Bigg(1-\frac{\epsilon}{r_{h}}\Bigg)^{-2\Delta}\hspace{0.2cm},\hspace{0.2cm}\Delta =1+\nu \label{defT}
\eea
and 
\bea
\alpha = \frac{1}{2}\Big(1-\frac{i\omega L^{2}}{r_{h}}+\frac{iJL}{r_{h}}+\nu\Big)\hspace{0.2cm},\hspace{0.2cm} \beta = \frac{1}{2}\Big(1-\frac{i\omega L^{2}}{r_{h}}-\frac{iJL}{r_{h}}+\nu\Big),\hspace{0.2cm} y=\frac{r_{h}^{2}}{r^{2}}
\eea

In Appendix \ref{AppHam} we discuss the Hamiltonian of this system. A crucial feature of the Hamiltonian is that the natural time coordinate associated to it is the boundary time.

\section{Emergence of the Bulk Hartle-Hawking Correlator}

The goal of this section is to demonstrate the emergence of the bulk Hartle-Hawking correlator from the bulk stretched horizon correlator in the black hole sliver. In order to do this, we will have to be careful about some details, so we begin with a discussion of the conventional Hartle-Hawking correlator corresponding to the smooth horizon. We loosely follow \cite{Festuccia-Liu, Festuccia}, but derive some of the technical details we need, in Appendix \ref{AppSmooth}.

To calculate the Bulk HH correlator, we need to impose free wave boundary conditions at the horizon \cite{Festuccia-Liu}. We will use the near-horizon solution found in $\phi_{\omega,J}$ from \cite{Burman,Das}
\bea\label{nearhorizonexpansion}
\phi_{hor}(r)\approx C_{1}\Bigg(P_{1}\Big(\frac{r}{r_{h}}-1\Big)^{\frac{-i\omega L^2}{2r_{h}}}+Q_{1}\Big(\frac{r}{r_{h}}-1\Big)^{\frac{i\omega L^2}{2r_{h}}}\Bigg)
\eea
where $P_{1}$ and $Q_{1}$ have known expressions \cite{Burman}. We will write the above expression in tortoise coordinates via
\bea\label{tortoisecoordinatedefinition}
z = -\int_{r}^{\infty}\frac{dr}{f(r)}
\eea
Where $f(r)$ is the blackening factor of the BTZ metric. At the horizon $r\rightarrow r_{h}$, $z$ becomes 
\bea\label{tortoise near horizon}
z\approx \frac{\beta_H}{4\pi}\log(r-r_{h})\rightarrow -\infty
\eea
In terms of tortoise coordinates, equation \eqref{nearhorizonexpansion} becomes 
\bea
\phi_{hor}(z)\approx C_{1}\Bigg(P_{1}e^{-i\omega z}r_{h}^{\frac{i\omega L^2}{2r_{h}}}+Q_{1}e^{i\omega z}r_{h}^{-\frac{i\omega L^2}{2r_{h}}}\Bigg) 
\eea
We have used $\beta_H = \frac{2\pi L^2}{r_{h}}$. We will find it convenient to write this as 
\bea\label{phihorfinal}
\phi_{hor}(z)\approx |C_{1}||P_{1}|\Bigg(e^{-i\omega z+i\alpha+i\theta+i\frac{\omega L^2}{2r_{h}}\log r_{h}}+e^{i\omega z+i\alpha+i\theta-i\frac{\omega L^2}{2r_{h}}\log r_{h}}\Bigg)
\eea
Where we have used $|P_{1}|=|Q_{1}|$. Here $\alpha\equiv {\rm Arg}(P_{1})$, $\beta\equiv {\rm Arg}(Q_{1})$ and $\theta\equiv {\rm Arg}(C_{1})$. The calculation of $\alpha$ and $\beta$ is done in Appendix \ref{AppSmooth}. For the massless case ($\nu=1$) these take the form
\bea\label{alpha}
\alpha = -2\pi-\frac{\omega L^2}{2r_{h}}\log 2 -\frac{L}{r_{h}}(\omega L+J)\log r_{h} - 2\Arg\Gamma\big[\frac{iL}{2r_{h}}(\omega L-J)\big]-\Arg\Gamma\big[\frac{iJL}{r_{h}}\big]+\Arg\Gamma\big[\frac{i\omega L^2}{r_{h}}\big]\nonumber\\ 
\eea
and 
\bea\label{beta}
\beta = \frac{\omega L^2}{r_{h}}\log 2 - \frac{L}{r_{h}}(\omega L+J)\log r_{h}+2\Arg\Gamma\big[\frac{iL}{2r_{h}}(\omega L+J)\big]-\Arg\Gamma\big[\frac{iJL}{r_{h}}\big]-\Arg\Gamma\big[\frac{i\omega L^2}{r_{h}}\big]\nonumber\\
\eea
Using the reality condition of the field $\phi_{\omega,J}(r)^{*} = \phi_{\omega,J}(r)$, we get (See \cite{Burman})
\bea
e^{i\theta} = - r_{h}^{\frac{iL}{r_{h}}(\omega L+J)}\Bigg[\frac{\Gamma(\beta)\Gamma(\alpha^{*})\Gamma(\frac{iJL}{r_{h}})}{\Gamma(\beta^{*})\Gamma(\alpha)\Gamma(\frac{-iJL}{r_{h}})}\Bigg]^{\frac{1}{2}}
\eea
Using some Gamma functions formulas (see Appendix \ref{AppSmooth} for more details), we can find the expression for $\theta$
\bea\label{thetha}
\theta = \pi+\frac{L}{r_{h}}(\omega L+J)\log r_{h}-\Arg\Gamma\Big[\frac{iL}{2r_{h}}(\omega L+J)\Big]+\Arg\Gamma\Big[\frac{iL}{2r_{h}}(\omega L-J)\Big]+\Arg\Gamma\Big[\frac{iJL}{r_{h}}\Big]
\eea
Putting \eqref{alpha},\eqref{beta} and \eqref{thetha} in \eqref{phihorfinal} we finally get
\bea\label{phihornormalizable}
\phi_{hor}(z)\approx |C_{1}||P_{1}|\big(-e^{i\omega z + i\delta_{\omega}}-e^{-i\omega z - i\delta_{\omega}}\big)
\eea
where,
\bea
\delta_{\omega}\equiv \frac{\omega L^2}{r_{h}}\log 2+\Arg\Gamma\Big[\frac{iL}{2r_{h}}(\omega L-J)\Big]+\Arg\Gamma\Big[\frac{iL}{2r_{h}}(\omega L+J)\Big]-  
\Arg\Gamma\Big[\frac{i\omega L^2}{r_{h}}\Big]- \frac{\omega L^2}{2r_{h}}\log r_{h}\nonumber \\
\eea
We normalize \ref{phihornormalizable} by demanding $|C_{1}||P_{1}|=1$ \cite{Festuccia-Liu}, which
gives
\bea
C_{1} = |C_{1}|e^{i\theta} = \frac{1}{|P_{1}|}e^{i\theta}
\eea
and so  we have
\bea
\hspace{-0.4cm}C_{1}\hspace{-0.1cm}=\hspace{-0.1cm}-\Gamma\Big[\frac{iJL}{r_{h}}\Big]\hspace{-0.1cm}\Bigg[\frac{\Gamma[1-\frac{iL}{2r_{h}}(\omega L+J)]\Gamma[1+\frac{iL}{2r_{h}}(\omega L-J)]}{\Gamma[1+\frac{iL}{2r_{h}}(\omega L+J)]\Gamma[1-\frac{iL}{2r_{h}}(\omega L-J)]}\Bigg]^{\frac{1}{2}}\hspace{-0.1cm}\Big(\frac{\omega L^2}{\pi r_{h}^{2}}\sinh{\big(\frac{\pi\omega L^{2}}{r_{h}}\big)}\Big)^{\frac{1}{2}}\hspace{-0.1cm}\frac{e^{\frac{\pi L}{2r_{h}}(\omega L+J)}}{r_{h}^{-\frac{iL}{r_{h}}(\omega L+J)}}
\eea
Using equation (2.9) of \cite{Burman}, we get the final form for the field
\bea\label{phinormalized}
\phi_{\omega,J}(r) = r^{-\frac{3}{2}}\Big(1-\frac{r_{h}^2}{r^2}\Big)^{-\frac{i\omega L^{2}}{2r_{h}}}L^{2}\sqrt{\omega\pi}\sqrt{\Tilde{f}(\omega,J)}{}_2F_{1}[\alpha,\beta;2;\frac{r_{h}^2}{r^2}]
\eea
Here $\Tilde{f}(\omega,J)$ is the expression of $\Tilde{f}(\omega_{n,J})$ given in eqn (5.5) of \cite{Burman} with the replacement $\omega_{n,J}\rightarrow\omega$.

Now to calculate the Hartle-Hawking correlation function, we use the notation of \cite{Festuccia}
\bea\label{HHcorrelatordefinition}
\mathcal{G}_{+}\equiv {}_{HH}\langle 0|\Phi(r,t,\psi)\Phi(r',t'\psi')|0\rangle_{HH}
\eea
where
\bea\label{modeexpansion}
\Phi(r,t,\psi) = \sum_{i=1}^{2}\int\frac{d\omega}{2\pi}\sum_{J}(H_{\omega J}^{(i)}b_{\omega J}^{(i)}+H_{\omega J}^{(i)*}b_{\omega J}^{(i)\dagger})
\eea
and $H_{\omega,J}^{(1)}$ and $H_{\omega,J}^{(2)}$ are defined as 
\bea
H_{\omega,J}^{(1)}=\cosh\theta_{\omega}\varphi_{\omega,J}^{(1)}+\sinh\theta_{\omega}\varphi_{\omega,J}^{(2)}\\
H_{\omega,J}^{(2)}=\cosh\theta_{\omega}\varphi_{\omega,J}^{(2)*}+\sinh\theta_{\omega}\varphi_{\omega,J}^{(1)*}
\eea
and 
\begin{equation}\label{varphi1}
\varphi_{\omega,J}^{(1)}=\begin{cases}
\frac{1}{\sqrt{2\omega}}e^{-i\omega t+iJ\psi}\frac{1}{\sqrt{r}}\phi_{\omega J}(r)\hspace{1cm}, & R \\ 
0, & L 
\end{cases}
\end{equation}
and
\begin{equation}\label{varphi2}
\varphi_{\omega,J}^{(1)}=\begin{cases}
\frac{1}{\sqrt{2\omega}}e^{-i\omega t+iJ\psi}\frac{1}{\sqrt{r}}\phi_{\omega J}(r)\hspace{1cm}, & L\\ 
0, & R
\end{cases}
\end{equation}
with
\bea\label{commutationrelation}
[b_{\omega J}^{(i)},b_{\omega' J'}^{(j)\dagger}]=\delta(\omega-\omega')\delta_{JJ'}\delta_{ij}
\eea
Using \eqref{HHcorrelatordefinition}, we have
\bea
\mathcal{G}_{HH}(\omega,J;r,r') = \frac{1}{4\pi\omega}\frac{e^{\beta\omega}}{e^{\beta\omega}-1}(rr')^{-1/2}\phi_{\omega J}^{*}(r)\phi_{\omega J}(r')
\eea
Plugging \eqref{phinormalized} into the above equation we get
\bea\label{smoothbulkHHcorrelator}
\mathcal{G}_{HH}(\omega,J;r,r')  =\frac{L^4}{4}\frac{e^{\beta\omega}}{e^{\beta\omega}-1}(rr')^{-2}\Big(1-\frac{r_{h}^2}{r^2}\Big)^{\frac{i\omega L^2}{2r_{h}}}\Big(1-\frac{r_{h}^2}{r'^2}\Big)^{\frac{-i\omega L^2}{2r_{h}}}\Tilde{f}(\omega,J)\times \\ \nonumber
\times {}_2F_{1}[\alpha^*,\beta^*;2;\frac{r_{h}^2}{r^2}]{}_2F_{1}[\alpha,\beta;2\frac{r_{h}^2}{r'^2}]
\eea
This is the final expression of the bulk HH correlator when the horizon is smooth. We will reproduce this from the sliver correlator in the next subsection.

\subsection{The Classical Limit of the Quantum Horizon}

We start with the expression for the bulk 2-point sliver correlator in the excited state \cite{Burman}
\bea\label{bulkSH2pointcorrelator}
\mathcal{G}_{c}^{+}\equiv \sum_{n,J}\frac{(rr')^{-\frac{1}{2}}}{4\pi\omega_{n,J}}\Big[(N_{n,J}+1)e^{-i\omega_{n,J}(t-t')+iJ(\psi-\psi')}+N_{n,J}e^{i\omega_{n,J}(t-t')-iJ(\psi-\psi')} \Big]\phi_{n,J}^{*}(r)\phi_{n,J}(r')\nonumber\\
\eea 
We have the explicit expression of $\phi_{n,J}(r)$ in \eqref{normfield}. 
Since $T$ as defined by \eqref{defT} is real, $|T|^{2}=T^{2}$. For reducing clutter, we define
\begin{align}
    f(\omega_{n,J},J) \equiv \frac{\Gamma(\beta^{*})\Gamma(\alpha)}{\Gamma\frac{i\omega L^2}{r_{h}}}
\end{align}
so that \cite{Burman}
\begin{align}
   |T|^{2}\equiv |f(\omega_{n,J},J)|^{2} = \frac{\pi L^{4}}{r_{h}^{3}} \frac{\omega_{n,J}\left(\frac{(\omega_{n,J}^2 L^{2}-J^2)}{4}\right)\sinh[\frac{\pi\omega_{n,J}L^{2}}{r_{h}}]}{\sinh[\frac{\pi L(\omega_{n,J}L+J)}{2r_{h}}]\sinh[\frac{\pi L(\omega_{n,J}L-J)}{2r_{h}}]}\equiv\frac{\pi\omega_{n,J}L^{4}}{r_{h}^{3}}\Tilde{f}(\omega_{n,J},J)
\end{align}
Eqn. \eqref{bulkSH2pointcorrelator} can be written as
\begin{align}
   L^{2} \mathcal{G}_{c}^{+}= L^{4}r_{h}K\sum_{n,J}\frac{(rr')^{-2}}{4}\Big[(N_{n,J}+1)e^{-i\omega_{n,J}(t-t')+iJ(\psi-\psi')}+N_{n,J}e^{i\omega_{n,J}(t-t')-iJ(\psi-\psi')} \Big]\Tilde{F}(\omega_{n,J},J)
\end{align}
where
\begin{align}\label{Ftilde}
\Tilde{F}(\omega_{n,J},J)\equiv \Tilde{f}(\omega_{n,J},J)\Bigg(1-\frac{r_{h}^2}{r^2}\Bigg)^{\frac{i\omega_{n,J}L^2}{2r_{h}}}\Bigg(1-\frac{r_{h}^2}{r'^2}\Bigg)^{-\frac{i\omega_{n,J}L^2}{2r_{h}}}
{}_2F_{1}[\alpha^{*},\beta^{*};\Delta;\frac{r_{h}^2}{r^2}] {}_2F_{1}[\alpha,\beta;\Delta;\frac{r_{h}^2}{r'^2}]
\end{align}
and $K\equiv \frac{1}{\log(\frac{r_{h}}{2\epsilon})}\Bigg(1-\frac{\epsilon}{r_{h}}\Bigg)^{-4}$. We can reshuffle the above equation and write it as 
\begin{align}\label{3.31}
    L^{2}\mathcal{G}_{c}^{+}= L^{4}r_{h}K\sum_{n,J}\frac{(rr')^{-2}}{4}N_{n,J}\Bigg[e^{-i\omega_{n,J}(t-t')}e^{iJ(\psi-\psi')}+e^{i\omega_{n,J}(t-t')}e^{-iJ(\psi-\psi')}\Bigg]\Tilde{F}(\omega_{n,J},J)+\nonumber\\
    +L^{4}r_{h}K\sum_{n,J}\frac{(rr')^{-2}}{4}\Tilde{F}(\omega_{n,J},J)e^{-i\omega_{n,J}(t-t')}e^{iJ(\psi-\psi')}
\end{align}

At this stage, we make the replacement of $N_{n, J}$ with its mean value $\langle N_{n, J}\rangle.$ This essentially means that we are considering typical states. This replacement is only valid if the variance of the two-point function is sufficiently suppressed. This was argued in \cite{Burman}, and the same argument applies here. Making the replacement\footnote{Here, we are assuming that $N_{n,J}$ is independent of $J$, because $\omega_{n,J}$ is approximately degenerate in $J$. See \cite{Burman} for a discussion.}
\begin{align}
   \langle N_{n}\rangle = \frac{1}{e^{\beta_H\omega_{n,J}}-1} 
\end{align}
and plugging it back into equation \eqref{3.31}, we get
\begin{align}\label{3.13}
   L^{2} \mathcal{G}_{c}^{+}= L^{4}r_{h}K\sum_{n=0}^{\infty}\sum_{J=-J_{cut}}^{J_{cut}}\frac{(rr')^{-2}}{4}&\frac{1}{e^{\beta_H\omega_{n,J}}-1}\Bigg[e^{-i\omega_{n,J}(t-t')}e^{iJ(\psi-\psi')}+e^{i\omega_{n,J}(t-t')}e^{-iJ(\psi-\psi')}\Bigg]\Tilde{F}(\omega_{n,J},J)+\nonumber\\
   & +L^{4}r_{h}K\sum_{n=0}^{\infty}\sum_{J=-J_{cut}}^{J_{cut}}\frac{(rr')^{-2}}{4}\Tilde{F}(\omega_{n,J},J)e^{-i\omega_{n,J}(t-t')}e^{iJ(\psi-\psi')}
\end{align}
Up to this stage, other than the fact that we are carrying around the bulk $r$ and $r'$ dependencies (and that therefore the expressions are more complicated), there are no extra subtleties compared to the boundary correlator calculation in \cite{Burman}.

Now, let us focus on the second term inside the square bracket 
\begin{align}
    II \equiv  L^{4}r_{h}K\sum_{n=0}^{\infty}\sum_{J=-J_{cut}}^{J_{cut}}\frac{(rr')^{-2}}{4}&\frac{1}{e^{\beta_H\omega_{n,J}}-1}e^{i\omega_{n,J}(t-t')}e^{-iJ(\psi-\psi')}\Tilde{F}(\omega_{n,J},J)\label{startman}
\end{align}
Putting \eqref{Ftilde} into the above equation 
\begin{align}
    II =  L^{4}r_{h}K\sum_{n=0}^{\infty}\sum_{J=-J_{cut}}^{J_{cut}}&\frac{(rr')^{-2}}{4}\frac{1}{e^{\beta\omega_{n,J}}-1}e^{i\omega_{n,J}(t-t')}e^{-iJ(\psi-\psi')}\Tilde{f}(\omega_{n,J},J)\Bigg(1-\frac{r_{h}^2}{r^2}\Bigg)^{\frac{i\omega_{n,J}L^2}{2r_{h}}}\times\nonumber\\
    &\times\Bigg(1-\frac{r_{h}^2}{r'^2}\Bigg)^{-\frac{i\omega_{n,J}L^2}{2r_{h}}}
    {}_2F_{1}[\alpha^{*},\beta^{*};\Delta;\frac{r_{h}^2}{r^2}] {}_2F_{1}[\alpha,\beta;\Delta;\frac{r_{h}^2}{r'^2}]
\end{align}
and changing $n\rightarrow-n$, $J\rightarrow-J$ and using $\Tilde{f}(-\omega_{n,J},-J) = -\Tilde{f}(\omega_{n,J},J)$, we get
\begin{align}\label{3.16}
    II = L^{4}r_{h}K\sum_{n=0}^{-\infty}\sum_{J=J_{cut}}^{-J_{cut}}&\frac{(rr')^{-2}}{4}\frac{e^{\beta\omega_{n,J}}}{e^{\beta\omega_{n,J}}-1}e^{-i\omega_{n,J}(t-t')}e^{iJ(\psi-\psi')}\Tilde{f}(\omega_{n,J},J)\Bigg(1-\frac{r_{h}^2}{r^2}\Bigg)^{\frac{-i\omega_{n,J}L^2}{2r_{h}}}\times\nonumber\\
    &\times\Bigg(1-\frac{r_{h}^2}{r'^2}\Bigg)^{\frac{i\omega_{n,J}L^2}{2r_{h}}}
    {}_2F_{1}[\beta,\alpha;\Delta;\frac{r_{h}^2}{r^2}] {}_2F_{1}[\beta^{*},\alpha^{*};\Delta;\frac{r_{h}^2}{r'^2}]
\end{align}
Using the following Hypergeometric identity
\begin{align}\label{Hypergeometricidentity}
    {}_2F_{1}(a,b,c,z) = (1-z)^{c-a-b}{}_2F_{1}(c-a,c-b,c,z)
\end{align}
the product of two hypergeometric functions can be written as
\begin{align}
   {}_2F_{1}[\beta,\alpha;\Delta;\frac{r_{h}^2}{r^2}] {}_2F_{1}[\beta^{*},\alpha^{*};\Delta;\frac{r_{h}^2}{r'^2}] = \Bigg(1-\frac{r_{h}^{2}}{r^{2}}\Bigg)^{\frac{i\omega_{n,J}L^{2}}{r_{h}}}\Bigg(1-\frac{r_{h}^{2}}{r'^{2}}\Bigg)^{\frac{-i\omega_{n,J}L^{2}}{r_{h}}}\times\nonumber\\
   \times{}_2F_{1}(\beta^{*},\alpha^{*},\Delta, \frac{r_{h}^2}{r^2}){}_2F_{1}(\beta,\alpha;\Delta; \frac{r_{h}^2}{r'^2})
\end{align}
Since ${}_2F_{1}(a,b,c,z)={}_2F_{1}(b,a,c,z)$, equation \eqref{3.16} becomes 
\begin{align}
     II &= L^{4}r_{h}K\sum_{n=0}^{-\infty}\sum_{J=J_{cut}}^{-J_{cut}}\frac{(rr')^{-2}}{4}\frac{e^{\beta\omega_{n,J}}}{e^{\beta\omega_{n,J}}-1}e^{-i\omega_{n,J}(t-t')}e^{iJ(\psi-\psi')}\Tilde{f}(\omega_{n,J},J)\Bigg(1-\frac{r_{h}^2}{r^2}\Bigg)^{\frac{i\omega_{n,J}L^2}{2r_{h}}}\times\nonumber\\
    &\times\Bigg(1-\frac{r_{h}^2}{r'^2}\Bigg)^{\frac{-i\omega_{n,J}L^2}{2r_{h}}}
    {}_2F_{1}[\alpha^{*},\beta^{*};\Delta;\frac{r_{h}^2}{r^2}] {}_2F_{1}[\alpha,\beta;\Delta;\frac{r_{h}^2}{r'^2}]\nonumber\\
    & = L^{4}r_{h}K\sum_{n=0}^{-\infty}\sum_{J=J_{cut}}^{-J_{cut}}\frac{(rr')^{-2}}{4}\frac{e^{\beta\omega_{n,J}}}{e^{\beta\omega_{n,J}}-1}e^{-i\omega_{n,J}(t-t')}e^{iJ(\psi-\psi')}\Tilde{F}(\omega_{n,J},J)
\end{align}
With this, equation \eqref{3.13} simplifies to
\begin{align}
   L^{2}\mathcal{G}_{c}^{+} =  L^{4}r_{h}K\sum_{n=-\infty}^{-\infty}\sum_{J=J_{cut}}^{-J_{cut}}\frac{(rr')^{-2}}{4}\frac{e^{\beta\omega_{n,J}}}{e^{\beta\omega_{n,J}}-1}e^{-i\omega_{n,J}(t-t')}e^{iJ(\psi-\psi')}\Tilde{F}(\omega_{n,J},J)
\end{align}

From this point on, the argument is parallel to that in \cite{Burman}. The summation over $n$ can be converted into integration over $\omega$ when we take $\epsilon\rightarrow 0$. The bulk two-point correlator in the stretched horizon geometry becomes
\begin{align}
    \mathcal{G}^{+}_{c} = \frac{L^{4}}{4}\int\frac{d\omega}{2\pi}\sum_{J=-\infty}^{+\infty}(rr')^{-2}\frac{e^{\beta\omega}}{e^{\beta\omega}-1}e^{-i\omega(t-t')}e^{iJ(\psi-\psi')}\Tilde{F}(\omega,J)
\end{align}
The momentum-space version of the above equation is
\begin{align}
    \mathcal{G}_{c}^{+}(\omega,J;r,r') = \frac{L^{4}}{4}(rr')^{-2}\frac{e^{\beta\omega}}{e^{\beta\omega}-1}\Tilde{F}(\omega,J)
\end{align}
Putting the value of $\Tilde{F}(\omega,J)$ from equation \eqref{Ftilde} into the above equation, we get

\bea\label{continummlimitofcutoffcorrelator}
\mathcal{G}_{c}^{+}(\omega,J;r,r')=\frac{L^4}{4}\frac{e^{\beta\omega}}{e^{\beta\omega}-1}(rr')^{-2}\Big(1-\frac{r_{h}^2}{r^2}\Big)^{\frac{i\omega L^2}{2r_{h}}}\Big(1-\frac{r_{h}^2}{r'^2}\Big)^{-\frac{i\omega L^2}{2r_{h}}}\Tilde{f}(\omega,J)\times \\ \nonumber
\times {}_2F_{1}[\alpha^{*},\beta^{*};2;\frac{r_{h}^2}{r^2}] {}_2F_{1}[\alpha,\beta;2;\frac{r_{h}^2}{r'^2}]
\eea
Comparison with \eqref{smoothbulkHHcorrelator} establishes what we set out to show, namely that
\bea
\mathcal{G}_{c}^{+}\xrightarrow{\epsilon\rightarrow0}\mathcal{G}_{HH}
\eea

The emergence of the HH correlator in the bulk is not unreasonable, considering our boundary results in \cite{Burman}. But there are technical and conceptual points worthy of note here. The technical point is that the manipulations starting around \eqref{startman} are intrinsically bulk-based, and were invisible from the boundary. The conceptual reason is that smoothness is fundamentally a bulk idea, and therefore it is natural to try and identify its bulk origins -- instead of only working with the boundary correlators as done in \cite{Burman}. We explore this further in the next section.

\section{Emergence of the Second Tensor Factor}

In this section, we will observe that there exists a natural mechanism for universal code subspaces to appear as emergent tensor factors in thermal systems. This follows the rough structure of the discussion in \cite{Onkar}, see also related previous work in \cite{Bala1, Bala2}. Our set up provides an explicit mechanism for the philosophy presented in Section 3.2 of \cite{Onkar}\footnote{The discussion there applies to large classes of thermal systems, we will try to clarify what is special about black holes.}, with some refinements. One ingredient that plays a role (as we will demonstrate in detail in this chapter) is approximate factorization of correlators on typical states in the sliver. In our setting, thermal factorization emerges via the large-$N$ limit,  which is the vanishing stretched horizon limit. These observations will be crucial ingredients for us, in realizing the proposal of \cite{Onkar}. 

Before we start however, let us be clear about a few things. Our model for a light CFT single trace operator is a free scalar field in the bulk. The stretched horizon on the other hand directly captures a finite-$N$ effect. The fact that we are installing the stretched horizon as a regulator rather than deriving it from a self-contained UV complete calculation means that in a more fundamental theory, we expect the stretched horizon location and the $1/N$ suppression in multi-trace correlators to be controlled by the same (or at the least closely related) parameters. But since we are only interested in non-perturbative in $1/N$ corrections to the leading result, this will be sufficient for us.

Because we are working with single trace operators, factorization (around the vacuum and the thermal state) in itself is unsurprising. What is interesting for us is the way it emerges in our bulk calculation -- we will show that the correlators of light fields on the microstates of the sliver Hilbert space are not dependent on the microstates themselves (up to variance corrections). Instead they are thermal. This will play a role in realizing the argument of \cite{Onkar} that there is a second emergent tensor factor, a universal code subspace.

While this is interesting, this does not fully address the question of what makes black holes different from generic (say, weakly coupled) thermal systems. Thermal factorization is certainly not a property exclusive to black holes. In this paper, we will give two answers to this question. Firstly, it is worth remembering that the emergence of ``free" particles is a {\em result} of the thermodynamic (large-$N$) limit in holographic theories. Factorization is a {\em feature} of the large-$N$ limit. This is distinct from weakly coupled theories where the thermodynamic limit (say, a large volume limit) and the notion of a free particle are largely independent. 

But we will also point out a second and perhaps more important way in which the black hole system is different -- in the large-$N$ limit, the correlators in the sliver exhibit the emergence of a {\em new}  analytic continuation. This is the Kruskal analytic continuation, that appears because of the emergence of the bulk Hartle-Hawking correlator of the previous section. We show this explicitly for our BTZ HH correlator in an appendix, for completeness. Note that at any finite value of the stretched horizon, this analytic continuation is unavailable. Of course, if we wish we can simply declare that we continue the radial coordinate beyond the stretched horizon in a correlator with a finite stretched horizon. But this object has no clear meaning. The analogy here is with the trajectory of an infalling object on the surface of the Earth\footnote{We thank Samir Mathur for this analogy.}. We could simply declare that this trajectory should be analytically continued into the interior instead of being splattered on the crust -- but there is nothing particularly compelling or natural about this operation. The Hartle-Hawking correlator on the other had allows a new geometric interpretation via Kruskal coordinates, once you analytically continue\footnote{Note that the situation with the stretched horizon is slightly worse than in the surface-of-the-Earth analogy. One could imagine the analytically continued trajectories on Earth to be the trajectories of a smaller Earth of the same mass. In the stretched horizon correlator, even this interpretation is not directly available because the normal modes that go in the correlator depend on the stretched horizon location \cite{Burman}. Technically, this fact is obvious. But conceptually, we believe this is a fairly deep fact -- the black hole horizon radius is the smallest radius an object can take with a given mass.}.  Part of the reason for this, is that the HH correlator has no $\epsilon$-dependence. This emergent analytic continuation can therefore be viewed as the emergence of the interior in the large-$N$ limit. It is a {\em bulk} realization of the transition to Type III algebras noted in \cite{Leutheusser}. This is intrinsically a large-$N$ phenomenon for which we are not aware of an analogue in weakly coupled thermal systems. 

An important feature of this emergent analytic continuation is that the analytically continued correlator is of course a solution of the (linear) wave equation in the two-sided eternal black hole geometry. Factorization of multi-point correlators, together with the existence of a geometry that can reproduce the two point function via linear wave equations, is a natural reason to think of such a geometry as being emergent. This was emphasized in \cite{ElShowk}. In \cite{ElShowk}, this perspective was used to argue (see their Section 5) that the light single trace sector of a large-$N$ conformal gauge theory implies the emergence of a bulk AdS. In precisely the same sense, the black hole interior is emergent in our construction. But there is also an interesting difference in detail. In AdS/CFT, the correlators with which one starts are the correlators on the boundary, the bulk spacetime is simply postulated \cite{ElShowk} -- the virtue of doing this is that it manifests the linearity of the wave equations that determine the (non-trivial) 2-point functions. In the present case however, we started with correlators in the exterior, and the new coordinates in the interior are obtained by an analytic continuation, and so the wave equations are automatically linear.

\subsection{The Big Picture}

As we mentioned, thermal factorization plays a role in both the universality of the code subspace as well as in providing a natural ``emergent interior" picture analogous to some of the discussions in \cite{ElShowk}. Now we will outline the philosophy of \cite{Onkar} while expanding some details that play a role in our story. In the next sub-section we will present the details of the thermal factorization, which readers not interested in minutiae can skip.

The basic picture of \cite{Onkar} is that one considers a dense sliver of states around the heavy microstate and then picks a subset of suitably random states which are chosen as ``base" states. If we are working with a microcanonical sliver, if the subset is small, such a random subset will be nearly orthogonal. In our case, we can choose a set of base states from the eigenstates in the sliver (which will guarantee exact orthogonality), or we can pick a base set (as mentioned above) from the microcanonical sliver constructed from these eigenstates. We can act on these states with a not too large (so that we can ignore backreaction) number of light operators and this creates the ``small" Hilbert space around a base state. We will denote the union of these small Hilbert spaces as $\mathcal{H}$:
\bea
\mathcal{H} = \bigoplus_\alpha \mathcal{H}_\alpha \label{Hilbert}
\eea
Here $\alpha$ labels the base state. This $\mathcal{H}$ can be viewed as the Hilbert space of independent small fluctuations.

The small operators that we can act with in our case, are naturally the light single trace operators in the CFT. These are captured by our scalar field $\phi$. Small operators are constructed from products of fields and their derivatives. As long as they are not too heavy, they are candidate small fluctuations and we will generically denote them by $O_{simple}$. On general grounds from thermalization-type arguments, we expect that \cite{Onkar}
\bea
\langle \alpha' | O_{simple} | \alpha \rangle \approx \delta_{\alpha\alpha'} \langle O_{simple} \rangle \label{ETH?}
\eea
with $\langle O_{simple} \rangle$ independent of the microstates $\alpha$ and dependent only on the energy of the sliver. We are suppressing small off-diagonal corrections in the above expression. The approximate diagonal structure of \eqref{ETH?} implies that we can write a general simple operator as \cite{Onkar}
\bea
O_{simple}=\bigoplus_\alpha O_{simple, \alpha}.
\eea
But even more is true, because on the right hand side of \eqref{ETH?}, the quantity $\langle O_{simple} \rangle $ is in fact independent of $\alpha$ in the sliver. This allows us to represent the algebra of simple operators on a {\em universal} code subspace, which can be viewed as the (approximate) Hilbert space built around {\em any} base state. We will denote it by $\mathcal{H}_{univ}$ following \cite{Onkar}. In our set up, the scalars $\phi$ naturally allow such a universality, because there is a notion of typicality in the sliver states we constructed. The correlators we computed (even at finite $\epsilon$) are indistinguishable from each other on typical states, up to suppressed variance corrections.

Following the notation of \cite{Onkar}, we  have denoted a base state in our discussion by the label $\alpha$. The action by light operators generates a Fock space and we label the states in it by $|\alpha, i_\alpha \rangle$, where $i_\alpha$ captures a level information. But the universality implies that the level structure is universal (up to subleading corrections) for any $\alpha$. So we are effectively working with $|\alpha \rangle \otimes |i\rangle$. In other words $\alpha$ is in effect a superselection sector label, and the light operators are acting on the $i$-label. This means that $O_{simple}$ acts trivially (as an identity operator) on the $\alpha$-label -- it keeps track of fine-grained information. We can write this as \cite{Onkar}
\bea
\mathcal{H} \approx \mathcal{H}_{SS} \otimes \mathcal{H}_{univ},
\eea
with $\mathcal{O}_{simple}= \mathcal{I} \otimes \mathcal{O}_{univ}$ where $\mathcal{I}$ stands for the identity operator. Since excited states are obtained by the repeated action of simple operators, demonstrating the above facts in our setting amounts to demonstrating universal factorization on typical states. This is intuitive, and we will demonstrate it explicitly in the next section. In the large-$N$ limit, this manifests itself as thermal factorization of HH correlators. 

In effect, the entire structure suggested in \cite{Onkar} has a straightforward realization in our stretched horizon sliver. But our set up also fixes some of the specifics, concretely. We will typically take the eigenstates we defined earlier as the base states. The light (scalar) operators and their low derivatives will take the place of the light (simple) operators. They have a universal action on any base state, and the leading result is thermal\footnote{The one point functions are zero in our setting, but the statement applies for two point functions.}, it is the Hartle-Hawking correlator. But to completely establish the parallel, we also need to show that multi-point correlators are also independent of the choice of base state -- acting with multiple operators is the analogue of working at a higher level $i$ on a given base state labelled by $\alpha$. We will accomplish this by demonstrating factorization on typical states, in detail, in the next section. The universality of factorization will indicate the independence on $\alpha$ and the universality of the code subspace. We work with 4-pt functions for concreteness, but the structures are general.

\subsection{Factorization}

We will consider the 4-point function on a heavy state as a representative example of a multi-point correlator:
\bea\label{fourpointfunctioninSH}
\langle N|\Phi(r_{1},t_{1},\psi_{1})\Phi(r_{2},t_{2},\psi_{2})\Phi(r_{3},t_{3},\psi_{3})\Phi(r_{4},t_{4},\psi_{4})|N\rangle
\eea
The heavy ket $| N\rangle$ and the field $\Phi(r,t,\psi)$ are as we defined in the previous sections. Introducing $X_{i}\equiv\{r_{i},t_{i},\psi_{i}\}$ as a convenient short-hand, we find
\begin{align}
&\langle N|\Phi(X_{1})\Phi(X_{2})\Phi(X_{3})\Phi(X_{4})|N\rangle \equiv G_4 =\nonumber\\
&=\hspace{-0.2cm}\sum_{\{n\},\{J\}}K\langle N|\Big[b_{n_{1}J_{1}}\mathcal{U}_{n_{1}J_{1}}(X_{1})
+b_{n_{1}J_{1}}^{\dagger}\mathcal{U}_{n_{1}J_{1}}^{*}(X_{1})\Big]
\Big[b_{n_{2}J_{2}}\mathcal{U}_{n_{2}J_{2}}(X_{2})+ b_{n_{2}J_{2}}^{\dagger}\mathcal{U}_{n_{2}J_{2}}^{*}(X_{2})\Big]\times\nonumber\\
&\times\Big[b_{n_{3}J_{3}}\mathcal{U}_{n_{3}J_{3}}(X_{3})+ b_{n_{3}J_{3}}^{\dagger}\mathcal{U}_{n_{3}J_{3}}^{*}(X_{3})\Big]
\Big[b_{n_{4}J_{4}}\mathcal{U}_{n_{4}J_{4}}(X_{4})+ b_{n_{4}J_{4}}^{\dagger}\mathcal{U}_{n_{4}J_{4}}^{*}(X_{4})\Big]|N\rangle 
\end{align}
where we have defined $K\equiv \frac{1}{(4\pi)^{2}\sqrt{\prod_{i=1}^{4}\omega_{n_{i}J{i}}}}$ to save space. There are six non-zero terms in the above expression.
\begin{align}
&\hspace{-2.3cm}G_4=\sum_{\{n\}\{J\}}K\langle N|\Big[b_{n_{1}J_{1}}b_{n_{2}J_{2}}b_{n_{3}J_{3}}^{\dagger}b_{n_{4}J_{4}}^{\dagger}\mathcal{U}_{n_{1}J_{1}}(X_{1})\mathcal{U}_{n_{2}J_{2}}(X_{2})\mathcal{U}_{n_{3}J_{3}}^{*}(X_{3})\mathcal{U}_{n_{4}J_{4}}^{*}(X_{4}) + \nonumber \\
&+b_{n_{1}J_{1}}b_{n_{2}J_{2}}^{\dagger}b_{n_{3}J_{3}}b_{n_{4}J_{4}}^{\dagger}\mathcal{U}_{n_{1}J_{1}}(X_{1})\mathcal{U}_{n_{2}J_{2}}^{*}(X_{2})\mathcal{U}_{n_{3}J_{3}}(X_{3})\mathcal{U}_{n_{4}J_{4}}^{*}(X_{4}) + \nonumber \\
&+b_{n_{1}J_{1}}b_{n_{2}J_{2}}^{\dagger}b_{n_{3}J_{3}}^{\dagger}b_{n_{4}J_{4}}\mathcal{U}_{n_{1}J_{1}}(X_{1})\mathcal{U}_{n_{2}J_{2}}^{*}(X_{2})\mathcal{U}_{n_{3}J_{3}}^{*}(X_{3})\mathcal{U}_{n_{4}J_{4}}(X_{4}) + \nonumber \\
&+b_{n_{1}J_{1}}^{\dagger}b_{n_{2}J_{2}}b_{n_{3}J_{3}}b_{n_{4}J_{4}}^{\dagger}\mathcal{U}_{n_{1}J_{1}}^{*}(X_{1})\mathcal{U}_{n_{2}J_{2}}(X_{2})\mathcal{U}_{n_{3}J_{3}}(X_{3})\mathcal{U}_{n_{4}J_{4}}^{*}(X_{4}) + \nonumber \\
&+b_{n_{1}J_{1}}^{\dagger}b_{n_{2}J_{2}}b_{n_{3}J_{3}}^{\dagger}b_{n_{4}J_{4}}\mathcal{U}_{n_{1}J_{1}}^{*}(X_{1})\mathcal{U}_{n_{2}J_{2}}(X_{2})\mathcal{U}_{n_{3}J_{3}}^{*}(X_{3})\mathcal{U}_{n_{4}J_{4}}(X_{4}) + \nonumber \\
&+b_{n_{1}J_{1}}^{\dagger}b_{n_{2}J_{2}}^{\dagger}b_{n_{3}J_{3}}b_{n_{4}J_{4}}\mathcal{U}_{n_{1}J_{1}}^{*}(X_{1})\mathcal{U}_{n_{2}J_{2}}^{*}(X_{2})\mathcal{U}_{n_{3}J_{3}}(X_{3})\mathcal{U}_{n_{4}J_{4}}(X_{4}) \Big]|N\rangle \label{G4}
\end{align}
There are two ways in which each term in the above expression can be non-zero. This implies that each term is a sum of two possibilities. We evaluate these quantities in Appendix \ref{AppD} systematically.  Assembling all the terms from Appendix \ref{AppD}, we get
\begin{align}
&\langle N|\Phi(X_{1})\Phi(X_{2})\Phi(X_{3})\Phi(X_{4})|N\rangle =\nonumber\\ &=\sum_{n_{1},n_{2},J_{1},J_{2}}K(N_{n_{1}J_{1}}+1)(N_{n_{2}J_{2}}+1)\mathcal{U}_{n_{1}J_{1}}(X_{1})\mathcal{U}_{n_{2}J_{2}}(X_{2})\mathcal{U}_{n_{2}J_{2}}^{*}(X_{3})\mathcal{U}_{n_{1}J_{1}}^{*}(X_{4})
+ \nonumber\\
&+\sum_{n_{1},n_{2},J_{1},J_{2}}K(N_{n_{1}J_{1}}+1)(N_{n_{2}J_{2}}+1)\mathcal{U}_{n_{1}J_{1}}(X_{1})\mathcal{U}_{n_{2}J_{2}}(X_{2})\mathcal{U}_{n_{1}J_{1}}^{*}(X_{3})\mathcal{U}_{n_{2}J_{2}}^{*}(X_{4}) + \nonumber\\
&+\sum_{n_{1},n_{2},J_{1},J_{2}}K(N_{n_{1}J_{1}}+1)(N_{n_{2}J_{2}})\mathcal{U}_{n_{1}J_{1}}(X_{1})\mathcal{U}_{n_{2}J_{2}}^{*}(X_{2})\mathcal{U}_{n_{2}J_{2}}(X_{3})\mathcal{U}_{n_{1}J_{1}}^{*}(X_{4}) + \nonumber\\
&+ \sum_{n_{1},n_{2},J_{1},J_{2}}K(N_{n_{1}J_{1}}+1)(N_{n_{2}J_{2}}+1)\mathcal{U}_{n_{1}J_{1}}(X_{1})\mathcal{U}_{n_{1}J_{1}}^{*}(X_{2})\mathcal{U}_{n_{2}J_{2}}(X_{3})\mathcal{U}_{n_{2}J_{2}}^{*}(X_{4}) + \nonumber\\
&+\sum_{n_{1},n_{2},J_{1},J_{2}}K(N_{n_{1}J_{1}}+1)(N_{n_{2}J_{2}})\mathcal{U}_{n_{1}J_{1}}(X_{1})\mathcal{U}_{n_{2}J_{2}}^{*}(X_{2})\mathcal{U}_{n_{1}J_{1}}^{*}(X_{3})\mathcal{U}_{n_{2}J_{2}}(X_{4}) + \nonumber\\
&+ \sum_{n_{1},n_{2},J_{1},J_{2}}K(N_{n_{1}J_{1}}+1)(N_{n_{2}J_{2}})\mathcal{U}_{n_{1}J_{1}}(X_{1})\mathcal{U}_{n_{1}J_{1}}^{*}(X_{2})\mathcal{U}_{n_{2}J_{2}}^{*}(X_{3})\mathcal{U}_{n_{2}J_{2}}(X_{4}) + \nonumber\\
&+\sum_{n_{1},n_{2},J_{1},J_{2}}K(N_{n_{1}J_{1}})(N_{n_{2}J_{2}}+1)\mathcal{U}_{n_{1}J_{1}}^{*}(X_{1})\mathcal{U}_{n_{2}J_{2}}(X_{2})\mathcal{U}_{n_{1}J_{1}}(X_{3})\mathcal{U}_{n_{2}J_{2}}^{*}(X_{4}) + \nonumber\\
&+ \sum_{n_{1},n_{2},J_{1},J_{2}}K(N_{n_{1}J_{1}})(N_{n_{2}J_{2}}+1)\mathcal{U}_{n_{1}J_{1}}^{*}(X_{1})\mathcal{U}_{n_{1}J_{1}}(X_{2})\mathcal{U}_{n_{2}J_{2}}(X_{3})\mathcal{U}_{n_{2}J_{2}}^{*}(X_{4}) + \nonumber\\
&+\sum_{n_{1},n_{2},J_{1},J_{2}}K(N_{n_{1}J_{1}})(N_{n_{2}J_{2}}+1)\mathcal{U}_{n_{1}J_{1}}^{*}(X_{1})\mathcal{U}_{n_{2}J_{2}}(X_{2})\mathcal{U}_{n_{2}J_{2}}^{*}(X_{3})\mathcal{U}_{n_{1}J_{1}}(X_{4}) + \nonumber\\
&+  \sum_{n_{1},n_{2},J_{1},J_{2}}K(N_{n_{1}J_{1}})(N_{n_{2}J_{2}})\mathcal{U}_{n_{1}J_{1}}^{*}(X_{1})\mathcal{U}_{n_{1}J_{1}}(X_{2})\mathcal{U}_{n_{2}J_{2}}^{*}(X_{3})\mathcal{U}_{n_{2}J_{2}}(X_{4}) + \nonumber\\
&+\sum_{n_{1},n_{2},J_{1},J_{2}}K(N_{n_{1}J_{1}})(N_{n_{2}J_{2}})\mathcal{U}_{n_{1}J_{1}}^{*}(X_{1})\mathcal{U}_{n_{2}J_{2}}^{*}(X_{2})\mathcal{U}_{n_{2}J_{2}}(X_{3})\mathcal{U}_{n_{1}J_{1}}(X_{4}) + \nonumber\\
&+ \sum_{n_{1},n_{2},J_{1},J_{2}}K(N_{n_{1}J_{1}})(N_{n_{2}J_{2}})\mathcal{U}_{n_{1}J_{1}}^{*}(X_{1})\mathcal{U}_{n_{2}J_{2}}^{*}(X_{2})\mathcal{U}_{n_{1}J_{1}}(X_{3})\mathcal{U}_{n_{2}J_{2}}(X_{4})
\end{align}

The goal now is to show that in a typical state, these terms can be re-arranged into a factorized form. The factorized form of the 4-pt function contains three pieces. We will illustrate the emergence of one of these pieces below by considering the sum of the $1$-st, $11$-th, $3$-th and $9$-th terms above. The rest proceed in a similar fashion. 

We first add $1$-st and $11$-th terms:
\begin{align}
&\hspace{-1cm}=\sum_{n_{1},n_{2},J_{1},J_{2}}K\Big[(N_{n_{1}J_{1}}+1)(N_{n_{2}J_{2}}+1)\mathcal{U}_{n_{1}J_{1}}(X_{1})\mathcal{U}_{n_{2}J_{2}}(X_{2})\mathcal{U}_{n_{2}J_{2}}^{*}(X_{3})\mathcal{U}_{n_{1}J_{1}}^{*}(X_{4}) + \nonumber\\
&\hspace{3.5cm}+(N_{n_{1}J_{1}})(N_{n_{2}J_{2}})\mathcal{U}_{n_{1}J_{1}}^{*}(X_{1})\mathcal{U}_{n_{2}J_{2}}^{*}(X_{2})\mathcal{U}_{n_{2}J_{2}}(X_{3})\mathcal{U}_{n_{1}J_{1}}(X_{4})\Big]
\end{align}
Now we do the replacement $N_{n,J} = \langle N_{n,J}\rangle = \frac{1}{e^{\beta\omega_{n,J}}-1}$. As argued in \cite{Mukund, Burman}, this replacement is valid up to suppressed variance corrections. The result is,
\begin{align}
&=\sum_{n_{1},n_{2}=0}^{\infty}\sum_{J_{1},J_{2}=-J_{cut}}^{+J_{cut}}\frac{1}{(4\pi)^{2}\omega_{n_{1}J_{1}}\omega_{n_{2}J_{2}}}\Bigg[\frac{e^{\beta\omega_{n_{1}J_{1}}}}{e^{\beta\omega_{n_{1}J_{1}}}-1}\frac{e^{\beta\omega_{n_{2}J_{2}}}}{e^{\beta\omega_{n_{2}J_{2}}}-1}\mathcal{U}_{n_{1}J_{1}}(X_{1})\mathcal{U}_{n_{2}J_{2}}(X_{2})\times\nonumber\\
&\times\mathcal{U}_{n_{2}J_{2}}^{*}(X_{3})\mathcal{U}_{n_{1}J_{1}}^{*}(X_{4}) 
+\frac{1}{e^{\beta\omega_{n_{1}J_{1}}}-1}\frac{1}{e^{\beta\omega_{n_{2}J_{2}}}-1}
\times\mathcal{U}_{n_{1}J_{1}}^{*}(X_{1})\mathcal{U}_{n_{2}J_{2}}^{*}(X_{2})\mathcal{U}_{n_{2}J_{2}}(X_{3})\mathcal{U}_{n_{1}J_{1}}(X_{4}) \Bigg]
\end{align}
In the second term inside the square bracket, we can do the replacement $n_{1}\rightarrow-n_{1},n_{2}\rightarrow-n_{2}$ and $J_{1}\rightarrow-J_{1},J_{2}\rightarrow-J_{2}.$ The expression now becomes,
\begin{align}\label{1}
&\hspace{-1cm}=\Big[\sum_{n_{1}=0}^{\infty}\sum_{n_{2}=0}^{\infty}+\sum_{n_{1}=-\infty}^{0}\sum_{n_{2}=-\infty}^{0}\Big]\sum_{J_{1},J_{2}=-J_{cut}}^{+J_{cut}}\frac{1}{(4\pi)^{2}\omega_{n_{1}J_{1}}\omega_{n_{2}J_{2}}}\frac{e^{\beta\omega_{n_{1}J_{1}}}}{e^{\beta\omega_{n_{1}J_{1}}}-1}\frac{e^{\beta\omega_{n_{2}J_{2}}}}{e^{\beta\omega_{n_{2}J_{2}}}-1}\times\nonumber\\
&\hspace{6cm}\times\mathcal{U}_{n_{1}J_{1}}(X_{1})\mathcal{U}_{n_{2}J_{2}}(X_{2})\mathcal{U}_{n_{2}J_{2}}^{*}(X_{3})\mathcal{U}_{n_{1}J_{1}}^{*}(X_{4})
\end{align}
Now we turn our attention to the sum of the $3$-rd and $9$-th terms:
\begin{align}
&\hspace{-0.5cm}=\sum_{n_{1}n_{2}=0}^{\infty}\sum_{J_{1},J_{2}=-J_{cut}}^{+J_{cut}}\frac{1}{(4\pi)^{2}\omega_{n_{1}J_{1}}\omega_{n_{2}J_{2}}}\Bigg[(N_{n_{1}J_{1}})(N_{n_{2}J_{2}}+1)\mathcal{U}_{n_{1}J_{1}}^{*}(X_{1})\mathcal{U}_{n_{2}J_{2}}(X_{2})\mathcal{U}_{n_{2}J_{2}}^{*}(X_{3})\times\nonumber\\
&\hspace{1cm}\times\mathcal{U}_{n_{1}J_{1}}(X_{4}) + 
(N_{n_{1}J_{1}}+1)(N_{n_{2}J_{2}})\mathcal{U}_{n_{1}J_{1}}(X_{1})\mathcal{U}_{n_{2}J_{2}}^{*}(X_{2})\mathcal{U}_{n_{2}J_{2}}(X_{3})\mathcal{U}_{n_{1}J_{1}}^{*}(X_{4})\Bigg]
\end{align}
Setting $N_{n,J}=\langle N_{n,J}\rangle = \frac{1}{e^{\beta\omega_{n,J}}-1}$ we find
\begin{align} &=\sum_{n_{1},n_{2}=0}^{\infty}\sum_{J_{1},J_{2}=-J_{cut}}^{+J_{cut}}\frac{1}{(4\pi)^{2}\omega_{n_{1},J_{1}}\omega_{n_{2},J_{2}}}\Bigg[\frac{1}{e^{\beta\omega_{n_{1}J_{1}}}-1}\frac{e^{\beta\omega_{n_{2}J_{2}}}}{e^{\beta\omega_{n_{2}J_{2}}}-1}\mathcal{U}_{n_{1}J_{1}}^{*}(X_{1})\mathcal{U}_{n_{2}J_{2}}(X_{2}) \times\nonumber\\ &\times\mathcal{U}_{n_{2}J_{2}}^{*}(X_{3})\mathcal{U}_{n_{1}J_{1}}(X_{4})+\frac{e^{\beta\omega_{n_{1}J_{1}}}}{e^{\beta\omega_{n_{1}J_{1}}}-1}\frac{1}{e^{\beta\omega_{n_{2}J_{2}}}-1}\mathcal{U}_{n_{1}J_{1}}(X_{1})\mathcal{U}_{n_{2}J_{2}}^{*}(X_{2})\mathcal{U}_{n_{2}J_{2}}(X_{3})\mathcal{U}_{n_{1}J_{1}}^{*}(X_{4}) \Bigg] \end{align}
Upon making the replacement $n_{1}\rightarrow-n_{1},J_{1}\rightarrow-J_{1}$ in the first bracket and $n_{2}\rightarrow-n_{2}, J_{2}\rightarrow-J_{2}$ in the second bracket, this reduces to
\begin{align}\label{2}
&=\Big[\sum_{n_{1}=-\infty}^{0}\sum_{n_{2}=0}^{+\infty} + \sum_{n_{1}=0}^{+\infty}\sum_{n_{2}=-\infty}^{0}\Big]\sum_{J_{1},J_{2}=-J_{cut}}^{+J_{cut}}\frac{1}{(4\pi)^{2}\omega_{n_{1}J_{1}}\omega_{n_{2}J_{2}}}\frac{e^{\beta\omega_{n_{1}J_{1}}}}{e^{\beta\omega_{n_{1}J_{1}}}-1}\frac{e^{\beta\omega_{n_{2}J_{2}}}}{e^{\beta\omega_{n_{2}J_{2}}}-1}\times\nonumber\\
&\hspace{6cm}\times\mathcal{U}_{n_{1}J_{1}}(X_{1})\mathcal{U}_{n_{2}J_{2}}(X_{2})\mathcal{U}_{n_{2}J_{2}}^{*}(X_{3})\mathcal{U}_{n_{1}J_{1}}^{*}(X_{4}) 
\end{align}
Now we add \eqref{1} and \eqref{2}
\begin{align}
&\hspace{-0.5cm}= \sum_{n_{1}=-\infty}^{+\infty}\sum_{J_{1}=-J_{cut}}^{+J_{cut}}\Bigg(\frac{1}{4\pi\omega_{n_{1}J_{1}}}\frac{e^{\beta\omega_{n_{1}J_{1}}}}{e^{\beta\omega_{n_{1}J_{1}}}-1}\mathcal{U}_{n_{1}J_{1}}(X_{1})\mathcal{U}_{n_{1}J_{1}}^{*}(X_{4})\Bigg)\times\nonumber\\
&\hspace{4cm}\times \sum_{n_{2}=-\infty}^{+\infty}\sum_{J_{2}=-J_{cut}}^{+J_{cut}}\Bigg(\frac{1}{4\pi\omega_{n_{2}J_{2}}}\frac{e^{\beta\omega_{n_{2}J_{2}}}}{e^{\beta\omega_{n_{2}J_{2}}}-1}\mathcal{U}_{n_{2}J_{2}}(X_{2})\mathcal{U}_{n_{2}J_{2}}^{*}(X_{3})\Bigg)
\end{align}
This is precisely the product of two 2-point functions. Note that this result holds at finite-$\epsilon$ and is valid up to suppressed variance corrections. An entirely analogous argument can be used to write down the other two terms in the factorized result.

Now we consider the limit $\epsilon\rightarrow 0$. The summation over $\omega$ becomes an integration as before and $J_{cut}\rightarrow\infty$. The expression in the above equation becomes the product of two thermal Hartle-Hawking correlators. Explicitly, we have
\begin{align}
&1+11+3+9={}_{HH}\langle 0|\Phi(r_{1},t_{1},\psi_{1})\Phi(r_{4},t_{4},\psi_{4})|0\rangle_{HH}\langle 0|\Phi(r_{2},t_{2},\psi_{2})\Phi(r_{3},t_{3},\psi_{3})|0\rangle_{HH}
\end{align}
We have written the relevant numbers of the terms that we are summing over, on the left hand side.
Similarly, if we add the $5$-th, $7$-th, $2$nd and $12$-th terms and take the limit $\epsilon\rightarrow 0$, we get the second thermally factorized piece:
\begin{align}
&5+7+2+12 = {}_{HH}\langle 0|\Phi(r_{1},t_{1},\psi_{1})\Phi(r_{3},t_{3},\psi_{3})|0\rangle_{HH}
\langle 0|\Phi(r_{2},t_{2},\psi_{2})\Phi(r_{4},t_{4},\psi_{4})|0\rangle_{HH}
\end{align}
and if we add the $4$-th, $10$-th, $6$-th and $8$-th terms we get the third thermally factorized term:
\begin{align}
&\hspace{-1cm}4+10+6+8 = \nonumber\\
&{}_{HH}\langle 0|\Phi(r_{1},t_{1},\psi_{1})\Phi(r_{2},t_{2},\psi_{2})|0\rangle_{HH}
\langle 0|\Phi(r_{3},t_{3},\psi_{3})\Phi(r_{4},t_{4},\psi_{4})|0\rangle_{HH}
\end{align}
Overall in the limit $\epsilon\rightarrow 0$ the final expression of the four-point function becomes 
\begin{align}
&\langle N|\Phi(r_{1},t_{1},\psi_{1})\Phi(r_{2},t_{2},\psi_{2})\Phi(r_{3},t_{3},\psi_{3})\Phi(r_{4},t_{4},\psi_{4})|N\rangle
\xrightarrow{\epsilon\rightarrow 0} \nonumber\\
&{}_{HH}\langle 0|\Phi(r_{1},t_{1},\psi_{1})\Phi(r_{2},t_{2},\psi_{2})|0\rangle_{HH}
\langle 0|\Phi(r_{3},t_{3},\psi_{3})\Phi(r_{4},t_{4},\psi_{4})|0\rangle_{HH} 
+ \nonumber\\
&+{}_{HH}\langle 0|\Phi(r_{1},t_{1},\psi_{1})\Phi(r_{3},t_{3},\psi_{3})|0\rangle_{HH}
\langle 0|\Phi(r_{2},t_{2},\psi_{2})\Phi(r_{4},t_{4},\psi_{4})|0\rangle_{HH} +\nonumber\\
&+{}_{HH}\langle 0|\Phi(r_{1},t_{1},\psi_{1})\Phi(r_{4},t_{4},\psi_{4})|0\rangle_{HH}
\langle 0|\Phi(r_{2},t_{2},\psi_{2})\Phi(r_{3},t_{3},\psi_{3})|0\rangle_{HH}
\end{align}

This final result is just the 4-pt Hartle-Hawking correlator -- the latter factorizes into a sum of products of 2-pt functions as we demonstrate in Appendix \ref{FactorHH}. What is worthy of note here is firstly, the fact that there indeed exists operators that qualify as ``simple" operators in the sense of \cite{Onkar}. These operators (and our microstate Hilbert space) satisfy all of the requirements suggested in \cite{Onkar} for the emergence of an interior code subspace. The second noteworthy thing is that in the strict large-$N$ (small-$\epsilon$) limit, the correlators (being Hartle-Hawking) exhibit a new analytic continuation into the interior. 

\section{Page Windows}

In this paper, our focus is on fairly conventional correlators. We will be interested in probing the system at some time $t=0$ and reading off the responses at later times\footnote{It will be interesting to study out-of-time-ordered correlators to probe chaos and scrambling, but we will not undertake it here.}. As mentioned earlier, one of our results is that correlators which are separated by more than a Page time, are indistinguishable from noise. We call such a window of time, a Page window. Our calculations make us suspect that notions of smoothness that make sense after a Page window, cannot be based on probe fields (and notions like infalling boundary conditions).

For multi-point correlators, this statement needs a bit of refinement, because of factorization. Depending on the temporal distance between any two of the insertions, we will have various cases where pieces of the correlator are indistinguishable from variance noise. 
These cases correspond to whether the time difference between two inserted operators in the multi-point function is greater than or less than the page time ($t_{p}$). 

Let us consider the 4-point function for illustration. We denote the stretched horizon 4-point function to be $C_{1234}$, where the indices are supposed to capture the times associated to the insertions. Here the ordering of $``1234"$ is important because we only consider\footnote{Our clock begins at the insertion of the first operator and therefore, one can think of $t_1$ as being the same as the zero of the clock.} $t_{1}<t_{2}<t_{3}<t_{4}.$  Similarly, we denote the stretched horizon two-point function to be $C_{ij}.$ Now, we can write the factorized form of this four-point function to be 
\bea\label{C1234}
C_{1234}= C_{12}C_{34}+C_{13}C_{24}+C_{14}C_{23} \label{mnemo}
\eea
and this result is valid up to exponentially suppressed corrections. When the separation between insertions becomes greater than the Page time, this noise will become dominant. 
To understand the behavior of the above form \eqref{C1234}  for different cases of time separation, we write the above expression suggestively as
\bea\label{C1234new}
C_{1234}= C_{12}C_{34}+C_{12+23}C_{23+34}+C_{12+23+34}C_{23}
\eea
by exploiting the time-translation invariance of the background. Different cases arise when considering the time difference between adjacent 
operators in the four-point function $C_{1234}.$ Because the times are sequential, the cases can be organized by considering only the time differences -- these are the three numbers $t_{12},t_{23},t_{34}$. In this simple setting, we can provide an exhaustive set of behaviors for the 4-point function by enumerating the various cases. Let us first consider one of the simplest cases.

\noindent\textbf{Case 1:
$\hspace{0.5cm}t_{12}>t_{p},\hspace{0.5cm}t_{23}>t_{p}\hspace{0.5cm}t_{34}>t_{p}$}

\noindent
In this case, the contributions of $C_{12}$ and $C_{34}$ are clearly $\mathcal{O}(e^{-S/2})$ because $t_{12}$ and $t_{34}$ are both bigger than $t_p$. So the first term is exponentially suppressed. The second term in \eqref{C1234} is $C_{13}C_{24}$. The term $C_{13}$ contains the time difference $t_{13}$ which can be written as $t_{12}+t_{23}$. Since, $t_{12}>t_{p}$ it is clear that when we add $t_{23}$ (which is a positive quantity) to $t_{12}$ the resulting quantity will still be $>t_{p}.$ So $t_{13}>t_{p}$. We can likewise conclude that $t_{23}+t_{34}>t_{p}$, and therefore the second term $C_{13}C_{24}$ is again exponentially suppressed in the entropy of the black hole. Similarly, we can argue that the third term in \eqref{C1234} is again exponentially suppressed. So in Case 1, the stretched horizon four-point function $C_{1234}$ in \eqref{C1234} has the behavior
\bea
C_{1234}\sim \mathcal{O}(e^{-S/2})
\eea
\textbf{Case 2: $\hspace{0.5cm}t_{12}<t_{p},\hspace{0.5cm}t_{23}>t_{p}\hspace{0.5cm}t_{34}>t_{p}$}

\noindent
Again, in this case, we can see from the behavior of the individual terms in \eqref{C1234} that $C_{1234}$ is exponentially suppressed in the entropy of the black hole.
\bea
C_{1234}\sim \mathcal{O}(e^{-S/2})
\eea
These cases where the overall behavior of the four-point function is exponentially suppressed are trivial. Now, we will quickly list all the trivial cases and then focus on the non-trivial cases.\\
\begin{equation*}
\begin{rcases}
&\textbf{Case 3:$\hspace{0.5cm}t_{12}>t_{p},\hspace{0.5cm}t_{23}<t_{p}\hspace{0.5cm}t_{34}>t_{p}$}\\
&\textbf{Case 4:$\hspace{0.5cm}t_{12}>t_{p},\hspace{0.5cm}t_{23}>t_{p}\hspace{0.5cm}t_{34}<t_{p}$}\\
&\textbf{Case 5:$\hspace{0.5cm}t_{12}<t_{p},\hspace{0.5cm}t_{23}<t_{p}\hspace{0.5cm}t_{34}>t_{p}$}\\
&\textbf{Case 6:$\hspace{0.5cm}t_{12}>t_{p},\hspace{0.5cm}t_{23}<t_{p}\hspace{0.5cm}t_{34}<t_{p}$}\\
\end{rcases}\text{$C_{1234}\sim \mathcal{O}(e^{-S/2})$}
\end{equation*}\\

\noindent\textbf{Case 7:$\hspace{0.5cm}t_{12}<t_{p},\hspace{0.5cm}t_{23}>t_{p},\hspace{0.5cm}t_{34}<t_{p}$}

\noindent
Doing the same analysis as we did above, we can see that the four-point function behaves as 
\bea
C_{1234}\sim C_{12}C_{34}+\mathcal{O}(e^{-S/2})
\eea
The first term $C_{12}C_{34}$ is an $\mathcal{O}(1)$ quantity.\\

\noindent\textbf{Case 8:$\hspace{0.5cm}t_{12}<t_{p},\hspace{0.5cm}t_{23}<t_{p},\hspace{0.5cm}t_{34}<t_{p}$}

\noindent
This is an interesting case because it has sub-cases. Since $t_{12}<t_{p},t_{23}<t_{p},t_{34}<t_{p}$, it seems that all the time intervals are below the page time. But this is not the case. Even though $t_{12}<t_{p},t_{23}<t_{p},t_{34}<t_{p}$, $t_{13}\equiv t_{12}+t_{23}$ can be $>t_{p}$ or $< t_{p}$ , $t_{24}\equiv t_{23}+t_{34}$ can be $<t_{p}$ or $>t_{p}$ and $t_{14}\equiv t_{12}+t_{23}+t_{34}$ can be $>t_{p}$ or $<t_{p}.$ We consider these sub-cases one by one below.

\noindent\textbf{Sub-case 1:$\hspace{0.5cm}t_{12}+t_{23}<t_{p},\hspace{0.5cm}t_{23}+t_{34}>t_{p},\hspace{0.5cm}t_{12}+t_{23}+t_{34}>t_{p}$}
\bea
C_{1234}\sim C_{12}C_{34}+\mathcal{O}(e^{-S/2})
\eea
\noindent\textbf{Sub-case 2:$\hspace{0.5cm}t_{12}+t_{23}>t_{p},\hspace{0.5cm}t_{23}+t_{34}<t_{p},\hspace{0.5cm}t_{12}+t_{23}+t_{34}>t_{p}$}
\bea
C_{1234}\sim C_{12}C_{34}+\mathcal{O}(e^{-S/2})
\eea
\noindent\textbf{Sub-case 3:$\hspace{0.5cm}t_{12}+t_{23}>t_{p},\hspace{0.5cm}t_{23}+t_{34}>t_{p},\hspace{0.5cm}t_{12}+t_{23}+t_{34}>t_{p}$}
\bea
C_{1234}\sim C_{12}C_{34}+\mathcal{O}(e^{-S/2})
\eea
\noindent\textbf{Sub-case 4:$\hspace{0.5cm}t_{12}+t_{23}<t_{p},\hspace{0.5cm}t_{23}+t_{34}<t_{p},\hspace{0.5cm}t_{12}+t_{23}+t_{34}<t_{p}$}
\bea
C_{1234}\sim C_{12}C_{34}+C_{13}C_{24}+C_{14}C_{23}
\eea
\noindent\textbf{Sub-case 5:$\hspace{0.5cm}t_{12}+t_{23}<t_{p},\hspace{0.5cm}t_{23}+t_{34}<t_{p},\hspace{0.5cm}t_{12}+t_{23}+t_{34}>t_{p}$}
\bea
C_{1234}\sim C_{12}C_{34}+C_{13}C_{24}+\mathcal{O}(e^{-S/2})
\eea
In all the above expressions, $C_{ij}$'s are $\mathcal{O}(1)$ quantities. Of course, if all the relevant time windows are smaller than $t_p$, then all quantities on the right hand side of \eqref{mnemo} are $\mathcal{O}(1)$ quantities.

The goal of the simple exercise above was to emphasize that a probe correlator seems to get unavoidably mixed with the microstate of the black hole, starting about a Page time. This is physical and is expected in conventional quantum mechanical systems -- the absence of this feature in (eternally) smooth horizon geometries was the origin of Maldacena's information paradox. As we will discuss in the next section and also in Appendix \ref{State-flip}, there is some evidence to suggest that the black hole interior (if left undisturbed) may grow even after the Page time for an exponentially long time. This is attributed to the fact that the circuit complexity\footnote{See eg. \cite{Chapman, Jefferson, Rifath} for an introductory discussion of circuit complexity in a field theory context, as well as references.} of a (perturbed) dual CFT state  can increase for an exponentially long time \cite{action, Action?}. The existence and growth of the interior is sometimes viewed as an indicator of smoothness of the black hole horizon. We will argue that the two notions are conceptually distinct -- the infalling notion of smoothness is naturally associated to a probe up to a Page time after its insertion, whereas complexity is a property of the evolving microstate even long after the Page time. 

Perturbing or inserting a light operator reduces the complexity of a state, so that we can effectively view ourselves as being at the $t=0$ slice of the Penrose diagram \cite{SusskindMP}. Time evolution afterwards results in increase in complexity. But correlations die out only until the Page time, when they become indistinguishable from variance noise. After the Page time, the complexity keeps increasing (unless another perturbation acts on the black hole) but the infalling boundary condition and the associated exponential decay of correlations is absent. We will comment more on this in the next section.

\section{Dialogue Concerning the Two Chief Black Hole Systems}

We have proposed a ``bottom-up" model for black hole microstates in this paper and \cite{Burman, Pradipta}. It is bottom-up, because we have introduced the stretched/quantum horizon location as a free parameter instead of deriving it from a UV-complete description -- it is a UV regulator that captures the finiteness of $N$. The claim is that black hole microstates are modeled by highly excited states of a quantum field theory with a stretched horizon. It is crucial that we are not simply using the stretched horizon {\em vacuum} state as a model for the black hole. That idea (while containing hints of promising features) has multiple drawbacks. Perhaps the most important ones are that (a) there is no explanation for the entropy or the thermodynamics in this approach, and (b) that there is a divergent negative stress energy near the horizon. Working with a highly excited state at the mass of the black hole removes both these problems in one fell swoop. It also naturally ties in with the intuition that a large AdS black hole is in equilibrium with its own Hawking radiation. We view this as a bulk Hilbert space implementation (within AdS-CFT) of the Israel-Mukohyama suggestion that one should dress the brick-wall with a radiation fluid.

By appropriately fixing the one-parameter freedom, we could reproduce both the black hole entropy and temperature \cite{Burman,Pradipta} including the precise coefficients. This improved upon the old calculation of 't Hooft both conceptually and technically.  At the conceptual level the improvement is that we are not working with the semi-classical modes trapped behind the angular momentum barrier (which are unaware of tunneling). We are instead working with the genuine normal modes of the system. It is the approximate degeneracy in the angular quantum numbers that is directly responsible for the area-scaling of entropy (instead of the volume scaling that one expects from a Planckian black body calculation). The technical improvement was that we computed the normal modes explicitly \cite{Pradipta}. Because his calculation was more indirect, 't Hooft needed both the temperature and mass to be specified before he could compute the entropy. The explicit knowledge of normal modes allowed us to specify only the mass (or temperature) to determine the ensemble, and reproduce the other thermodynamic quantities via a conventional statistical mechanics calculation.

Along with these thermodynamic results, it was also demonstrated in \cite{Burman} that the boundary correlator associated to a typical state on the stretched horizon reproduces the boundary Hartle-Hawking correlator in the large-$N$ (vanishing stretched horizon) limit. At finite-$N$ the correlator has exponentially small corrections in the black hole entropy, which become important at the Page time. This is a strong hint that the smoothness of the horizon in the bulk EFT (up to the Page time) is consistent with the presence of a Planckian stretched horizon in the UV complete bulk description. We will have more to say about post-Page timescales later in this section.

In this paper, we discussed these questions from a more bulk-oriented perspective, because we wanted to understand the nature of smoothness more directly from the bulk. After all, smoothness is ultimately a {\em bulk} statement. To this end, we first demonstrated the emergence of the bulk Hartle-Hawking correlator in the large-$N$ limit from the bulk stretched horizon correlator. The bulk  HH correlator has the property that it allows an analytic continuation into (what has an interpretation as) the black hole interior in this limit. This is a {\em bulk} realization of the transition to Type III algebra from our Type I system, with the role of large-$N$ played by the vanishing limit of the stretched horizon. The emergent analytic continuation can therefore be viewed as due to the appearance of extended half-sided modular translations in Type III \cite{Leutheusser}. We further demonstrated that the correlators exhibit a type of thermal factorization in this limit, which was used to argue (along the lines of \cite{Onkar}) for the appearance of a universal code subspace as a second tensor factor. The fact that ``free" particles (aka., generalized free fields) and the analytic continuation across the horizon are both emergent features of a large-$N$ limit, seems to be a distinction between strongly coupled and weakly coupled thermal systems. In weakly coupled systems, free particles are presumably present even far from the thermodynamic limit, and we are not aware of a natural analytic continuation into an ``interior" that emerges in such a limit. 

A key message of our calculations in \cite{Burman} and here, is that changing the UV structure of the geometry at around a Planck length near the horizon does {\em not} affect many of the usual expectations from black holes, due to the dressing from the excited states in the sliver. Apparently, these are enough to avoid the splattering into the brick-wall, in the sense of external Hartle-Hawking correlators. Features like thermodynamics and the apparent smoothness of the horizon as perceived by low-point correlators all remain intact (the latter, at least till the Page time). The precise implication of this for infallers, is currently under investigation. 

Underlying our results is the fact that the Principle of Equivalence is a feature of bulk EFT. The presence of a stretched/quantum horizon in the UV-complete description is {\em not} in immediate violation of PoE if the effective correlators retain enough of the smoothness properties. Our calculations show that at least some aspects of such an effective smoothness can indeed be preserved.

We conclude the paper with some comments relating our work to various extant ideas. The goal is to place our results appropriately in the current landscape of ideas and also to make some comments about future directions.

\begin{itemize}
\item {\bf Bottom-Up vs Top-Down Microstates:} As we noted previously, our construction is best viewed as a ``bottom-up" approach. Let us discuss some basic points that would be more satisfying, if we had a top-down approach. 

1. We have not offered any dynamical mechanism for stabilizing the stretched horizon. We have simply assumed that it is a feature of the UV complete theory. Note however that the problem is in fact somewhat milder than what it might seem superficially. A quantum version of the difficulty of installing structures above the horizon is the observation that the Boulware vacuum has a divergent negative stress tensor flux at the horizon. Our stretched horizon vacuum can be viewed as a regulated version of the Boulware vacuum -- but we are not working with the vacuum itself, but with excited states at the mass of the black hole. Israel and Mukohyama \cite{Israel-Mukohyama} argued based on heuristic estimates that such states are better defined than the vacuum itself, see \cite{Israel-Mukohyama} as well as the discussion in \cite{Burman}. These arguments apply for us as well. While it would undoubtedly be useful to understand the origin of this picture from a UV complete setting, it is worth emphasizing that the sliver states on the stretched horizon are {\em not} violating the principle of equivalence in the same sense that the stretched horizon vacuum is. The smooth HH correlator we find (when the dust settles) is a detailed realization of this fact.

2. Our picture of a microstate crucially relied on the black hole solution of bulk EFT (with the stretched horizon installed around the horizon). The size of a black hole of mass $M$ in general relativity increases with $M$ and our construction inherits it because it modifies the geometry only near the horizon. It would clearly be much more satisfying to derive the ``size" of a microstate from a UV-complete description without any reliance on bulk EFT. It is believed that in string theory, the sizes of D-brane bound states scale in the correct way \cite{MathurFuzz}.

3. A related point is that of backreaction. As long as we are working with a small number (compared to a suitable positive power of $N$) of light excitations around a given microstate, our calculations give satisfactory answers. But if we act with operators or combinations of operators with sufficiently high scaling dimension, then our description will break down. The transition between the dynamics of fluctuations around a background and the dynamics of the background itself has always been a difficult one in quantum field theory in curved space and AdS/CFT. This is a manifestation of that problem.

The root cause of all the above problems is the same. We have proposed a way of implementing finite-$N$ aspects of the horizon in the bulk which has many reasonable features, but we would like a clear understanding of this as a finite-$N$ feature of heavy states in the CFT as well as in string theory. Progress on this question will be the key to a better understanding of black hole horizons. 

\item{\bf Why did the calculation work?:} The calculations of \cite{Burman} revealed that the normal modes that contribute to the thermodynamics of the black hole are the low-lying normal modes. Conveniently, the approximate form of the spectrum became linear in $n$ and essentially degenerate in $J$, in that limit. By treating the spectrum as degenerate in $J$ (and introducing a cut-off in $J$) the calculation reduced to a type of quantum gas/oscillator calculation. Fortunately, there was only a one parameter freedom in the calculation (related to our ignorance of the UV cut-off), so we could do meaningful checks. 

The quasi-degeneracy in $J$ was absolutely crucial for multiple aspects of the physics. In particular, this degeneracy is what carried the entropy. In the next item, we will also note that the weak $J$-dependence contains features of chaos and random matrices. 

It is worth pointing out here that a naive 1+1 d Rindler calculation misses this $J$-dependence and associated physics, because there are no transverse dimensions. It is easy enough to get a linear in $n$ spectrum in 1+1 dimensional Rindler if you introduce a stretched horizon. This is because a massless scalar wave equation does not change its form after a conformal transformation and therefore has sinusoidal solutions in 1+1 d Rindler\footnote{See \cite{Ishibashi, Suchetan} for an interesting ``strongly coupled" version of the 1+1 Rindler result, where a similar set of linear in $n$ normal modes were identified for general 2D CFTs without the assumption of a free theory. We suspect that this may have connections to RT surface calculations in the AdS$_3$ bulk via normal modes, as speculated in \cite{Pradipta}.}. Imposing a stretched horizon (together with a suitable asymptotic boundary condition) therefore leads to standing waves and a linear in $n$ spectrum. Note also that this quantization is independent of the precise location of the stretched horizon. But in higher dimensions, something more interesting happens -- there is a non-trivial $J$-dependence in the normal mode spectrum which is almost (but not quite) degenerate, together with the linear $n$. This happens only if the stretched horizon is sufficiently close to the horizon, and this structure is crucial in many places in our calculations. Exact form of the spectrum (using numerical and semi-analytical methods) was computed in \cite{Pradipta}.

\item {\bf Early and Late Time Chaos:} In \cite{Burman} and in this paper, we explored the thermodynamics and boundary/bulk correlators of our stretched horizon quantum field theory. Part of the goal was to clarify aspects of the emergence of the interior. To do this, as we noted above, we could make the approximation that the spectrum was degenerate in $J$. But if we want to study more dynamical aspects like thermalization, we will need to keep track of more detailed aspects of the spectrum. Considering the fact that black holes are believed to have features of chaotic systems, it will be interesting to see if any of these features can be obtained from our model. 

We expect that early time chaos like OTOCs and scrambling will be relatively easy to understand from our construction, because they depend only on the redshift at the horizon. Note that the normal modes are controlled by $\omega_0$, eqn.\eqref{omega0}, which is set by the scrambling timescale. It will be interesting to study the single-sided versions of the OTOC calculations in \cite{StanfordShenker}, in our setting. 

To study late-time chaos on the other hand, we need more detailed properties of the spectrum. Preliminary steps towards this were taken in \cite{synth, Das, Riemann, ArnabSuman, Pradipta} (see also closely related papers in  \cite{Murdia, Diptarka, Souvik1, Souvik2, Su1, Su2}) where it was noted that the single particle spectrum $\omega_{n,J}$ had two crucial properties that are usually associated to random matrices. The first was that the system exhibited a ramp of slope 1 \cite{synth} in a log-log plot of the Spectral Form Factor (SFF) vs. time \cite{Cotler}. The second was that there were strong hints of level repulsion in the spectrum \cite{Das}. These facts are surprising because both these features are believed to be associated to random matrices\footnote{While various forms of ramps are often found in various theories, the {\em linear} ramp is often viewed as a smoking gun of RMT behavior.}, while we do not have random matrices anywhere in our construction. In fact, we are working with free modes and the only potential source of ``chaos" in our spectrum is the stretched horizon boundary conditions\footnote{It was emphasized in \cite{Das} that it is crucial for these results that the stretched horizon is sufficiently close to the horizon -- a hole in empty flat space, for example, does not exhibit the linearity of the ramp.} As far as we are aware, this is the only bulk construction that can reproduce a linear ramp with fluctuations, together with level repulsion. In ensemble-averaged calculations like JT gravity, while the ramp is visible once one introduces wormholes, the fluctuations on the ramp are not accessible. We view these calculations as starting from $N=\infty$ and adding $e^{-N}$ corrections to see the ramp, while our approach to be a direct prescription to see finite-$N$ effects.
Efforts to study the second quantized spectrum in our system are currently under way. It will be interesting to see if all the relevant scales (in say, the spectral form factor) can be correctly reproduced by choosing our one free parameter.

A related comment that is perhaps worthy of note is that the RMT like features that we see in the spectrum \cite{synth, Das, Pradipta} are present only when the black hole is {\em away} from extremality \cite{PradiptaNew}. It will be interesting to understand if this fact has connections to recent discussions about microstates of BPS black holes and the fortuity of black hole microstates \cite{ChiMing1}.

\item {\bf Fuzzballs:} The last comment above brings us to the connection between our construction and the conventional fuzzball program. Clearly the fuzzball proposal has moral similarities to what we are discussing here, for the reason that both pictures suggest that the UV complete description of a black hole microstate has no interior. But there are also substantial differences. Below we make some comparisons between the two approaches.

1. The conventional fuzzball program aims to construct microstates of BPS black holes in supergravity. There is evidence to think that the superstrata \cite{superstrata} solutions that have been constructed so far, are more like BPS multigraviton states than genuine black hole microstates \cite{ChiMing1, ChiMing2}. The comment about the spectrum that we made in the previous item regarding stretched horizons of extremal black holes may have an understanding in this light. For generic (finite temperature) black holes, the near-horizon geometry is of a Rindler form. In Rindler ($\times$ compact space) we can check that the normal modes exhibit the quasi-degeneracy in the angular quantum number that was so crucial for the success of our calculations. However for extremal black holes, the near horizon geometry is an AdS$_2$ ($\times$ compact space) form. If one computes the normal modes with a stretched horizon near the Poincare horizon\footnote{The location of the stretched horizon is (interestingly) not crucial in this result. This is because if one puts a stretched horizon at a generic location in  spacetime, whether AdS or flat space, the spectrum is roughly linear in $J$. It seems (though more work is needed to fully settle this because normal mode computations are complicated and often numerical) that a single-zero horizon is crucial for the quasi-degeneracy in $J$.}, one finds that the $J$-dependence is not quasi-degenerate, it is quasi-linear \cite{PradiptaNew}. (Quasi-)linear $J$-dependence is familiar in flat space and global AdS as well, and is insufficient to explain black hole entropy, let alone chaos, from a brickwall-type calculation. 

2. That said, the fuzzball program (broadly defined) is something we feel is likely to be an important ingredient in our final picture of microstates. In particular, unlike our ``bottom-up" approach, the fuzzball paradigm offers mechanisms for the sizes of bound states in string theory to get stabilized at parametrically larger values than what one might expect naively. The 3-charge geometric fuzzball microstates that have been constructed so far (at least the ones that are far from typicality) are legitimate constructions in supergravity which demonstrate that there are mechanisms for evading no-hair theorems in string theory.

3. An interesting distinction between our viewpoint and the conventional fuzzball program is that in our construction, the stretched horizon enters very directly as a finite-$N$ effect. The role of finite-$N$ in the BPS supergravity solutions on the other hand, is more indirect. Naively, one might think that being supergravity solutions, these solutions only capture infinite $N$. But it turns out that one can define a symplectic form on the space of these supergravity solutions, and upon quantization it leads to a natural notion of finite-$N$ \cite{Rychkov, Grant}. This approach is sufficient to explain the entropy \cite{Rychkov} of the two-charge D1-D5 black hole (which has vanishing horizon area at leading order), including the precise numerical coefficient \cite{CK-Avinash}. In light of the recent developments involving fortuitous states \cite{ChiMing2}, it is tempting to speculate that supergravity solutions where the quantization of the symplectic form is responsible for the finiteness of $N$ capture only BPS multi-graviton states. To capture the fortuitous states, we need a new bulk finite-$N$ ingredient which is inaccessible in supergravity. Our stretched horizon is a candidate for such an ingredient. The fact that we are able to explain the entropy and thermodynamics of finite horizon black holes, we take as an encouraging hint in this direction.

4. Our microstates are naturally states in the Hilbert space of a field theory with a stretched horizon. Fuzzball microstates are typically constructed in supergravity and therefore are best viewed as coherent (classical) states. This distinction was useful for us in setting up a correlator calculation, where the smooth correlators could be understood as a precise and natural limit. It will be interesting to see if there is some sense in which our calculations here can have some connections to ideas like fuzzball complementarity \cite{FuzzComp}.

\item  {\bf State-dependence:} In \cite{Vyshnav-State}, some comments were made about the ambiguities \cite{MarolfPolchinskiSD, HarlowSD} associated to state-dependence \cite{PR} in thermofield double constructions of the interior. A key point is that after Page time, a smooth horizon (HH) correlator is losing information. The construction of \cite{PR} hard-wires a bulk Kruskal time for the entire history of the black hole, even at finite-$N$ (which implies finite Page time). If one assumes a conventional extrapolate AdS/CFT dictionary, this results in CFT correlators that decay exponentially after the Page time and lose information. This is in tension with both general principles, and explicit results about Virasoro blocks \cite{Kaplan} and SYK spectral form factors \cite{Cotler}. Note that the \cite{Leutheusser} construction bypasses this problem, because they are working with infinite-$N$, and therefore the Page time is also infinite.

That the HH correlator is ``too thermal" at late times is the crux of Maldacena's information paradox \cite{MaldacenaEternal}. Indeed, it was shown in \cite{Vyshnav-State} that when the Page time is infinite, the construction of \cite{PR} can be made state-{\em in}dependent using crossed products \cite{Witten} and modular translations \cite{Leutheusser}. This is consistent -- the argument for Type III algebras in \cite{Leutheusser} relies on the spectrum of the Hartle-Hawking correlator, and makes sense only when the latter is meaningful\footnote{See \cite{Roji} for closely related comments.}. When the Page time is finite however, \cite{Vyshnav-State} argued that the post-Page construction of \cite{PR} needs to be re-evaluated because it crucially relies on the Hartle-Hawking correlator in the information-losing regime \cite{MaldacenaEternal}\footnote{Note that this argument does not have much to say directly about the state-specific constructions of \cite{Akers}, because these constructions do not directly rely on the thermofield double or the thermal correlator. We believe new ideas (or at least new interpretations) are required to completely resolve the issue of smoothness after the Page time.}. 

Our work in this paper and \cite{Burman} demonstrates these points quite explicitly. We find that the correlator can be approximated very well by a smooth-horizon HH correlator before the Page time, but after the Page time, information restoration effects do not allow this possibility. 

\item {\bf Emergent Interior vs Infall:} What we have shown is that there exists a natural bulk model for heavy states in a holographic CFT, which captures many expectations from black holes, but without leading to Maldacena's version of information loss. Exterior correlators in this set up are indistinguishable from the smooth horizon correlator in the large-$N$ limit. At finite-$N$ there are small corrections to the HH correlator. The HH correlator has an analytic continuation which has a natural interpretation as the black hole interior, but its existence is only meaningful in this approximate sense. We expect that bulk EFT should make approximate sense within such a Page window. It is worth noting here that the infall time (measured on the boundary clock) from the AdS boundary to a Planckian stretched horizon is in fact the scrambling time. It is also worth noting that the fine-grained information in the microstate  acted as a super-selection label in our thermal factorization calculation and provided the second tensor factor. This makes it plausible that information spreading into the fine-grained degrees of freedom has an interpretation as propagation in the second tensor factor. 

Because we have quite an explicit set-up, it will be interesting to compute the evolution of an infalling bulk wave packet (in perhaps a suitable eikonal limit) on a typical excited microstate. What is its evolution before and after it has reached the stretched horizon? It will be very interesting to see if the quantum spreading of this state into the huge number of quasi-degenerate $J$-modes can be given a simple ``analytic continuation into the interior" interpretation in terms of the superselection sector label. Can we obtain an infalling geodesic interpretation from a parametrically large (but not scaling with $N$) mass limit of such a calculation?

\item {\bf Post-Page Smoothness and Quantum Charts:} As far as we are aware, unambiguous evidence for smoothness of the horizon exists only till the Page time\footnote{We discus the current status of smoothness after the Page time in the replica wormhole calculation of the Page curve, in Appendix \ref{State-flip} to emphasize the ambiguities.}. It may be that we need a better operational {\em definition} of post-Page smoothness before we can make a clear statement about its status. Let us clarify what we mean by this. 

In terms of probe fields, the usual definition of smoothness is that fields have infalling boundary conditions at the horizon. This leads to the HH correlator, which exponentially decays beyond the Page time, and therefore to Maldacena's information paradox. On a typical heavy microstate, if we perturb the state with a light probe field at time $t=0$, within a diffusion time, we start seeing the HH correlator. This is what we saw in our calculations, and this result is valid until a Page time after the insertion of the initial probe scalar. As we saw in our Page windows discussion, infalling boundary conditions  simply aren't meaningful for a probe inserted at $t=0$, a Page time later. It is natural to view probes and their correlators as a way to define a ``quantum chart" for the eternal black hole, with each chart covering a Page window.

But we suspect that there is a notion of interior associated to an evolving microstate itself (and not to infalling probes), that may be meaningful even after the Page time\footnote{Note that this goes beyond Hawking's calculation which deals only with perturbative fields.}.  This belief stems from the success of certain calculations that take the Penrose diagrams of (interiors of) black holes seriously even after the Page time. We will phrase the discussion below in terms of the double-sided eternal AdS black hole Penrose diagram. A quantity that is often considered is the complexity of the quantum state as captured by the size of the Einstein-Rosen bridge \cite{action, Action?}. The linear growth of the bridge has an interpretation as the linear growth of complexity expected of the state. The growth of entanglement entropy noted in \cite{Hartman} and the QES calculation of the Page curve \cite{Penington, Almheiri} also indicate that the bulk interior may be meaningful after the Page time\footnote{The QES calculation in effect excludes the interior via the island mechanism. But to set up the calculation geometrically, one needs to assume that the interior is well-defined.}. In \cite{SusskindMP} the perspective is taken that this increasing complexity phase is a {\em definition} of a smooth horizon. Note however that this is distinct from the notion of smoothness as experienced by probes via an infalling boundary condition, which is the Hawking notion of smoothness. Every time we perturb the state, an effectively infalling boundary condition becomes valid for the duration of a Page window, because the complexity of the state gets reset to a low value (roughly, we are at the $t=0$ slice of the eternal black hole Penrose diagram after a perturbation). The complexity calculations suggest that classical Kruskal geometry is more than what a light probe can see. 

Reconciling these two notions of smoothness in a coherent overall picture seems necessary to form a complete picture of the black hole interior. Our calculation in this paper directly only deals with the pre-Page phase. It illustrates that the notion of smoothness of the horizon in the sense of infall and the notion of ``smoothness" in the sense of the existence (and growth) of the interior, are not identical. 

\item{\bf A non-Argument for post-Page Smoothness:} We consider the observations above about complexity (and QES) to be plausibility arguments for thinking that the black hole interior may make sense even after the Page time, in some suitable sense in the bulk EFT. As we discussed, there are good reasons to think that if such a late-time notion of the interior exists, it must be conceptually different from probe-based ideas like smooth infall. 

But we should add here, that there is a conflation of pre- and post-Page smoothness in many discussions on this topic, that is a bit more quick than we would like. It is important to remember that for any macroscopic region, the Page time that would be associated to a black hole of that size or larger would be many orders of magnitudes bigger than the age of the Universe. So we cannot appeal to the statement that ``we could be passing through the horizon of a very large black hole right now, and not know it" as an argument for {\em post-Page} smoothness. All such black holes would be in their pre-Page era. Note also the more fundamental problem in justifying this type of argument -- perturbing by a probe re-sets the clock, so we expect to see a smooth horizon anyway for another Page window. 

The key issue that makes post-Page discussions confusing is that we are dealing very directly with finite-$N$ physics. So how much we can trust the intuition from classical gravity is unclear. On the other hand, we feel that any quantum model  of the black hole should be able to reproduce the classical black hole interior as an approximation, in the pre-Page era.

\item{\bf Gravitational Waves and Modifications at the Horizon:} Ever since the discovery of gravitational waves from colliding black holes, there has been an effort to study potentially new signatures of corrections to general relativity from near the horizon. One such effort imposes new boundary conditions for these waves near the horizon (see e.g. \cite{2105.12313}). Our results suggest that it may be instructive to study new boundary conditions that are dressed with radiation as well, since these may be better models for black hole horizons. The emergence of the smooth horizon correlator is a suggestion that it may be difficult to get hints of beyond-GR effects. Another place where our perspective may play a role is in the efforts to connect fuzzballs with observations -- this is a big subject with much recent activity that we are not experts on, so we refer the reader to \cite{Mayerson} and follow ups. 

\item {\bf Other Boundary Conditions and UV Sensitivity:} In \cite{Burman} and here, we have worked with a simple model for the stretched horizon, which is to  impose a Dirichlet boundary condition for the probe scalar. This raises the question -- does changing this boundary condition affect our conclusions? We suspect that our basic messages remain intact even with other boundary conditions (as long as they respect some broad requirements, like non-dissipativity, etc.). One reason for this, is that we are not considering the vacuum state, we are interested in {\em generic} excited states. This means that it is plausible that the variations in the boundary conditions can be captured by  a different choice in the excited pure state. In fact, angle-dependent stretched horizon profiles for the scalar field were studied in \cite{Das} and it was noticed that the normal mode structure was qualitatively similar, but with some noise in the spectrum. While for thermodynamics and smoothness of the horizon we do not expect this to make a difference, for studying chaos and related aspects it may be worth studying this in more detail.

\item {\bf More General Black Holes:} We have in fact been able to generalize \cite{PradiptaNew} the calculations of \cite{Pradipta} to more general black holes -- namely, Kerr-Newman and Cvetic-Youm -- and reproduce the detailed thermodynamics of these black holes. This is possible because the wave equations in these black holes naturally allow a holomorphically factorized interpretation (as has been noted before \cite{CveticLarsen1, CMS, CK-KerrCFT}). Together with the left and right temperatures identified in \cite{CMS, CK-KerrCFT}, the one-parameter freedom we noted earlier becomes the choice of the central charge. The calculation works, because the choice of the central charge is all that is needed to make Kerr-CFT work. Our stretched horizon approach therefore gives a {\em mechanical understanding} of the Kerr-CFT central charge in terms of the stretched horizon normal modes. This may have the advantage that the usual Kerr-CFT approach to the central charge has to make appeals to the extremal limit, where the black hole often has instabilities. A second point of interest is that the Kerr-CFT central charge is often proportional to the angular momentum of the black hole -- this turns out to also have a natural understanding in our setting due to the relationship between the central charge and $J_{cut}$, a hint of this can be seen already in \eqref{Jcut}. The price of these results is of course that we are unable to fix the one-parameter freedom completely (and independently) in our approach. That seems to require knowledge of the UV.

\item{\bf Evaporating Black Holes:} The explicit identification of the microstate degrees of freedom implies that our model for black holes effectively turns the system into a burning piece of coal -- albeit with a different normal mode spectrum than say, a black body. This means that even though our present discussion is for the large-AdS black hole, a similar picture coupled to a heat bath (or equivalently a flat space or small AdS black hole) can naturally explain the emergence of the Page curve. But like in recent discussions of the Page curve, we do not have a dynamical mechanism for connecting between the various epochs during a black hole's evaporation. In the JT gravity case, this is because the ensemble is hard-wired \cite{Saad} at each epoch, in our case it is because we will have to change the background geometry as the black hole evaporates. It will be very interesting to understand the connection between this perspective and the quantum extremal surface perspective. 
  
\end{itemize}

\section{Acknowledgments}

We are very grateful to Suchetan Das and Pradipta Pathak for numerous discussions and related collaborations. We thank Diptarka Das, Suman Das, Sumit Garg, Daniel Grumiller, Sachin Jain, Nirmalya Kajuri, Arnab Kundu, Alok Laddha, Raghu Mahajan, Samir Mathur, Shiraz Minwalla, Vyshnav Mohan, K. Narayan, Neils Obers, Sarthak Parikh, Sidharth Prabhu, Suvrat Raju,  Debajyoti Sarkar, Ronak Soni and Sandip Trivedi for comments and/or questions. CK thanks the organizers and participants of conferences at IIT Mandi, BITS Goa, IISER Pune and ICTS Bangalore for lively interactions during talks based on this material.

\appendix

\section{Replica Wormholes and Post-Page Smoothness}\label{State-flip}

In this Appendix, we would like to spell out the trade-off that is being made when one accepts the replica wormhole/QES prescription to obtain the Page curve. Our goal is to phrase the discussion as simply as possible, but without losing the essential content, so that we can assess what it might mean to salvage smoothness after the Page time\footnote{The phrase ``factorization puzzle" is often used in this context, but that sanitizes the underlying challenge to smoothness, so we will refrain from using it in this Appendix.}. Even though the ideas we present here have appeared before in various places (see in particular \cite{Liu-Vardhan, Bousso, Vyshnav, Mehta}), they do not juxtapose some of the aspects that we believe are relevant. The discussion in \cite{Vyshnav} is closest to what we present here, but the presentation there was technical and obscured the simple underlying message. So we will first summarize the statements in \cite{Vyshnav} in a simplified setting considered in \cite{Mehta}. Our perspective here will be different from that of \cite{Mehta} however\footnote{We thank Samir Mathur for discussions.}. Our first goal will be to emphasize that the main idea in \cite{PSSY} can be viewed as a prescription to switch the (maximally entangled) smooth horizon state with the UV-complete state at a crucial stage in the calculation. Not making that switch is tantamount to working with an averaged state from the UV-complete perspective\footnote{That the bulk EFT state has features of an averaged state has been noted before \cite{Bousso,Vyshnav}.}, and gives rise to the information paradox in the form of the ever-rising Page curve. Making the switch does bring down the Page curve, but now the state is no longer maximally entangled (and therefore more thought is required about the status of smoothness, after the Page time). 

We will first state the story in words, and then present the calculation that goes with this story.

A key observation of \cite{Liu-Vardhan, Vyshnav} was that if one considers a UV-complete pure state that is in approximate local thermal equilibrium, one can compute entanglement entropy\footnote{We will be working with the second Renyi entropy here for transparency. Working with entanglement entropy makes things technically more complicated \cite{Liu-Vardhan, Vyshnav} without adding any conceptual nuances of interest to us.} via a conventional ensemble average. This ensemble average has nothing to do with the subtleties of gravity or black holes -- the ensemble is simply the eigenstate ensemble (sometimes called Berry's ensemble), and the  average is simply a proxy for a time average. This is quite palatable to anyone familiar with ensembles in statistical mechanics and has nothing to do with averages over theories. The underlying theory is still unitary. In the gravitational language, one can write the relevant contractions that emerge from this equilibrium ensemble average in terms of replica wormholes. This amounts to  filling in the contraction loops in the black hole tensor factor. This observation is important, because it emphasizes that it is not the replica wormholes themselves that are problematic -- the real question is whether the averaging operation can be consistently associated to a conventional statistical mechanics average.

The trouble is that the UV-complete equilibrium pure state that one starts with in \cite{Liu-Vardhan,Vyshnav} is {\em not} maximally entangled. Ergo, its reduced density matrix is not exactly thermal. Ergo, the smoothness of the horizon is not manifest. The maximally entangled state that Hawking would like us to start with, which is presumably a bulk EFT state, requires the switch to the UV-complete pure state, before the arguments of \cite{Liu-Vardhan, Vyshnav} can apply. A related interesting fact that is worthy of note here is that if one does the (eigenstate) ensemble average on the density matrix associated to the UV complete state, then the resulting state is the density matrix of the maximally entangled Hawking EFT state. We find this fact to be quite striking. 

Let us make these comments a bit more explicit. We will loosely follow the notation of \cite{Mehta}. We start with the assumption \cite{Liu-Vardhan,Vyshnav} that the total UV-complete state is an entangled state between the black hole and the radiation subsystems, at a given epoch of Hawking radiation. This assumption is nowadays called the central dogma. (An epoch is a period during which the black hole's mass can be treated as roughly constant.) The UV-complete state takes the general form
\bea
|\Psi \rangle = \sum_{i,a}c_{ia}|i\rangle |a\rangle \label{UVpure}
\eea
Here and henceforth we will take $i, i', j \in B$ and $a, a',b \in R$ where $B$ and $R$ denote the black hole and radiation subsystems respectively. The density matrix associated to this state is
\bea
\rho_\Psi \equiv | \Psi \rangle \langle \Psi|= \sum_{ii'aa'}c_{ia}c^*_{i'a'}|i\rangle |a\rangle \langle i'| \langle a'| \label{purestatedensitymatrix}
\eea
The reduced density matrix of radiation can be obtained by tracing out the black hole system 
\bea
\rho_R \equiv \sum_{j} \langle j | \rho_\Psi | j\rangle =\sum_{iaa'}c_{ia}c^*_{ia'}|a\rangle \langle a'| \label{red-pure}
\eea
We will sometimes find it convenient to write $(\rho_R)_{aa'}=\sum_i c_{ia}c^*_{ia'}$. Our goal is to compute the (second Renyi) entropy of this state:
\bea
e^{-S_2}= {\rm Tr} (\rho_R^2)=\sum_{a,a'} (\rho_R)_{aa'} (\rho_R)_{a'a}=\sum_{ii'aa'} c_{ia}c^*_{ia'} c_{i'a'}c^*_{i'a}.\label{UV-rad-reduced}
\eea
The idea of \cite{Liu-Vardhan, Vyshnav} is that when the system is in approximate equilibrium, this quantity can be determined using an ensemble average over Berry's (eigenstate) ensemble. In other words, 
\bea
e^{-S_2}=\sum_{ii'aa'} \langle\langle c_{ia}c^*_{ia'} c_{i'a'}c^*_{i'a} \rangle\rangle,
\eea
where the $\langle\langle \  \rangle\rangle$ denotes an average over Berry's ensemble. The latter is a Gaussian ensemble and is completely determined via contractions once we specify the mean and variance:
\bea
\langle\langle c_{ia} \rangle\rangle = \langle\langle c^*_{ia} \rangle\rangle =0, \ \ 
\langle\langle c_{ia}c^*_{i'a'} \rangle\rangle
 = \frac{1}{d_B d_R} \delta_{ii'}\delta_{aa'} \label{BerryRules}
 \eea
 where $d_B$ and $d_R$ are the dimensionalities of the black hole and radiation systems. We treat them as finite dimensional -- note that the black hole has finite entropy, so this may not be unreasonable. 

 In the 4-point object that arises in the Renyi entropy above, the Berry ensemble replacement results in two kinds of contractions. The result is
 \bea
 e^{-S_2}&=&\sum_{ii'aa'} \langle\langle c_{ia}c^*_{ia'}\rangle\rangle \langle\langle c_{i'a'}c^*_{i'a} \rangle\rangle+ \sum_{ii'aa'} \langle\langle c_{ia}c^*_{i'a}\rangle\rangle \langle\langle c_{i'a'}c^*_{ia'} \rangle\rangle \label{BerryCont}\\
 & =& \frac{1}{d_B^2 d_R^2} (d_B^2 d_R +d_B d_R^2) = \frac{1}{d_R} +\frac{1}{d_B} \label{PageCorrect}
 \eea
 We see that at early epochs when the radiation dimensionality is smaller, the first term dominates. This is what gives rise to the rising part of the Page curve\footnote{More precisely, its analogue for the second Renyi entropy.}. But in the later stages when the black hole is smaller, the second term dominates and we have the return of the Page curve. 

 This is very satisfying, but it also has the problem that it has nothing specifically to do with black holes. An exactly analogous calculation applies for a burning piece of coal (or a hard sphere gas leaking from one box to another), with $B$ replaced by the coal (or gas). This was emphasized in \cite{Vyshnav} to point out that ensemble averages can naturally arise in calculations of {\em unitary} Page curves without any reference to black holes or ensembles of theories. 

 The trouble is that the reduced density matrix \eqref{red-pure} of the pure UV-state \eqref{UVpure} is not exactly thermal. The Hawking state on the other hand is maximally entangled and can be taken in the form \cite{PSSY}
 \bea
 |\Psi_H\rangle = \frac{1}{\sqrt{d_R}}\sum_{a} |\psi_a\rangle |a \rangle,
 \eea
where the state $|\psi_a\rangle$ is a state living in the Hilbert space $B$. The ever-rising Page curve and the information paradox arises, if one assumes that $|\psi_a \rangle$ are orthonormal. To see this, we simply compute the reduced density matrix of the Hawking state:
\bea
\rho_R^H \equiv {\rm Tr}_B(\rho_{\Psi_H}) =\frac{1}{d_R}\sum_{iab}\langle i|\psi_a\rangle |a\rangle \langle b| \langle\psi_b|i\rangle=\frac{1}{d_R}\sum_{ab}\langle\psi_b |\psi_a\rangle |a\rangle \langle b| = \frac{1}{d_R} \sum_a |a\rangle \langle a|
\eea
where in the last step we have assumed that $|\psi_a\rangle$ are orthonormal. The resulting density matrix is maximally mixed, and its 2nd Renyi entropy is 
\bea
e^{-S_2}={\rm Tr}((\rho^H_{R})^2)=\frac{1}{d_R}.
\eea
We see that the second term in \eqref{PageCorrect} is missing -- this is the Page curve that loses information for second Renyi entropy. 

A noteworthy fact about the Hawking density matrix $\rho_R^H$ is that it can be understood as the Berry/eigenstate ensemble average of the radiation density matrix of the UV-complete state \eqref{red-pure}. This was emphasized in \cite{Bousso, Vyshnav}. This is trivial to check using the Berry contraction rules \eqref{BerryRules}, but it suggests the interesting interpretation that bulk EFT is a theory for the averaged state, as mentioned earlier.  

The replica wormhole prescription suggested in \cite{PSSY} is, in our opinion, hard to justify directly with the maximally entangled Hawking state that they use there. What the prescription there really amounts to, is interpreting (replacing) the Hawking state as (with) the UV-complete state. This means that we should write the $|\psi_a\rangle$ in terms of the true orthonormal basis states of the black hole tensor factor. Once that is done, the conventional Berry's ensemble average does the job and we have a derivation of the Page curve from conventional statistical mechanics as described above, without any averages over theories etc. This point has been  made previously, but it seems not well appreciated. Part of the goal of this Appendix is to strip down the argument of \cite{Vyshnav} to its bare essentials and to make a very direct comparison with the results of \cite{PSSY}. It is crucial that the contractions resulting from the UV-state replacement lead precisely to the replica wormhole contractions of \cite{PSSY} and not to some generic structures which share partial features of those contractions. This is what we check explicitly, below.

Let us first write the state $|\psi_a\rangle$ in terms of the true orthonormal (UV) basis states.
\bea
|\psi_a \rangle = \sqrt{d_R}\sum_{i} c_{ia} |i\rangle 
\eea
where $|i \rangle$ is the orthonormal basis. We have made choices of conventions in the above equation so that with these choices, the Hawking state $|\Psi_H \rangle$ becomes the UV-complete state $|\Psi \rangle$. Note that in effect, constructions like \cite{Akers} which use non-isometric codes to map bulk to the boundary, are simply prescriptions for mapping the Hawking state to the UV-complete equilibrium state. It is clear that after the Page time this will involve various challenges, because the dimensionality of $|\psi_a\rangle$ increases with time as the radiation leaks out, while the dimensionality of $|i\rangle$ keeps decreasing. 

In any event, with the UV-complete replacement, the second Renyi entropy of the reduced density matrix of the radiation is simply given by what we found in \eqref{UV-rad-reduced}. It is convenient to introduce a diagrammatic notation for the contractions of this object. We write it as
\begin{figure}[H]
    \begin{minipage}{0.5\textwidth}
    \begin{equation*}
        \sum_{ii'aa'} c_{ia}c^*_{ia'} c_{i'a'}c^*_{i'a}\hspace{1cm}= \nonumber
    \end{equation*}
\end{minipage}
\begin{minipage}{0.5\textwidth}
   \centering
   \includegraphics[width=0.7\textwidth,height=0.4\textwidth]{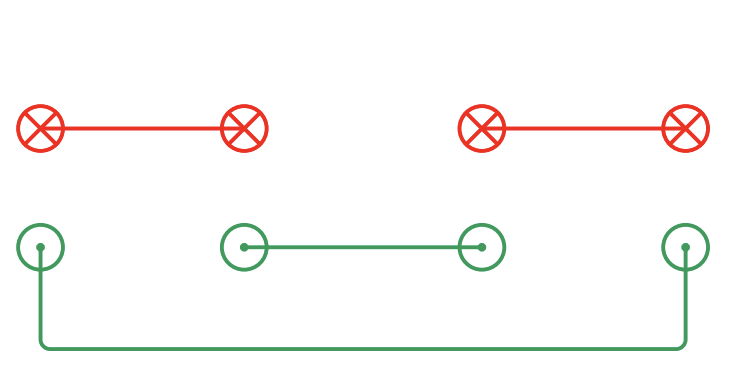}
\end{minipage}
\end{figure}
\noindent
Now, since we are in a UV-complete equilibrium state, we should also do the contractions arising from Berry's ensemble. We will denote the first contraction in \eqref{BerryCont} by 
\begin{figure}[H]
    \begin{minipage}{0.5\textwidth}
        \begin{equation*}
            \sum_{ii'aa'} \langle\langle c_{ia}c^*_{ia'}\rangle\rangle \langle\langle c_{i'a'}c^*_{i'a} \rangle\rangle \hspace{1cm}=\nonumber
        \end{equation*}
    \end{minipage}
    \begin{minipage}{0.5\textwidth}
        \centering
        \includegraphics[width=0.7\textwidth,height=0.4\textwidth]{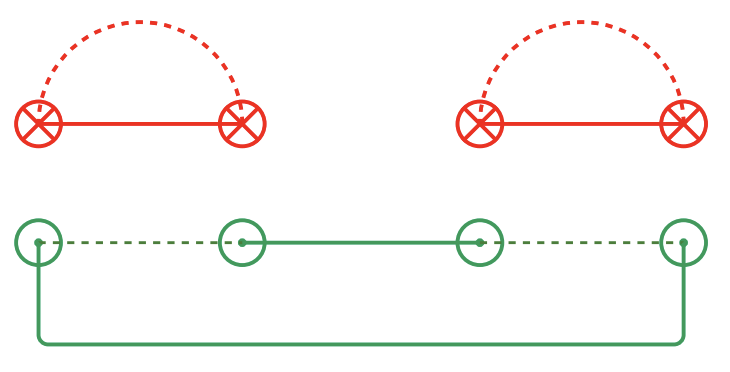}
    \end{minipage}
\end{figure}
\noindent
Similarly for the second Berry contraction, we have
\begin{figure}[H]
    \begin{minipage}{0.5\textwidth}
        \begin{equation*}
            \sum_{ii'aa'} \langle\langle c_{ia}c^*_{i'a}\rangle\rangle \langle\langle c_{i'a'}c^*_{ia'} \rangle\rangle\hspace{1cm} =\nonumber
        \end{equation*}
    \end{minipage}
    \begin{minipage}{0.5\textwidth}
        \centering
        \includegraphics[width=0.7\textwidth,height=0.4\textwidth]{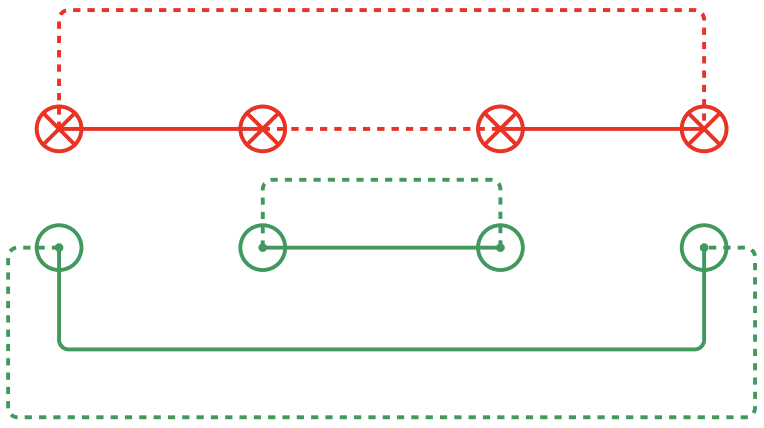}
    \end{minipage}
\end{figure}
\noindent
Replica wormholes are simply the instruction to fill in the black hole ($i, i'$) index loops above, and interpret them as the bulk. Once we do that, the last two figures take the form 
\begin{figure}[H]
    \begin{minipage}{0.5\textwidth}
        \centering
        \includegraphics[width=0.7\textwidth,height=0.4\textwidth]{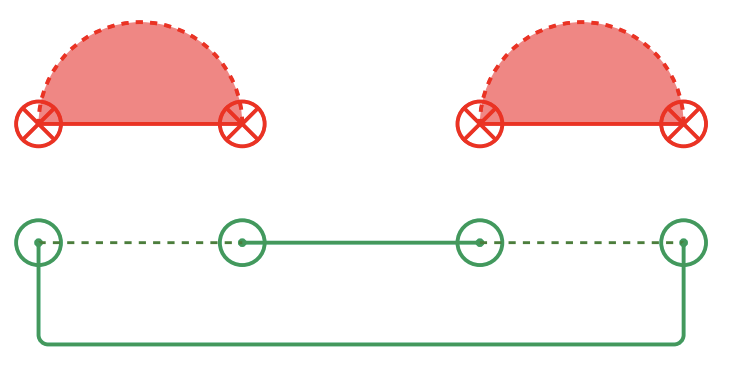}
    \end{minipage}
    \begin{minipage}{0.5\textwidth}
        \centering
        \includegraphics[width=0.7\textwidth,height=0.4\textwidth]{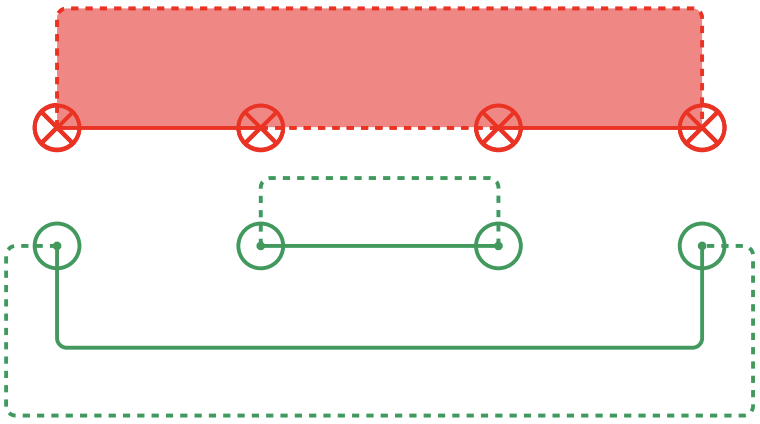}
    \end{minipage}
\end{figure}
It is easy to check that topologically these last two diagrams are identical to the two diagrams on the right of Figure 3 in \cite{PSSY}, once we join the pair of $i$ (and $j$) indices there. The right diagram above is therefore a replica wormhole. What we have done here, establishes that the replica wormhole contractions can be obtained from a conventional statistical mechanics calculation, from a UV-complete (ie., micro) state. Conventional quantum statistical mechanics of equilibrium states, then provides a derivation of replica wormholes and one does not have to deal with the unpleasantness associated to ensemble averages over couplings. 

Even though we gave up manifest smoothness in the derivation of the Page curve, there are reasons to suspect there may be some sense to the interior geometry even in the post-Page era. One can view the replica wormhole calculation as deriving the QES formula \cite{EngelhardtWall} which goes into the reproduction of the Page curve \cite{Penington, Almheiri}. The general ideas behind this calculation have been argued to generalize to flat space as well \cite{Jude, Matsuo, Thor}. The QES derivation of the Page curve, especially in the doubly holographic setting \cite{Mahajan, Critical, Kausik}, seems to give geometric content to the black hole interior even after the Page time. This, together with the complexity/Einstein-Rosen bridge connection raises the possibility that black hole microstates may allow an appropriate geometric interpretation, even in the post-Page era. Understanding this unambiguously, seems like a central problem in the field.


\section{The Bulk Hamiltonion} \label{AppHam}

The Lagrangian density of the free massless scalar field $\Phi(t,r,\psi)$ is 
\bea\label{Lagrangiandensity}
\mathcal{L} = -\frac{1}{2}\sqrt{-g}g^{\mu\nu}\partial_{\mu}\Phi\partial_{\nu}\Phi
\eea
Using the metric \ref{BTZmetric} into \ref{Lagrangiandensity}, this becomes
\bea
\mathcal{L} = -\frac{r}{2}\Bigg[-\frac{L^2}{r^2-r_{h}^2}(\partial_{t}\Phi)^2 + \frac{r^2-r_{h}^2}{L^2}(\partial_{r}\Phi)^2 + \frac{1}{r^2}(\partial_{\psi}\Phi)^2  \Bigg]
\eea
From the conjugate momentum to $\Phi$ 
\bea
\Pi = \frac{\partial\mathcal{L}}{\partial\Dot{\Phi}}
\eea
the Hamiltonian density can be computed as  
\bea\label{Hamiltoniandensity}
\mathcal{H} = \Pi\Dot{\Phi}-\mathcal{L} =\frac{rL^2}{2(r^2-r_{h}^2)}(\partial_{t}\Phi)^2 + \frac{r(r^2-r_{h}^2)}{2L^2}(\partial_{r}\Phi)^2+\frac{1}{2r}(\partial_{\psi}\Phi)^2
\eea
The Hamiltonian is the integral of this object over a spatial slice, $H = \int drd\psi\mathcal{H}$. The result is
\bea
H = \int drd\psi\frac{rL^2}{2(r^2-r_{h}^2)}(\partial_{t}\Phi)^2 + \int drd\psi\frac{r(r^2-r_{h}^2)}{2L^2}(\partial_{r}\Phi)^2+\int drd\psi\frac{1}{2r}(\partial_{\psi}\Phi)^2
\eea
After integration by parts and going on-shell, this becomes
\bea\label{Hamiltonian}
H = \int dr d\psi\frac{rL^2}{2(r^2-r_{h}^2)}(\Dot{\Phi}^2-\Phi\Ddot{\Phi})
\eea
Putting the mode expansion of $\Phi$\eqref{modeexpansionofthefield} into \eqref{Hamiltonian} and using the orthonormality conditions for $\phi$'s, the expression for the Hamiltonian is
\bea
H = \frac{1}{2}\sum_{n,J}\omega_{n,J}(b_{n,J}b^{\dagger}_{n,J}+b^{\dagger}_{n,J}b_{n,J})
\eea
which after normal ordering becomes
\bea
:H: = \sum_{n,J}\omega_{n,J}b^{\dagger}_{n,J}b_{n,J}
\eea
The expectation value of this Hamiltonian on a general highly excited state $|N\rangle$
\eqref{excitedstate} built upon the stretched horizon vacuum is
\bea
\langle N|:H:|N\rangle = \sum_{n,J}\omega_{n,J}\langle N|b^{\dagger}_{n,J}b_{n,J}|N\rangle = \sum_{n,J}\omega_{n,J}N_{n,J} = E
\eea
where $E$ is to be viewed as an energy sliver corresponding to the mass of the black hole.

\section{Some Details of the Smooth Horizon Calculation}\label{AppSmooth}

In this appendix, first, we will find out the expressions for $\alpha\equiv \Arg(P_{1})$ and $\beta\equiv \Arg(Q_{1})$ and then find out the normalization of the field $\phi_{\omega,J}(r)$ for the smooth horizon where we will assume the free wave boundary conditions at the horizon. 

\subsection{Calculation of $\Arg(P_{1})$}

Here we will use the expression of $P_{1}$ from the Appendix B (specifically equation B.2) of \cite{Burman}
\bea\label{appendixexpressionforP_{1}}
P_{1} = \frac{-i2^{-\frac{i\omega L^{2}}{2r_{h}}}e^{-\frac{\pi L}{2r_{h}}(\omega L+J)}e^{-\frac{i\pi}{2}(1+\nu)}\pi r_{h}^{\frac{1}{2}-\frac{iL}{r_{h}}(\omega L+J)}\csc h\Big(\frac{\pi\omega L^{2}}{r_{h}}\Big)\Gamma[\frac{1}{2}(1-\frac{iL}{r_{h}}(\omega L-J)+\nu)]} { \Gamma[\frac{iJL}{r_{h}}]\Gamma[1-\frac{i\omega L^{2}}{r_{h}}]\Gamma[\frac{1}{2}(1+\frac{iL}{r_{h}}(\omega L-J)+\nu)]  }
\eea
Taking the argument of both sides of the above equation and using the fact that the argument of the real number is zero,  $\Arg(-i)=-\frac{\pi}{2}$ and $\Arg(a^{ib})=\Arg(e^{ib\log a})=b\log a$, the expression for $\Arg(P_{1})$ becomes
\bea\label{P1 simplified}
&\hspace{-0.6cm}\Arg(P_{1})=-\frac{\pi}{2}-\frac{\omega L^{2}}{2r_{h}}\log2-\frac{\pi}{2}(1+\nu)-\frac{L(\omega L+J)}{r_{h}}\log r_{h}+\Arg(\Gamma[\frac{1}{2}(1-\frac{iL}{r_{h}}(\omega L-J)]+\nu)-\nonumber\\
&\hspace{2cm}-\Arg(\frac{iJL}{r_{h}})-\Arg(1-\frac{i\omega L^{2}}{r_{h}})-\Arg(\Gamma[\frac{1}{2}(1+\frac{iL}{r_{h}}(\omega L-J)]+\nu)
\eea
For the massless scalar field, $\nu=1$. In order to simplify the above expression, we use the Gamma function identities 
\bea\label{gammaidentity3}
\Gamma(1+z)=z\Gamma(z)
\eea
and
\bea\label{gammaidentity4}
\Arg(\Gamma(iy))=-\Arg(\Gamma(-iy))
\eea
Taking the argument on both sides of \eqref{gammaidentity3}, we get
\bea
\Arg(\Gamma(1+z))=\Arg(z)+\Arg(\Gamma(z))
\eea
from which we find
\bea\label{C.7}
&\Arg\Big(\Gamma\Big(1-\frac{i\omega L^{2}}{r_{h}}\Big)\Big)=\Arg\Big(-\frac{i\omega L^{2}}{r_{h}}\Big)+\Arg\Big(\Gamma\Big(-\frac{i\omega L^{2}}{r_{h}}\Big)\Big)\nonumber\\
&\hspace{0.8cm}=-\frac{\pi}{2}-\Arg(\Gamma(\frac{i\omega L^{2}}{r_{h}}))
\eea
and 
\bea\label{C.8}
&\Arg\Big(\Gamma\Big(1-\frac{iL}{2r_{h}}(\omega L-J)\Big)\Big)=\Arg\Big(-\frac{iL}{2r_{h}}(\omega L-J)\Big)+\Arg\Big(\Gamma[-\frac{iL}{2r_{h}}(\omega L-J)]\Big)\nonumber\\
&\hspace{1.6cm}=-\frac{\pi}{2}-\Arg\Big(\Gamma\Big(\frac{iL}{2r_{h}}(\omega L-J)\Big)\Big)
\eea
Similarly,
\bea\label{C.9}
&\Arg\Big(\Gamma\Big(1+\frac{iL}{2r_{h}}(\omega L-J)\Big)\Big)=\Arg\Big(\frac{iL}{2r_{h}}(\omega L-J)\Big)+\Arg\Big(\Gamma\Big(\frac{iL}{2r_{h}}(\omega L-J)\Big)\Big)\nonumber\\
&\hspace{1.6cm}=\frac{\pi}{2}+\Arg\Big(\Gamma\Big(\frac{iL}{2r_{h}}(\omega L-J)\Big)\Big)
\eea
Putting \eqref{C.7},\eqref{C.8} and \eqref{C.9} into \eqref{P1 simplified}, the expression for $P_{1}$ becomes
\bea\label{expressionforalpha}
&\alpha\equiv \Arg(P_{1})=-2\pi-\frac{\omega L^{2}}{2r_{h}}\log2-\frac{L(\omega L+J)}{r_{h}}\log r_{h}-2\Arg(\Gamma[\frac{iL}{2r_{h}}(\omega L-J)])-\Arg(\frac{iJL}{r_{h}})+\nonumber\\
&+\Arg(\Gamma[\frac{i\omega L^{2}}{r_{h}}])
\eea

\subsection{Calculation of $\Arg(Q_{1})$}

Now using the expression of $Q_{1}$ from the Appendix B (specifically equation B.3) of \cite{Burman}
\bea
Q_{1} = \frac{i2^{\frac{i\omega L^{2}}{2r_{h}}}e^{-\frac{\pi L}{2r_{h}}(\omega L+J)}e^{-\frac{i\pi}{2}(1+\nu)}\pi r_{h}^{\frac{1}{2}-\frac{iL}{r_{h}}(\omega L+J)}\csc h\Big(\frac{\pi\omega L^{2}}{r_{h}}\Big)\Gamma[\frac{1}{2}(1+\frac{iL}{r_{h}}(\omega L+J)+\nu)]} { \Gamma[\frac{iJL}{r_{h}}]\Gamma[1+\frac{i\omega L^{2}}{r_{h}}]\Gamma[\frac{1}{2}(1-\frac{iL}{r_{h}}(\omega L+J)+\nu)]  }
\eea
Again, taking the argument of both sides, 
\bea\label{Q1simplified}
&\hspace{-0.6cm}\Arg(Q_{1})=\frac{\pi}{2}+\frac{\omega L^{2}}{2r_{h}}\log2-\frac{\pi}{2}(1+\nu)-\frac{L(\omega L+J)}{r_{h}}\log r_{h}+\Arg(\Gamma[\frac{1}{2}(1+\frac{iL}{r_{h}}(\omega L+J)]+\nu)-\nonumber\\
&\hspace{2cm}-\Arg(\frac{iJL}{r_{h}})-\Arg(1+\frac{i\omega L^{2}}{r_{h}})-\Arg(\Gamma[\frac{1}{2}(1-\frac{iL}{r_{h}}(\omega L+J)]+\nu).
\eea
Using the same gamma function identities \eqref{gammaidentity3} and \eqref{gammaidentity4} here, we can write 
\bea
\Arg\Big(\Gamma\Big(1+\frac{iL}{2r_{h}}(\omega L+J)\Big)\Big)=\frac{\pi}{2}+\Arg\Big(\Gamma\Big(\frac{iL}{2r_{h}}(\omega L+J)\Big)\Big)
\eea
\bea
\Arg\Big(\Gamma\Big(1+\frac{i\omega L^{2}}{r_{h}}\Big)\Big)=\frac{\pi}{2}+\Arg\Big(\Gamma\Big(\frac{i\omega L^{2}}{r_{h}}\Big)\Big)
\eea
\bea
\Arg\Big(\Gamma\Big(1-\frac{iL}{2r_{h}}(\omega L+J)\Big)\Big)=-\frac{\pi}{2}-\Arg\Big(\Gamma\Big(\frac{iL}{2r_{h}}(\omega L+J)\Big)\Big)
\eea
Putting these equations into the expression of $\Arg(Q_{1})$, we get
\bea\label{expressionforbeta}
&\hspace{-2cm}\beta\equiv \Arg(Q_{1})=\frac{\omega L^{2}}{2r_{h}}\log2-\frac{L(\omega L+J)}{r_{h}}\log r_{h}+2\Arg(\Gamma[\frac{iL}{2r_{h}}(\omega L+J)])-\Arg(\frac{iJL}{r_{h}})-\nonumber\\
&\hspace{9cm}-\Arg(\Gamma[\frac{i\omega L^{2}}{r_{h}}])
\eea

\subsection{Calculation of $\theta$ and the normalization of the field $\phi_{\omega,J}(r)$}

We write the expression for the field $\phi_{\omega,J}(r)$ from \cite{Burman}
\bea\label{appendixunnormalizedfield}
\phi_{\omega,J}(r) = Sr^{\frac{1}{2}-\frac{iJL}{r_{h}}}(r^{2}-r_{h}^{2})^{-\frac{i\omega L^{2}}{2r_{h}}}e^{-i\pi\beta}\Big(\frac{r}{r_{h}}\Big)^{-2\beta}{}_2F_{1}[\alpha,\beta;2;y]C_{1} 
\eea
where $S,\alpha,\beta$ and $y$ are defined in \cite{Burman}. Now using the reality condition $\phi_{\omega,J}(r)=\phi_{\omega,J}(r)^{*}$ and writing $C_{1}=|C_{1}|e^{i\theta}$, where $\theta$ is the argument of $C_{1}$, we get
\bea\label{realityconditionsimplified}
&\hspace{-2.5cm}1=r^{-\frac{2iJL}{r_{h}}}(r^{2}-r_{h}^{2})^{-\frac{i\omega L^{2}}{r_{h}}}e^{-2\pi i}(\frac{r}{r_{h}})^{\frac{2iL}{r_{h}}(\omega L+J)}e^{2i\theta}\frac{\Gamma[1+\frac{iL}{2r_{h}}(\omega L+J)]\Gamma[1-\frac{iL}{2r_{h}}(\omega L-J)] }{\Gamma[1-\frac{iL}{2r_{h}}(\omega L+J)]\Gamma[1+\frac{iL}{2r_{h}}(\omega L-J)]}\times\nonumber\\
&\hspace{9cm}\times\frac{\Gamma[-\frac{iJL}{r_{h}}]}{\Gamma[\frac{iJL}{r_{h}}]}\frac{{}_2F_{1}[\alpha,\beta;2;\frac{r_{h}^{2}}{r^{2}}]}{{}_2F_{1}[\alpha^{*},\beta^{*};2;\frac{r_{h}}{r^{2}}]}
\eea
Using the hypergeometric identity
\bea
{}_2F_{1}(a,b;c;z)=(1-z)^{c-a-b}{}_2F_{1}(c-a,c-b;c,z)
\eea
we can write
\bea
\frac{{}_2F_{1}[\alpha,\beta;2;\frac{r_{h}^{2}}{r^{2}}]}{{}_2F_{1}[\alpha^{*},\beta^{*};2;\frac{r_{h}^{2}}{r^{2}}]} = \Big(1-\frac{r_{h}^{2}}{r^{2}}\Big)^{\frac{i\omega L^{2}}{r_{h}}}
\eea
Putting the above equation into \eqref{realityconditionsimplified} we get
\bea\label{appendixexpoftheta }
e^{2i\theta}=e^{2i\pi}r_{h}^{\frac{2iL}{r_{h}}(\omega L+J)}\Bigg[\frac{\Gamma[1-\frac{iL}{2r_{h}}(\omega L+J)]\Gamma[1+\frac{iL}{2r_{h}}(\omega L-J)]\Gamma[\frac{iJL}{r_{h}}]}{\Gamma[1+\frac{iL}{2r_{h}}(\omega L+J)]\Gamma[1-\frac{iL}{2r_{h}}(\omega L-J)]\Gamma[-\frac{iJL}{r_{h}}]}\Bigg]
\eea
Using the gamma function identity $\Gamma(1+z)=z\Gamma(z)$ into the expression inside the big square bracket, the above equation further gets simplified to
\bea
e^{2i\theta}=e^{2i\pi}r_{h}^{\frac{2iL}{r_{h}}(\omega L+J)}\Bigg[\frac{\Gamma[-\frac{iL}{2r_{h}}(\omega L+J)]\Gamma[\frac{iL}{2r_{h}}(\omega L-J)]\Gamma[\frac{iJL}{r_{h}}]}{\Gamma[\frac{iL}{2r_{h}}(\omega L+J)]\Gamma[-\frac{iL}{2r_{h}}(\omega L-J)]\Gamma[-\frac{iJL}{r_{h}}]}\Bigg]
\eea
Taking the argument on both sides of the above equation, we find
\bea
&\hspace{-1.7cm}2\theta=2\pi+\frac{2L}{r_{h}}(\omega L+J)\log r_{h}+\Arg\Big(\Gamma\Big(\frac{-iL}{2r_{h}}(\omega L+J)\Big)\Big)+ \Arg\Big(\Gamma\Big(\frac{iL}{2r_{h}}(\omega L-J)\Big)\Big)+\nonumber\\
&\hspace{-1.3cm}+\Arg\Big(\Gamma\Big(\frac{iJL}{r_{h}}\Big)\Big)-\Arg\Big(\Gamma\Big(\frac{iL}{2r_{h}}(\omega L+J)\Big)\Big)-\Arg\Big(\Gamma\Big(\frac{-iL}{2r_{h}}(\omega L-J)\Big)\Big)-\Arg\Big(\Gamma\Big(\frac{iJL}{r_{h}}\Big)\Big)
\eea
Using $\Arg\Gamma(-iz)=-\Arg\Gamma(iz)$, the above equation becomes
\bea\label{expressionfortheta}
&\hspace{-3cm}\theta=\pi+\frac{L}{r_{h}}(\omega L+J)\log r_{h}-\Arg\Gamma\Big(\frac{iL}{2r_{h}}(\omega L+J)\Big)+ \Arg\Gamma\Big(\frac{iL}{2r_{h}}(\omega L-J)\Big)+\nonumber\\
&\hspace{10cm}+\Arg\Big(\Gamma\Big(\frac{iJL}{r_{h}}\Big)\Big)
\eea

Our goal is to identify the phases in the radial wave near the horizon. There are two conventions that appear in the literature, that differ by a sign. We demonstrate the origin of this difference below. First, note that adding \eqref{expressionfortheta},\eqref{expressionforalpha} and $\frac{\omega L^{2}}{2r_{h}}\log r_{h}$ we get
\bea
&\hspace{-1.5cm}\theta+\alpha+\frac{\omega L^{2}}{2r_{h}}\log r_{h}=-\pi-\Arg\Gamma\Big(\frac{iL}{2r_{h}}(\omega L+J)\Big)-\Arg\Gamma\Big(\frac{iL}{2r_{h}}(\omega L-J)\Big)-\frac{\omega L^{2}}{2r_{h}}\log 2+\nonumber\\
&\hspace{8cm}+\Arg\Big(\Gamma\Big(\frac{i\omega L^{2}}{r{h}}\Big)\Big)+\frac{\omega L^{2}}{2r_{h}}\log r_{h}
\eea
But adding \eqref{expressionfortheta},\eqref{expressionforbeta} and $-\frac{\omega L^{2}}{2r_{h}}\log r_{h}$ we get
\bea
&\hspace{-1.5cm}\theta+\beta-\frac{\omega L^{2}}{2r_{h}}\log r_{h}=\pi+\Arg\Gamma\Big(\frac{iL}{2r_{h}}(\omega L-J)\Big)+\Arg\Gamma\Big(\frac{iL}{2r_{h}}(\omega L+J)\Big)+\frac{\omega L^{2}}{2r_{h}}\log 2-\nonumber\\
&\hspace{8cm}-\Arg\Big(\Gamma\Big(\frac{i\omega L^{2}}{r{h}}\Big)\Big)-\frac{\omega L^{2}}{2r_{h}}\log r_{h}
\eea
If we define,
\bea
&\hspace{-1cm}\tilde \delta_{\omega}\equiv\theta+\beta-\frac{\omega L^{2}}{2r_{h}}\log r_{h}=\pi+\Arg\Gamma(\frac{iL}{2r_{h}}(\omega L-J))+\Arg\Gamma(\frac{iL}{2r_{h}}(\omega L+J))+\frac{\omega L^{2}}{2r_{h}}\log 2-\nonumber\\
&\hspace{8cm}-\Arg(\Gamma[\frac{i\omega L^{2}}{r{h}}])-\frac{\omega L^{2}}{2r_{h}}\log r_{h}
\eea
then 
$\theta+\alpha+\frac{\omega L^{2}}{2r_{h}}\log r_{h}=-\tilde \delta_{\omega}$.
Now, the expansion of the field $\phi$ near the horizon becomes
\bea
\phi_{hor}(z)\approx|C_{1}||P_{1}|\Big[e^{-i\omega z-i\tilde \delta_{\omega}}+ e^{i\omega z+i\tilde \delta_{\omega}}\Big]
\eea
This is one possible choice of sign for the radial wave near the horizon. 

On the other hand, if we define 
\bea
&\hspace{-1.2cm}\delta_{\omega} = \Arg\Gamma(\frac{iL}{2r_{h}}(\omega L-J))+\Arg\Gamma(\frac{iL}{2r_{h}}(\omega L+J))+\frac{\omega L^{2}}{2r_{h}}\log 2-\Arg(\Gamma[\frac{i\omega L^{2}}{r{h}}])-\frac{\omega L^{2}}{2r_{h}}\log r_{h}
\eea
then the expansion of the field $\phi$ near the horizon becomes
\bea\label{appendixphiexpansionnearthehorizon}
\phi_{hor}(z)\approx|C_{1}||P_{1}|\Big[-e^{-i\omega z-i\delta_{\omega}}-e^{i\omega z+i\delta_{\omega}}\Big]
\eea
We will work with this choice of signs for our radial wave.

Now, we demand the normalizability condition $|C_{1}||P_{1}|=1$. Since, we know that
\bea
|P_{1}|^{2}=P_{1}P_{1}^{*}
\eea
Using \eqref{appendixexpressionforP_{1}} into the above equation we get
\bea\label{appendixmodP_{1}}
|P_{1}|=\Big(\frac{Jr_{h}}{\omega L}\Big)^{1/2}e^{-\frac{\pi L}{2r_{h}}(\omega L+J)}\sqrt{\csc h(\frac{\pi\omega L^{2}}{r_{h}})\sinh({\frac{\pi JL}{r_{h}}})}
\eea
Now the normalization $C_{1}$ of the field $\phi_{\omega,J}(r)$ can be written as
\bea
C_{1}=|C_{1}|e^{i\theta}=\frac{1}{|P_{1}|}e^{i\theta}
\eea
Using \eqref{appendixmodP_{1}}, \eqref{expressionfortheta} and the gamma function formula $\Gamma[\frac{iJL}{r_{h}}]\Gamma[\frac{-iJL}{r_{h}}]=\frac{\pi r_{h}}{JL\sinh(\frac{\pi JL}{r_{h}})}$ into the above equation we finally get
\bea
C_{1}=-\Gamma[\frac{iJL}{r_{h}}]\Bigg[\frac{\Gamma[1-\frac{iL}{2r_{h}}(\omega L+J)]\Gamma[1+\frac{iL}{2r_{h}}(\omega L-J)]}{\Gamma[1+\frac{iL}{2r_{h}}(\omega L+J)]\Gamma[1-\frac{iL}{2r_{h}}(\omega L-J)]}\Bigg]^{\frac{1}{2}}\sqrt{\frac{\omega L^{2}}{\pi r_{h}^{2}}\sinh\Big(\frac{\pi\omega L^{2}}{r_{h}}\Big)}\times\nonumber\\
\times\frac{e^{\frac{\pi L}{2r_{h}}}(\omega L+J)}{r_{h}^{-\frac{iL}{r_{h}}(\omega L+J)}}
\eea
This is the normalization of the scalar field $\phi_{\omega,J}(r)$ for the free wave boundary condition at the horizon. Putting this normalization into \eqref{appendixunnormalizedfield}, we get the normalized field as
\bea
\phi_{\omega,J}(r)=r^{-3/2}\Big(1-\frac{r_h^2}{r^2}\Big)^{-\frac{i\omega L^2}{2r_h}}L^2\sqrt{\pi\omega}\sqrt{\Tilde{f}(\omega,J)}{}_2F_{1}[\alpha,\beta;2;\frac{r_{h}^{2}}{r^{2}}]
\eea

\section{Details of the 4-point Finite-$\epsilon$ Correlator}\label{AppD}

Here we outline the evaluation of the 4-point correlator in \eqref{G4}. The order of the terms here is the same  as in \eqref{G4} and we denote 
\bea
G_4 = I +II +III + IV + V + VI 
\eea
for convenience of reference. We start with 
\bea\label{Stretchedhorizonfirstterm}
I = \sum_{\{n\},\{J\}}K \langle N|b_{n_{1}J_{1}}b_{n_{2}J_{2}}b_{n_{3}J_{3}}^{\dagger}b_{n_{4}J_{4}}^{\dagger}|N\rangle \mathcal{U}_{n_{1}J_{1}}(X_{1})\mathcal{U}_{n_{2}J_{2}}(X_{2})\mathcal{U}_{n_{3}J_{3}}^{*}(X_{3})\mathcal{U}_{n_{4}J_{4}}^{*}(X_{4})
\eea
Note that this is only non-zero if either the excitation created by $b_{n_{4}J_{4}}^{\dagger}$ is annihilated by $b_{n_{1}J_{1}}$ and the excitation creation by $b_{n_{3}J_{3}}^{\dagger}$ is annihilated by $b_{n_{2}J_{2}}$, or if the excitation created by $b_{n_{4}J_{4}}^{\dagger}$ is annihilated by $b_{n_{2}J_{2}}$ and the excitation creation by $b_{n_{3}J_{3}}^{\dagger}$ is annihilated by $b_{n_{1}J_{1}}$. These two possibilities give rise to two different terms and the $I$ is the sum of these two possibilities. So we find
\begin{align}\label{firsttermtwopossibilities}
&\hspace{-0.4cm}I=\sum_{\{n\},\{J\}}K\Big[\sqrt{N_{n_{1}J_{1}}+1}\sqrt{N_{n_{2}J_{2}}+1}\sqrt{N_{n_{3}J_{3}}+1}\sqrt{N_{n_{4}J_{4}}+1}\delta_{n_{1}n_{4}}\delta_{n_{2}n_{3}}\delta_{J_{1}J_{4}}\delta_{J_{1}J_{4}}\mathcal{U}_{n_{1}J_{1}}(X_{1})\times\nonumber\\
&\times\mathcal{U}_{n_{2}J_{2}}(X_{2})\mathcal{U}_{n_{3}J_{3}}^{*}(X_{3})\mathcal{U}_{n_{4}J_{4}}^{*}(X_{4}) + \sqrt{N_{n_{1}J_{1}}+1}\sqrt{N_{n_{2}J_{2}}+1}\sqrt{N_{n_{3}J_{3}}+1}\sqrt{N_{n_{4}J_{4}}+1}\times\nonumber\\
&\hspace{4cm}\times\delta_{n_{2}n_{4}}\delta_{n_{1}n_{3}}\delta_{J_{2}J_{4}}\delta_{J_{1}J_{3}}\mathcal{U}_{n_{1}J_{1}}(X_{1})\mathcal{U}_{n_{2}J_{2}}(X_{2})\mathcal{U}_{n_{3}J_{3}}^{*}(X_{3})\mathcal{U}_{n_{4}J_{4}}^{*}(X_{4})\Big]
\end{align}
In \eqref{firsttermtwopossibilities}, we sum over $n_{3},n_{4}$ and $J_{3},J_{4}$. The final expression becomes
\begin{align}\label{finalformofSHfirstterm}
&I = \sum_{n_{1}n_{2},J_{1},J_{2}}K\Big[(N_{n_{1}J_{1}}+1)(N_{n_{2}J_{2}}+1)\mathcal{U}_{n_{1}J_{1}}(X_{1})\mathcal{U}_{n_{2}J_{2}}(X_{2})\mathcal{U}_{n_{2}J_{2}}^{*}(X_{3})\mathcal{U}_{n_{1}J_{1}}^{*}(X_{4}) + \nonumber\\
&\hspace{3cm}+(N_{n_{1}J_{1}}+1)(N_{n_{2}J_{2}}+1)\mathcal{U}_{n_{1}J_{1}}(X_{1})\mathcal{U}_{n_{2}J_{2}}(X_{2})\mathcal{U}_{n_{1}J_{1}}^{*}(X_{3})\mathcal{U}_{n_{2}J_{2}}^{*}(X_{4})\Big]
\end{align}
By a very similar (but not identical) argument, we find
\begin{align}\label{finalformofSHsecondterm}
&\hspace{-0.5cm}II = \sum_{n_{1},n_{2},J_{1},J_{2}}K(N_{n_{1}J_{1}}+1)(N_{n_{2}J_{2}})\mathcal{U}_{n_{1}J_{1}}(X_{1})\mathcal{U}_{n_{2}J_{2}}^{*}(X_{2})\mathcal{U}_{n_{2}J_{2}}(X_{3})\mathcal{U}_{n_{1}J_{1}}^{*}(X_{4}) + \nonumber \\
&\hspace{1cm}+ \sum_{n_{1},n_{2},J_{1},J_{2}}K(N_{n_{1}J_{1}}+1)(N_{n_{2}J_{2}}+1)\mathcal{U}_{n_{1}J_{1}}(X_{1})\mathcal{U}_{n_{1}J_{1}}^{*}(X_{2})\mathcal{U}_{n_{2}J_{2}}(X_{3})\mathcal{U}_{n_{2}J_{2}}^{*}(X_{4}),
\end{align}
\begin{align}\label{finalformofSHthirdterm}
&\hspace{-0.4cm}III = \sum_{n_{1},n_{2},J_{1},J_{2}}K(N_{n_{1}J_{1}}+1)(N_{n_{2}J_{2}})\mathcal{U}_{n_{1}J_{1}}(X_{1})\mathcal{U}_{n_{2}J_{2}}^{*}(X_{2})\mathcal{U}_{n_{1}J_{1}}^{*}(X_{3})\mathcal{U}_{n_{2}J_{2}}(X_{4}) + \nonumber \\
&\hspace{1.5cm}+ \sum_{n_{1},n_{2},J_{1},J_{2}}K(N_{n_{1}J_{1}}+1)(N_{n_{2}J_{2}})\mathcal{U}_{n_{1}J_{1}}(X_{1})\mathcal{U}_{n_{1}J_{1}}^{*}(X_{2})\mathcal{U}_{n_{2}J_{2}}^{*}(X_{3})\mathcal{U}_{n_{2}J_{2}}(X_{4}),
\end{align}
\begin{align}\label{finalformofSHfourthterm}
&\hspace{-0.7cm}IV = \sum_{n_{1},n_{2},J_{1},J_{2}}K(N_{n_{1}J_{1}})(N_{n_{2}J_{2}}+1)\mathcal{U}_{n_{1}J_{1}}^{*}(X_{1})\mathcal{U}_{n_{2}J_{2}}(X_{2})\mathcal{U}_{n_{1}J_{1}}(X_{3})\mathcal{U}_{n_{2}J_{2}}^{*}(X_{4}) + \nonumber \\
&\hspace{1cm}+ \sum_{n_{1},n_{2},J_{1},J_{2}}K(N_{n_{1}J_{1}})(N_{n_{2}J_{2}}+1)\mathcal{U}_{n_{1}J_{1}}^{*}(X_{1})\mathcal{U}_{n_{1}J_{1}}(X_{2})\mathcal{U}_{n_{2}J_{2}}(X_{3})\mathcal{U}_{n_{2}J_{2}}^{*}(X_{4}),
\end{align}
\begin{align}\label{finalformofSHfifthterm}
&\hspace{-1cm}V = \sum_{n_{1},n_{2},J_{1},J_{2}}K(N_{n_{1}J_{1}})(N_{n_{2}J_{2}}+1)\mathcal{U}_{n_{1}J_{1}}^{*}(X_{1})\mathcal{U}_{n_{2}J_{2}}(X_{2})\mathcal{U}_{n_{2}J_{2}}^{*}(X_{3})\mathcal{U}_{n_{1}J_{1}}(X_{4}) + \nonumber \\
&\hspace{1.5cm}+ \sum_{n_{1},n_{2},J_{1},J_{2}}K(N_{n_{1}J_{1}})(N_{n_{2}J_{2}})\mathcal{U}_{n_{1}J_{1}}^{*}(X_{1})\mathcal{U}_{n_{1}J_{1}}(X_{2})\mathcal{U}_{n_{2}J_{2}}^{*}(X_{3})\mathcal{U}_{n_{2}J_{2}}(X_{4}),
\end{align}
\begin{align}\label{finalformofSHsixthterm}
&\hspace{-1cm}VI = \sum_{n_{1},n_{2},J_{1},J_{2}}K\Big[(N_{n_{1}J_{1}})(N_{n_{2}J_{2}})\mathcal{U}_{n_{1}J_{1}}^{*}(X_{1})\mathcal{U}_{n_{2}J_{2}}^{*}(X_{2})\mathcal{U}_{n_{2}J_{2}}(X_{3})\mathcal{U}_{n_{1}J_{1}}(X_{4}) + \nonumber \\
&\hspace{3cm}+ (N_{n_{1}J_{1}})(N_{n_{2}J_{2}})\mathcal{U}_{n_{1}J_{1}}^{*}(X_{1})\mathcal{U}_{n_{2}J_{2}}^{*}(X_{2})\mathcal{U}_{n_{1}J_{1}}(X_{3})\mathcal{U}_{n_{2}J_{2}}(X_{4})\Big].
\end{align}

\section{Thermal Factorisation: Harmonic Oscillator}

In this paper, we have discussed various versions of large-$N$/thermal factorization results for correlation functions. The structure of these calculations can sometimes get involved, so it is useful to have a simple prototypical example in mind -- namely the harmonic oscillator \cite{CKReview}. 

Below we present the thermal factorization calculation of the four-point function in the harmonic oscillator. The result is expected on general grounds, our goal is merely to understand the structure of the calculation by explicitly evaluating the correlator. Specifically, we want to show that
\begin{align}\label{A.1}
&\frac{1}{Z}\Tr(e^{-\beta H}x(t_{1})x(t_{2})x(t_{3})x(t_{4})) =
\frac{1}{Z}\Tr(e^{-\beta H}x(t_{1})x(t_{2}))\frac{1}{Z}\Tr(e^{-\beta H}x(t_{3})x(t_{4}))+ \nonumber \\ 
&+\frac{1}{Z}\Tr(e^{-\beta H}x(t_{1})x(t_{3}))\frac{1}{Z}\Tr(e^{-\beta H}x(t_{2})x(t_{4}))+\frac{1}{Z}\Tr(e^{-\beta H}x(t_{1})x(t_{4}))\frac{1}{Z}\Tr(e^{-\beta H}x(t_{2})x(t_{3}))
\end{align}
where $Z$ is the harmonic oscillator partition function and $x(t)$ is a Heisenberg operator defined as
\bea
x(t) = \frac{1}{\sqrt{2\omega}}[e^{-i\omega t}a+e^{i\omega t}a^{\dagger}]
\eea
To avoid clutter, we define
\bea\label{A.3}
x(t) = x^{+}(t) + x^{-}(t)
\eea
where, $x^{+}(t) = \frac{1}{\sqrt{2\omega}}e^{-i\omega t}a$ , $x^{-}(t) = \frac{1}{\sqrt{2\omega}}e^{i\omega t}a^{\dagger}$ and $x^{+}(t)|0\rangle = 0$ , $\langle 0|x^{-}(t)=0.$\\

We will  start by evaluating the LHS. Later, we will also explicitly calculate the RHS and show that both are exactly equal. First we write
\bea\label{A.4}
\Tr(e^{-\beta H}x(t_{1})x(t_{2})x(t_{3})x(t_{4})) = \sum_{n=0}^{\infty}e^{-\beta(n+\frac{1}{2})\hslash\omega}\langle n|x(t_{1})x(t_{2})x(t_{3})x(t_{4})|n\rangle
\eea
where,
\bea
|n\rangle = \frac{1}{\sqrt{n!}}(a^{\dagger})^{n}|0\rangle
\eea
We put \eqref{A.3} into \eqref{A.4}. Only six terms will survive, and these are the terms that have an equal number of $x^{+}(t)$ and $x^{-}(t).$ We get
\begin{align}\label{A.6}
& \sum_{n=0}^{\infty}\frac{e^{-\beta(n+\frac{1}{2})\hslash\omega}}{n!}\langle 0|\big[(a)^{n}x^{+}(t_{1})x^{+}(t_{2})x^{-}(t_{3})x^{-}(t_{4})(a^{\dagger})^{n}+(a)^{n}x^{+}(t_{1})x^{-}(t_{2})x^{+}(t_{3})x^{-}(t_{4})(a^{\dagger})^{n}+\nonumber \\ 
&+ (a)^{n}x^{+}(t_{1})x^{-}(t_{2})x^{-}(t_{3})x^{+}(t_{4})(a^{\dagger})^{n}+(a)^{n}x^{-}(t_{1})x^{+}(t_{2})x^{+}(t_{3})x^{-}(t_{4})(a^{\dagger})^{n}+ \nonumber \\
&+(a)^{n}x^{-}(t_{1})x^{+}(t_{2})x^{-}(t_{3})x^{+}(t_{4})(a^{\dagger})^{n}+(a)^{n}x^{-}(t_{1})x^{-}(t_{2})x^{+}(t_{3})x^{+}(t_{4})(a^{\dagger})^{n}\big]|0\rangle \\
&\equiv I + II +III+IV+V+VI
\end{align}
We will evaluate each of these, term by term. The calculations are roughly of the same type, even though not identical, so we present Case $II$ as an illustrative example. 
\begin{align}
&II=\sum_{n=0}^{\infty}\frac{e^{-\beta(n+\frac{1}{2})\hslash\omega}}{n!}\langle 0|(a)^{n}x^{+}(t_{1})x^{-}(t_{2})x^{+}(t_{3})x^{-}(t_{4})(a^{\dagger})^{n}|0\rangle \nonumber \\
&= \frac{e^{-\beta\hslash\omega/2}}{(2\omega)^{2}}\sum_{n=0}^{\infty}e^{-\beta\hslash\omega n}(n+1)^{2}e^{-i\omega t_{1}+\omega t_{2}-i\omega t_{3}+i\omega t_{4}} \nonumber \\
&= \frac{e^{-\beta\hslash\omega/2}}{(2\omega)^{2}}\frac{e^{2\beta\hslash\omega}(1+e^{\beta\hslash\omega})}{(e^{\beta\hslash\omega}-1)^{3}}e^{-i\omega t_{1}+\omega t_{2}-i\omega t_{3}+i\omega t_{4}}
\end{align}
The rest of the terms proceed similarly. Combining all of them, and using  the harmonic oscillator partition function $Z = \frac{e^{\beta\hslash\omega/2}}{e^{\beta\hslash\omega}-1}$, LHS of eqn. \eqref{A.1} becomes: 
\begin{align}
&\frac{1}{Z}\Tr(e^{-\beta H}x(t_{1})x(t_{2})x(t_{3})x(t_{4})) = \frac{1}{(2\omega)^{2}(e^{\beta\hslash\omega}-1)^{2}}\Bigg[2e^{2\beta\hslash\omega}e^{-i\omega t_{1}-\omega t_{2}+i\omega t_{3}+i\omega t_{4}}+\nonumber\\
&+e^{\beta\hslash\omega}(1+e^{\beta\hslash\omega})e^{-i\omega t_{1}+\omega t_{2}-i\omega t_{3}+i\omega t_{4}}
+2e^{\beta\hslash\omega}e^{-i\omega t_{1}+\omega t_{2}+i\omega t_{3}-i\omega t_{4}}+2e^{\beta\hslash\omega}e^{i\omega t_{1}-\omega t_{2}-i\omega t_{3}+i\omega t_{4}}+\nonumber\\
&+(1+e^{\beta\hslash\omega})e^{i\omega t_{1}-\omega t_{2}+i\omega t_{3}-i\omega t_{4}}+2e^{i\omega t_{1}+\omega t_{2}-i\omega t_{3}-i\omega t_{4}}\Bigg] \label{LHS} \end{align}

Now, we turn to the evaluation of the RHS of \eqref{A.1}. There are three terms, which we denote by $I'$, $II'$ and $III'$. A straightforward calculation leads to 
\begin{align}\label{A.17}
I'&=\frac{1}{Z}\Tr(e^{-\beta H}x(t_{1})x(t_{2}))\frac{1}{Z}\Tr(e^{-\beta H}x(t_{3})x(t_{4}))=\nonumber\\
&=\frac{1}{(2\omega)^{2}(e^{\beta\hslash\omega}-1)^{2}}\Bigg[e^{2\beta\hslash\omega}e^{-i\omega t_{1}+i\omega t_{2}-i\omega t_{3}+i\omega t_{4}}
+e^{\beta\hslash\omega}e^{-i\omega t_{1}+i\omega t_{2}+i\omega t_{3}-i\omega t_{4}}+\nonumber\\
&+e^{\beta\hslash\omega}e^{i\omega t_{1}-i\omega t_{2}-i\omega t_{3}+i\omega t_{4}} + e^{i\omega t_{1}-i\omega t_{2}+i\omega t_{3}-i\omega t_{4}} \Bigg]
\end{align}
$II'$ and $III'$ can be obtained from \eqref{A.17} via 
$II' =I'(t_{1}\rightarrow t_{1},t_{2}\rightarrow t_{3},t_{3}\rightarrow t_{2},t_{4}\rightarrow t_{4})$, and $III'=I'(t_{1}\rightarrow t_{1},t_{2}\rightarrow t_{4},t_{3}\rightarrow t_{2},t_{4}\rightarrow t_{3})$. Adding all three up we finally get
\begin{align}
&\frac{1}{(2\omega)^{2}(e^{\beta\hslash\omega}-1)^{2}}\Bigg[2e^{2\beta\hslash\omega}e^{-i\omega t_{1}-\omega t_{2}+i\omega t_{3}+i\omega t_{4}}+e^{\beta\hslash\omega}(1+e^{\beta\hslash\omega})e^{-i\omega t_{1}+\omega t_{2}-i\omega t_{3}+i\omega t_{4}}+\nonumber \\
&+2e^{\beta\hslash\omega}e^{-i\omega t_{1}+\omega t_{2}+i\omega t_{3}-i\omega t_{4}}+2e^{\beta\hslash\omega}e^{i\omega t_{1}-\omega t_{2}-i\omega t_{3}+i\omega t_{4}}+(1+e^{\beta\hslash\omega})e^{i\omega t_{1}-\omega t_{2}+i\omega t_{3}-i\omega t_{4}}+\nonumber\\
&+2e^{i\omega t_{1}+\omega t_{2}-i\omega t_{3}-i\omega t_{4}}\Bigg]
\end{align}
which is identical to \eqref{LHS}. This prototypical calculation helps us to organize things even when there are many indices and fields around to complicate things.

\section{Thermal Factorisation: Hartle-Hawking}\label{FactorHH}

Here, we will show the thermal factorization of the four-point function in the Hartle-Hawking vacuum. We are not aware of an explicit calculation of this type in the literature, but it helps us to clearly see the parallels with our heavy typical state calculation. 

We will follow the notation of Appendix A of \cite{Festuccia}, but with the slight caveat that we are working with compact angular coordinates. The quantum number $J$ is an integer. 
The mode expansion and the definition of the vacuum are given in \eqref{modeexpansion} and the nearby equations. We can write the 4-pt function as
\begin{align}\label{fourpointfunction}
&{}_{HH}\langle 0|\Phi(X_{1})\Phi(X_{2})\Phi(X_{3})\Phi(X_{4})|0\rangle_{HH} = \int_{0}^{\infty}\prod_{i=1}^{4}\frac{d\omega_{i}}{(2\pi)}\sum_{J_{1},J_{2},J_{3},J_{4}}\nonumber\\
&{}_{HH}\Big\langle 0\Big|\Big[H^{(1)}_{\omega_{1}J_{1}}(X_{1})b^{(1)}_{\omega_{1}J_{1}} +
H^{(1)*}_{\omega_{1}J_{1}}(X_{1})b^{(1)\dagger}_{\omega_{1}J_{1}} + H^{(2)}_{\omega_{1}J_{1}}(X_{1})b^{(2)}_{\omega_{1}J_{1}} + H^{(2)*}_{\omega_{1}J_{1}}(X_{1})b^{(2)\dagger}_{\omega_{1}J_{1}} \Big]\nonumber\\
&\Big[H^{(1)}_{\omega_{2}J_{2}}(X_{2})b^{(1)}_{\omega_{2}J_{2}} + H^{(1)*}_{\omega_{2}J_{2}}(X_{2})b^{(1)\dagger}_{\omega_{2}J_{2}} + 
H^{(2)}_{\omega_{2}J_{2}}(X_{2})b^{(2)}_{\omega_{2}J_{2}} + H^{(2)*}_{\omega_{2}J_{2}}(X_{2})b^{(2)\dagger}_{\omega_{2}J_{2}}\Big]\nonumber\\
&\Big[H^{(1)}_{\omega_{3}J_{3}}(X_{3})b^{(1)}_{\omega_{3}J_{3}} + H^{(1)*}_{\omega_{3}J_{3}}(X_{3})b^{(1)\dagger}_{\omega_{3}J_{3}}
+H^{(2)}_{\omega_{3}J_{3}}(X_{3})b^{(2)}_{\omega_{3}J_{3}} + 
H^{(2)*}_{\omega_{3}J_{3}}(X_{3})b^{(2)\dagger}_{\omega_{3}J_{3}}\Big]\nonumber\\
&\Big[H^{(1)}_{\omega_{4}J_{4}}(X_{4})b^{(1)}_{\omega_{4}J_{4}} + H^{(1)*}_{\omega_{4}J_{4}}(X_{4})b^{(1)\dagger}_{\omega_{4}J_{4}} + H^{(2)}_{\omega_{4}J_{4}}(X_{4})b^{(2)}_{\omega_{4}J_{4}} + H^{(2)*}_{\omega_{4}J_{4}}(X_{4})b^{(2)\dagger}_{\omega_{4}J_{4}}\Big]\Big|0\Big\rangle_{HH}
\end{align}
where, $X_{i}\equiv(r_{i},x_{i})$. After expanding, ten terms survive in the expression \eqref{fourpointfunction} and \eqref{fourpointfunction} becomes
\begin{align}
&=\int\prod_{i=1}^{4}\frac{d\omega_{i}}{(2\pi)}\sum_{J_{1},J_{2},J_{3},J_{4}}{}_{HH}\Big\langle 0\Big|\Big[H^{(1)}_{\omega_{1}J_{1}}(X_{1})H^{(1)}_{\omega_{2}J_{2}}(X_{2})H^{(1)*}_{\omega_{3}J_{3}}(X_{3})H^{(1)*}_{\omega_{4}J_{4}}(X_{4})\times\nonumber\\
&\times b^{(1)}_{\omega_{1}J_{1}}b^{(1)}_{\omega_{2}J_{2}}b^{(1)\dagger}_{\omega_{3}J_{3}}b^{(1)\dagger}_{\omega_{4}J_{4}}
+ H^{(1)}_{\omega_{1}J_{1}}(X_{1})H^{(1)*}_{\omega_{2}J_{2}}(X_{2})H^{(1)}_{\omega_{3}J_{3}}(X_{3})H^{(1)*}_{\omega_{4}J_{4}}(X_{4})b^{(1)}_{\omega_{1}J_{1}}b^{(1)\dagger}_{\omega_{2}J_{2}}b^{(1)}_{\omega_{3}J_{3}}b^{(1)\dagger}_{\omega_{4}J_{4}} + \nonumber\\
&+H^{(1)}_{\omega_{1}J_{1}}(X_{1})H^{(1)*}_{\omega_{2}J_{2}}(X_{2})H^{(2)}_{\omega_{3}J_{3}}(X_{3})H^{(2)*}_{\omega_{4}J_{4}}(X_{4})b^{(1)}_{\omega_{1}J_{1}}b^{(1)\dagger}_{\omega_{2}J_{2}}b^{(2)}_{\omega_{3}J_{3}}b^{(2)\dagger}_{\omega_{4}J_{4}}+ 
H^{(1)}_{\omega_{1}J_{1}}(X_{1})H^{(2)}_{\omega_{2}J_{2}}(X_{2})\times\nonumber\\
&\times H^{(1)*}_{\omega_{3}J_{3}}(X_{3})H^{(2)*}_{\omega_{4}J_{4}}(X_{4})b^{(1)}_{\omega_{1}J_{1}}b^{(2)}_{\omega_{2}J_{2}}b^{(1)\dagger}_{\omega_{3}J_{3}}b^{(2)\dagger}_{\omega_{4}J_{4}}+ H^{(1)}_{\omega_{1}J_{1}}(X_{1})H^{(2)}_{\omega_{2}J_{2}}(X_{2})H^{(2)*}_{\omega_{3}J_{3}}(X_{3})H^{(1)*}_{\omega_{4}J_{4}}(X_{4})\times \nonumber\\
&\times b^{(1)}_{\omega_{1}J_{1}}b^{(2)}_{\omega_{2}J_{2}}b^{(2)\dagger}_{\omega_{3}J_{3}}b^{(1)\dagger}_{\omega_{4}J_{4}} + 
H^{(2)}_{\omega_{1}J_{1}}(X_{1})H^{(1)}_{\omega_{2}J_{2}}(X_{2})H^{(1)*}_{\omega_{3}J_{3}}(X_{3})H^{(2)*}_{\omega_{4}J_{4}}(X_{4}) b^{(2)}_{\omega_{1}J_{1}}b^{(1)}_{\omega_{2}J_{2}}b^{(1)\dagger}_{\omega_{3}J_{3}}b^{(2)\dagger}_{\omega_{4}J_{4}} + \nonumber \\
&+H^{(2)}_{\omega_{1}J_{1}}(X_{1}) H^{(1)}_{\omega_{2}J_{2}}(X_{2}) H^{(2)*}_{\omega_{3}J_{3}}(X_{3}) H^{(1)*}_{\omega_{4}J_{4}}(X_{4})
b^{(2)}_{\omega_{1}J_{1}}b^{(1)}_{\omega_{2}J_{2}}b^{(2)\dagger}_{\omega_{3}J_{3}}b^{(1)\dagger}_{\omega_{4}J_{4}} +H^{(2)}_{\omega_{1}J_{1}}(X_{1})H^{(2)}_{\omega_{2}J_{2}}(X_{2})\times\nonumber\\
&\times H^{(2)*}_{\omega_{3}J_{3}}(X_{3})H^{(2)*}_{\omega_{4}J_{4}}(X_{4})b^{(2)}_{\omega_{1}J_{1}}b^{(2)}_{\omega_{2}J_{2}}b^{(2)\dagger}_{\omega_{3}J_{3}}b^{(2)\dagger}_{\omega_{4}J_{4}} +H^{(2)}_{\omega_{1}J_{1}}(X_{1})H^{(2)*}_{\omega_{2}J_{2}}(X_{2})
H^{(1)}_{\omega_{3}J_{3}}(X_{3})H^{(1)*}_{\omega_{4}J_{4}}(X_{4})\times\nonumber\\
&\times b^{(2)}_{\omega_{1}J_{1}}b^{(2)\dagger}_{\omega_{2}J_{2}}b^{(1)}_{\omega_{3}J_{3}}b^{(1)\dagger}_{\omega_{4}J_{4}}+H^{(2)}_{\omega_{1}J_{1}}(X_{1})H^{(2)*}_{\omega_{2}J_{2}}(X_{2})H^{(2)}_{\omega_{3}J_{3}}(X_{3})H^{(2)*}_{\omega_{4}J_{4}}(X_{4}) b^{(2)}_{\omega_{1}J_{1}}b^{(2)\dagger}_{\omega_{2}J_{2}}b^{(2)}_{\omega_{3}J_{3}}b^{(2)\dagger}_{\omega_{3}J_{3}}
\Big]\Big|0\Big\rangle_{HH}
\end{align}
Each term can be simplified using the commutation relation \eqref{commutationrelation}. We can integrate the delta functions and the result is 
\bea\label{F.3}
\hspace*{-2cm}=\int\frac{d\omega_{3}d\omega_{4}}{(2\pi)^{4}}\sum_{J_{3},J_{4}}H^{(1)}_{\omega_{3}J_{3}}(X_{1})H^{(1)}_{\omega_{4}J_{4}}(X_{2})H^{(1)*}_{\omega_{3}J_{3}}(X_{3})H^{(1)*}_{\omega_{4}J_{4}}(X_{4})+ \nonumber\\
\hspace*{-2cm}+\int\frac{d\omega_{3}d\omega_{4}}{(2\pi)^{4}}\sum_{J_{3},J_{4}}H^{(1)}_{\omega_{4}J_{4}}(X_{1})H^{(1)}_{\omega_{3}J_{3}}(X_{2})H^{(1)*}_{\omega_{3}J_{3}}(X_{3})H^{(1)*}_{\omega_{4}J_{4}}(X_{4})+ \nonumber \\
\hspace*{-2cm}+\int\frac{d\omega_{2}d\omega_{4}}{(2\pi)^{4}}\sum_{J_{2},J_{4}}H^{(1)}_{\omega_{2}J_{2}}(X_{1})H^{(1)*}_{\omega_{2}J_{2}}(X_{2})H^{(1)}_{\omega_{4}J_{4}}(X_{3})H^{(1)*}_{\omega_{4}J_{4}}(X_{4})+ \nonumber \\
\hspace*{-2cm}+\int\frac{d\omega_{2}d\omega_{4}}{(2\pi)^{4}}\sum_{J_{2},J_{4}}H^{(1)}_{\omega_{2}J_{2}}(X_{1})H^{(1)*}_{\omega_{2}J_{2}}(X_{2})H^{(2)}_{\omega_{4}J_{4}}(X_{3})H^{(2)*}_{\omega_{4}J_{4}}(X_{4})+ \nonumber \\
\hspace*{-2cm}+\int\frac{d\omega_{3}d\omega_{4}}{(2\pi)^{4}}\sum_{J_{3},J_{4}}H^{(1)}_{\omega_{3}J_{3}}(X_{1})H^{(2)}_{\omega_{4}J_{4}}(X_{2})H^{(1)*}_{\omega_{3}J_{3}}(X_{3})H^{(2)*}_{\omega_{4}J_{4}}(X_{4})+\nonumber\\
\hspace*{-2cm}+\int\frac{d\omega_{3}d\omega_{4}}{(2\pi)^{4}}\sum_{J_{3},J_{4}}H^{(1)}_{\omega_{4}J_{4}}(X_{1})H^{(2)}_{\omega_{3}J_{3}}(X_{2})H^{(2)*}_{\omega_{3}J_{3}}(X_{3})H^{(1)*}_{\omega_{4}J_{4}}(X_{4})+ \nonumber \\
\hspace*{-2cm}+\int\frac{d\omega_{3}d\omega_{4}}{(2\pi)^{4}}\sum_{J_{3},J_{4}}H^{(2)}_{\omega_{4}J_{4}}(X_{1})H^{(1)}_{\omega_{3}J_{3}}(X_{2})H^{(1)*}_{\omega_{3}J_{3}}(X_{3})H^{(2)*}_{\omega_{4}J_{4}}(X_{4})+ \nonumber \\
\hspace*{-2cm}+\int\frac{d\omega_{3}d\omega_{4}}{(2\pi)^{4}}\sum_{J_{3},J_{4}}H^{(2)}_{\omega_{3}J_{3}}(X_{1})H^{(1)}_{\omega_{4}J_{4}}(X_{2})H^{(2)*}_{\omega_{3}J_{3}}(X_{3})H^{(1)*}_{\omega_{4}J_{4}}(X_{4})+ \nonumber \\
\hspace*{-2cm}+\int\frac{d\omega_{3}d\omega_{4}}{(2\pi)^{4}}\sum_{J_{3},J_{4}}H^{(2)}_{\omega_{3}J_{3}}(X_{1})H^{(2)}_{\omega_{4}J_{4}}(X_{2})H^{(2)*}_{\omega_{3}J_{3}}(X_{3})H^{(2)*}_{\omega_{4}J_{4}}(X_{4})+ \nonumber \\
\hspace*{-2cm}+\int\frac{d\omega_{3}d\omega_{4}}{(2\pi)^{4}}\sum_{J_{3},J_{4}}H^{(2)}_{\omega_{4}J_{4}}(X_{1})H^{(2)}_{\omega_{3}J_{3}}(X_{2})H^{(2)*}_{\omega_{3}J_{3}}(X_{3})H^{(2)*}_{\omega_{4}J_{4}}(X_{4})+ \nonumber \\
\hspace*{-2cm}+\int\frac{d\omega_{2}d\omega_{4}}{(2\pi)^{4}}\sum_{J_{2},J_{4}}H^{(2)}_{\omega_{2}J_{2}}(X_{1})H^{(2)*}_{\omega_{2}J_{2}}(X_{2})H^{(1)}_{\omega_{4}J_{4}}(X_{3})H^{(1)*}_{\omega_{4}J_{4}}(X_{4})+\nonumber \\
\hspace*{-2cm}+\int\frac{d\omega_{2}d\omega_{4}}{(2\pi)^{4}}\sum_{J_{2},J_{4}}H^{(2)}_{\omega_{2}J_{2}}(X_{1})H^{(2)*}_{\omega_{2}J_{2}}(X_{2})H^{(2)}_{\omega_{4}J_{4}}(X_{3})H^{(2)*}_{\omega_{4}J_{4}}(X_{4})
\eea

\noindent For the one-sided correlator
\bea
H_{\omega,J}^{(1)} = \cosh{\theta_{\omega}}\varphi^{(1)}_{\omega,J} = \frac{1}{\sqrt{1-e^{-\beta\omega}}}\varphi^{(1)}_{\omega,J}
\eea
and
\bea
H_{\omega,J}^{(2)} = \sinh{\theta_{\omega}}\varphi^{(1)*}_{\omega,J} = \frac{e^{-\beta\omega/2}}{\sqrt{1-e^{-\beta\omega}}}\varphi^{(1)*}_{\omega,J}
\eea
First, we combine $1^{st}$ and $9^{th}$ terms:
\begin{align}\label{1stand9thcombined}
&= \int_{0}^{\infty}\frac{d\omega_{3}}{(2\pi)}\frac{d\omega_{4}}{(2\pi)}\frac{1}{2\pi}\sum_{J_{3}=-\infty}^{+\infty}\frac{1}{2\pi}\sum_{J_{4}=-\infty}^{+\infty}\Bigg[\frac{e^{\beta\omega_{3}}}{e^{\beta\omega_{3}}-1}\frac{e^{\beta\omega_{4}}}{e^{\beta\omega_{4}}-1}\varphi^{(1)}_{\omega_{3}J_{3}}(X_{1})\varphi^{(1)*}_{\omega_{3}J_{3}}(X_{3})\varphi^{(1)}_{\omega_{4}J_{4}}(X_{2})\times\nonumber\\
&\times\varphi^{(1)*}_{\omega_{4}J_{4}}(X_{4})+ \frac{1}{e^{\beta\omega_{3}}-1}\frac{1}{e^{\beta\omega_{4}}-1}\varphi^{(1)*}_{\omega_{3}J_{3}}(X_{1})\varphi^{(1)}_{\omega_{3}J_{3}}(X_{3})\varphi^{(1)*}_{\omega_{4}J_{4}}(X_{2})\varphi^{(1)}_{\omega_{4}J_{4}}(X_{4}) \Bigg]
\end{align}
In the above expression, in the second term inside the square bracket, we do the replacement $\omega_{3}\rightarrow -\omega_{3}, J_{3}\rightarrow-J_{3}$ \& $\omega_{4}\rightarrow-\omega_{4}, J_{4}\rightarrow-J_{4}.$ 
With this replacement, the second term in the above expression becomes
\begin{align}
&= \int_{-\infty}^{0}\frac{d\omega_{3}}{(2\pi)}\frac{d\omega_{4}}{(2\pi)}\frac{1}{2\pi}\sum_{J_{3}=-\infty}^{+\infty}\frac{1}{2\pi}\sum_{J_{4}=-\infty}^{+\infty}
\frac{e^{\beta\omega_{3}}}{e^{\beta\omega_{3}}-1}\frac{e^{\beta\omega_{4}}}{e^{\beta\omega_{4}}-1}\varphi^{(1)}_{\omega_{3}J_{3}}(X_{1})\varphi^{(1)*}_{\omega_{3}J_{3}}(X_{3})\varphi^{(1)}_{\omega_{4}J_{4}}(X_{2})\times\nonumber\\
&\hspace{12cm}\times\varphi^{(1)*}_{\omega_{4}J_{4}}(X_{4})
\end{align}
The combined form of $1^{st}$ and $9^{th}$ terms of \eqref{F.3} becomes
\begin{align}\label{B.10}
&=\Bigg[\int_{0}^{\infty}\frac{d\omega_{3}}{2\pi}\int_{0}^{\infty}\frac{d\omega_{4}}{2\pi}+\int_{-\infty}^{0}\frac{d\omega_{3}}{2\pi}\int_{-\infty}^{0}\frac{d\omega_{4}}{2\pi}\Bigg]\frac{1}{2\pi}\sum_{J_{3}=-\infty}^{+\infty}\frac{1}{2\pi}\sum_{J_{4}=-\infty}^{+\infty}\frac{e^{\beta\omega_{3}}}{e^{\beta\omega_{3}}-1}\frac{e^{\beta\omega_{4}}}{e^{\beta\omega_{4}}-1}\varphi^{(1)}_{\omega_{3}J_{3}}(X_{1})\times \nonumber \\
&\hspace{9cm}\times\varphi^{(1)*}_{\omega_{3}J_{3}}(X_{3})\varphi^{(1)}_{\omega_{4}J_{4}}(X_{2})\varphi^{(1)*}_{\omega_{4}J_{4}}(X_{4})
\end{align}
Now, let's combine $5^{th}$ and $8^{th}$ term of \eqref{F.3}
\begin{align}
&=\int_{0}^{\infty}\frac{d\omega_{3}}{2\pi}\frac{d\omega_{4}}{2\pi}\frac{1}{2\pi}\sum_{J_{3}=-\infty}^{+\infty}\frac{1}{2\pi}\sum_{J_{4}=-\infty}^{+\infty}\Bigg[\frac{1}{e^{\beta\omega_{4}}-1}\frac{e^{\beta\omega_{3}}}{e^{\beta\omega_{3}}-1}\varphi_{\omega_{3}J_{3}}^{(1)}(X_{1})\varphi_{\omega_{3}J_{3}}^{(1)*}(X_{3})\varphi_{\omega_{4}J_{4}}^{(1)*}(X_{2})\varphi_{\omega_{4}J_{4}}^{(1)}(X_{4})+ \nonumber \\
&\hspace{3cm}+ \frac{e^{\beta\omega_{4}}}{e^{\beta\omega_{4}}-1}\frac{1}{e^{\beta\omega_{3}}-1}\varphi_{\omega_{3}J_{3}}^{(1)*}(X_{1})\varphi_{\omega_{3}J_{3}}^{(1)}(X_{3})\varphi_{\omega_{4}J_{4}}^{(1)}(X_{2})\varphi_{\omega_{4}J_{4}}^{(1)*}(X_{4}) 
 \Bigg]
\end{align}
In the above expression, in the first term inside the square bracket, we do the replacement $\omega_{4}\rightarrow-\omega_{4},J_{4}\rightarrow-J_{4}$, while in the second term inside the square bracket, we do the replacement $\omega_{3}\rightarrow-\omega_{3},J_{3}\rightarrow-J_{3}$. The above expression becomes
\begin{align}\label{B.12}
&\hspace{-1cm}=\Bigg[\int_{0}^{\infty}\frac{d\omega_{3}}{2\pi}\int_{-\infty}^{0}\frac{d\omega_{4}}{2\pi}+\int_{-\infty}^{0}\frac{d\omega_{3}}{2\pi}\int_{0}^{\infty}\frac{d\omega_{4}}{2\pi}\Bigg]\frac{1}{2\pi}\sum_{J_{3}=-\infty}^{+\infty}\frac{1}{2\pi}\sum_{J_{4}=-\infty}^{+\infty}\frac{e^{\beta\omega_{3}}}{e^{\beta\omega_{3}}-1}\frac{e^{\beta\omega_{4}}}{e^{\beta\omega_{4}}-1}\times \nonumber \\ 
&\hspace{6cm}\times\varphi^{(1)}_{\omega_{3}J_{3}}(X_{1})\varphi^{(1)*}_{\omega_{3}J_{3}}(X_{3})
 \varphi^{(1)}_{\omega_{4}J_{4}}(X_{2})\varphi^{(1)*}_{\omega_{4}J_{4}}(X_{4})
\end{align}
We combine \eqref{B.10} and \eqref{B.12}, which yields the result
\begin{align}
&=\int_{-\infty}^{+\infty}\frac{d\omega_{3}}{2\pi}\frac{1}{2\pi}\sum_{J_{3}=-\infty}^{+\infty}\frac{1}{2\pi}\sum_{J_{4}=-\infty}^{+\infty}\frac{e^{\beta\omega_{3}}}{e^{\beta\omega_{3}}-1}\frac{e^{\beta\omega_{4}}}{e^{\beta\omega_{4}}-1}\varphi^{(1)}_{\omega_{3}J_{3}}(X_{1})\varphi^{(1)*}_{\omega_{3}J_{3}}(X_{3})\varphi^{(1)}_{\omega_{4}J_{4}}(X_{2})\varphi^{(1)*}_{\omega_{4}J_{4}}(X_{4})
\end{align}
We can write the above expression as 
\begin{align}
&=\Bigg[\int_{-\infty}^{+\infty}\frac{d\omega_{3}}{2\pi}\sum_{J_{3}=-\infty}^{+\infty}\frac{1}{2\pi}\frac{e^{\beta\omega_{3}}}{e^{\beta\omega_{3}}-1}\varphi^{(1)}_{\omega_{3}J_{3}}(X_{1})\varphi^{(1)*}_{\omega_{3}J_{3}}(X_{3})\Bigg]\times \nonumber \\
&\times\Bigg[\int_{-\infty}^{+\infty}\frac{d\omega_{4}}{2\pi}\sum_{J_{4}=-\infty}^{+\infty}\frac{1}{2\pi}\frac{e^{\beta\omega_{4}}}{e^{\beta\omega_{4}}-1}\varphi^{(1)}_{\omega_{4}J_{4}}(X_{2})\varphi^{(1)*}_{\omega_{4}J_{4}}(X_{4})\Bigg]
\end{align}
In the above result, each term inside the square bracket is a position-space version of the 2-point Hartle-Hawking correlator \eqref{smoothbulkHHcorrelator}.
So, by combining $1^{st},9^{th},5^{th}$ and $8^{th}$ terms of \eqref{F.3}, we get
\bea\label{firstfactorizedterm}
 {}_{HH}\langle 0|\Phi(r_{1},x_{1})\Phi(r_{3},x_{3})|0\rangle_{HH}\langle 0|\Phi(r_{2},x_{2})\Phi(r_{4},x_{4})|0\rangle_{HH}
\eea
Similarly, we can combine $2^{nd},10^{th},6^{th}$ and $7^{th}$ terms into a single term and $3^{rd},12^{th},4^{th}$ and $11^{th}$ into another single term, which respectively, yield 
\bea\label{secondfactorizedterm}
 {}_{HH}\langle 0|\Phi(r_{1},x_{1})\Phi(r_{4},x_{4})|0\rangle_{HH}\langle 0|\Phi(r_{2},x_{2})\Phi(r_{3},x_{3})|0\rangle_{HH}
\eea
and 
\bea\label{thirdfactorizedterm}
 {}_{HH}\langle 0|\Phi(r_{1},x_{1})\Phi(r_{2},x_{2})|0\rangle_{HH}\langle 0|\Phi(r_{3},x_{3})\Phi(r_{4},x_{4})|0\rangle_{HH}
\eea
Adding \eqref{firstfactorizedterm},\eqref{secondfactorizedterm} and \eqref{thirdfactorizedterm}, the 4-point function \eqref{fourpointfunction} becomes \begin{align}\label{HHthermalfactorization}
&{}_{HH}\langle 0|\Phi(r_{1},x_{1})\Phi(r_{2},x_{2})\Phi(r_{3},x_{3})\Phi(r_{4},x_{4})|0\rangle_{HH} = 
{}_{HH}\langle 0|\Phi(r_{1},x_{1})\Phi(r_{2},x_{2})|0\rangle_{HH}\times\nonumber\\
&\times\langle 0|\Phi(r_{3},x_{3})\Phi(r_{4},x_{4})|0\rangle_{HH}
+ {}_{HH}\langle 0|\Phi(r_{1},x_{1})\Phi(r_{3},x_{3})|0\rangle_{HH}\langle 0|\Phi(r_{2},x_{2})\Phi(r_{4},x_{4})|0\rangle_{HH} + \nonumber \\
&+ {}_{HH}\langle 0|\Phi(r_{1},x_{1})\Phi(r_{4},x_{4})|0\rangle_{HH}\langle 0|\Phi(r_{2},x_{2})\Phi(r_{3},x_{3})|0\rangle_{HH}
\end{align}
which demonstrates factorization. The parallels with the factorization of the four-point function in the stretched horizon case is evident. We have not fully understood the origin of this parallel at a deep level -- in particular, the quasi-degeneracy of the $J$-quantum numbers seems to have been tailor-made for the success of the calculation.

\section{Analytic Continuation of the Smooth Horizon Correlator}

The goal of this section is to review the fact that the Hartle-Hawking correlator allows an analytic continuation into the interior regions of the Kruskal geometry. We will (along the way) present the HH correlator explicitly in Kruskal coordinates. 

We repeat the BTZ black hole metric here for the reader's convenience:
\bea\label{BTZmetricappendix}
ds^{2}=-\frac{r^2-r_{h}^{2}}{L^{2}}dt^{2}+\frac{L^{2}}{r^2-r_{h}^{2}}dr^{2}+r^{2}d\phi^{2}
\eea
We define BTZ Kruskal coordinates $U,V$ via 
\bea\label{kruskalcoordinatedefinition}
U=&-e^{-ku},\nonumber\\
V=&e^{kv},
\eea
where $k=\frac{r_{h}}{L^{2}}$ is the surface gravity and $u=t-z$, $v=t+z.$ Here $z(r)$ is the tortoise coordinate as defined in \eqref{tortoisecoordinatedefinition}
\bea
z=-\int_{r}^{\infty}\frac{dr}{f(r)} = -\int_{r}^{\infty}\frac{L^{2}}{r^{2}-r_{h}^{2}}dr = -\frac{L^{2}}{r_{h}}\tanh^{-1}{\Big(\frac{r_{h}}{r}\Big)}
\eea
Since $\tanh^{-1}{(x)}=\frac{1}{2}\log\Big(\frac{1+x}{1-x}\Big)$ tortoise coordinate can be written as
\bea\label{tortoisecoordinateappendix}
z = \frac{L^{2}}{2r_{h}}\log\Big(\frac{r-r_{h}}{r+r_{h}}\Big)
\eea
From \eqref{kruskalcoordinatedefinition} it is clear that 
\bea\label{UV}
UV = -e^{k(v-u)} = -e^{2kz}=-\frac{r-r_{h}}{r+r_{h}}
\eea
A useful expression is to write $r$ in terms of $UV$:
\bea
r=r_{h}\Big(\frac{1-UV}{1+UV}\Big) \label{rinK}
\eea
Now the BTZ metric \eqref{BTZmetricappendix} in Kruskal coordinates becomes
\bea\label{BTZmetricinKruskalcoordinateappendix}
ds_{kruskal}^{2} = -\frac{4L^{2}}{(1+UV)^{2}}dUdV+\frac{r_{h}^{2}(1-UV)^{2}}{(1+UV)^{2}}d\phi^{2}
\eea
In Kruskal coordinates, using \eqref{rinK} we can write the HH correlator \eqref{continummlimitofcutoffcorrelator} as
\bea
&\hspace{-1.5cm}\mathcal{G}^{+}_{c}(\omega,J;U,V) = \frac{L^{4}}{4}\frac{e^{\beta\omega}}{e^{\beta\omega}-1}r_{h}^{-4}\Big(\frac{1-UV}{1+UV}\Big)^{-2}\Big(\frac{1-U'V'}{1+U'V'}\Big)^{-2}\Big(1-\Big(\frac{1+UV}{1-UV}\Big)^{2}\Big)^{\frac{i\omega L^{2}}{2r_{h}}}\times\nonumber\\
&\hspace{-1.5cm}\times\Big(1-\Big(\frac{1+U'V'}{1-U'V'}\Big)^{2}\Big)^{\frac{-i\omega L^{2}}{2r_{h}}}\Tilde{f}(\omega,J){}_{2}F_{1}\Big[\alpha^{*},\beta^{*};2;\Big(\frac{1+UV}{1-UV}\Big)^{2}\Big]{}_{2}F_{1}\Big[\alpha,\beta;2;\Big(\frac{1+U'V'}{1-U'V'}\Big)^{2}\Big]
\eea
Note that this object is the bulk two-point correlator in Kruskal coordinates, and it has a natural analytic continuation into the interior. It has a well-defined Green function interpretation in the entire Penrose diagram. For any finite value of the stretched horizon on the other hand, a continuation of the radial coordinate into the interior will lead to an object with no clear meaning as a Green function in the interior. Note in particular, that such an object explicitly depends on the normal modes of the {\em finite} stretched horizon. In the $\epsilon \rightarrow 0$ limit, this dependence is gone, and we have a conventional Hartle-Hawking correlator. 



\begin{thebibliography}{99}



\bibitem{Burman}
V.~Burman, S.~Das and C.~Krishnan,
``A smooth horizon without a smooth horizon,''
JHEP \textbf{03}, 014 (2024)
doi:10.1007/JHEP03(2024)014
[arXiv:2312.14108 [hep-th]].

\bibitem{Hawking}
S.~W.~Hawking,
``Breakdown of Predictability in Gravitational Collapse,''
Phys. Rev. D \textbf{14}, 2460-2473 (1976)
doi:10.1103/PhysRevD.14.2460

\bibitem{Page}
D.~N.~Page,
``Information in black hole radiation,''
Phys. Rev. Lett. \textbf{71}, 3743-3746 (1993)
doi:10.1103/PhysRevLett.71.3743
[arXiv:hep-th/9306083 [hep-th]].

\bibitem{Mathur}
S.~D.~Mathur,
``The Information paradox: A Pedagogical introduction,''
Class. Quant. Grav. \textbf{26}, 224001 (2009)
doi:10.1088/0264-9381/26/22/224001
[arXiv:0909.1038 [hep-th]].

\bibitem{AMPS}
A.~Almheiri, D.~Marolf, J.~Polchinski and J.~Sully,
``Black Holes: Complementarity or Firewalls?,''
JHEP \textbf{02}, 062 (2013)
doi:10.1007/JHEP02(2013)062
[arXiv:1207.3123 [hep-th]].

\bibitem{Sen}
A.~Sen,
``Extremal black holes and elementary string states,''
Mod. Phys. Lett. A \textbf{10}, 2081-2094 (1995)
doi:10.1142/S0217732395002234
[arXiv:hep-th/9504147 [hep-th]].

\bibitem{Strominger-Vafa}
A.~Strominger and C.~Vafa,
``Microscopic origin of the Bekenstein-Hawking entropy,''
Phys. Lett. B \textbf{379}, 99-104 (1996)
doi:10.1016/0370-2693(96)00345-0
[arXiv:hep-th/9601029 [hep-th]].

\bibitem{Mathur1}
O.~Lunin and S.~D.~Mathur,
``AdS / CFT duality and the black hole information paradox,''
Nucl. Phys. B \textbf{623}, 342-394 (2002)
doi:10.1016/S0550-3213(01)00620-4
[arXiv:hep-th/0109154 [hep-th]].

\bibitem{BenaReview}
I.~Bena, E.~J.~Martinec, S.~D.~Mathur and N.~P.~Warner,
``Fuzzballs and Microstate Geometries: Black-Hole Structure in String Theory,''
[arXiv:2204.13113 [hep-th]].

\bibitem{ChiMing1}
C.~M.~Chang and Y.~H.~Lin,
``Words to describe a black hole,''
JHEP \textbf{02}, 109 (2023)
doi:10.1007/JHEP02(2023)109
[arXiv:2209.06728 [hep-th]].

\bibitem{ChiMing2}
C.~M.~Chang and Y.~H.~Lin,
``Holographic covering and the fortuity of black holes,''
[arXiv:2402.10129 [hep-th]].

\bibitem{WittenAdSCFT}
E.~Witten,
``Anti-de Sitter space and holography,''
Adv. Theor. Math. Phys. \textbf{2}, 253-291 (1998)
doi:10.4310/ATMP.1998.v2.n2.a2
[arXiv:hep-th/9802150 [hep-th]].

\bibitem{WittenAdSBH}
E.~Witten,
``Anti-de Sitter space, thermal phase transition, and confinement in gauge theories,''
Adv. Theor. Math. Phys. \textbf{2}, 505-532 (1998)
doi:10.4310/ATMP.1998.v2.n3.a3
[arXiv:hep-th/9803131 [hep-th]].

\bibitem{Gautam}
G.~Mandal and A.~Mohan,
``Exact lattice bosonization of finite N matrix quantum mechanics and c = 1,''
[arXiv:2406.07629 [hep-th]].

\bibitem{tHooft}
G.~'t Hooft,
``On the Quantum Structure of a Black Hole,''
Nucl. Phys. B \textbf{256}, 727-745 (1985)
doi:10.1016/0550-3213(85)90418-3

\bibitem{STU}
L.~Susskind, L.~Thorlacius and J.~Uglum,
``The Stretched horizon and black hole complementarity,''
Phys. Rev. D \textbf{48}, 3743-3761 (1993)
doi:10.1103/PhysRevD.48.3743
[arXiv:hep-th/9306069 [hep-th]].

\bibitem{MathurFuzz}
S.~D.~Mathur,
``The Fuzzball proposal for black holes: An Elementary review,''
Fortsch. Phys. \textbf{53}, 793-827 (2005)
doi:10.1002/prop.200410203
[arXiv:hep-th/0502050 [hep-th]].



\bibitem{Israel-Mukohyama}
S.~Mukohyama and W.~Israel,
``Black holes, brick walls and the Boulware state,''
Phys. Rev. D \textbf{58}, 104005 (1998)
doi:10.1103/PhysRevD.58.104005
[arXiv:gr-qc/9806012 [gr-qc]].

\bibitem{Pradipta}
C.~Krishnan and P.~S.~Pathak,
``Normal modes of the stretched horizon: a bulk mechanism for black hole microstate level spacing,''
JHEP \textbf{03}, 162 (2024)
doi:10.1007/JHEP03(2024)162
[arXiv:2312.14109 [hep-th]].

\bibitem{PradiptaNew}
C.~Krishnan and P.~S.~Pathak,
``Holomorphic Factorization at the Quantum Horizon: The Mechanics of Kerr-CFT,''
To appear.

\bibitem{Marolf-Polchinski}
D.~Marolf and J.~Polchinski,
``Gauge/Gravity Duality and the Black Hole Interior,''
Phys. Rev. Lett. \textbf{111}, 171301 (2013)
doi:10.1103/PhysRevLett.111.171301
[arXiv:1307.4706 [hep-th]].

\bibitem{MaldacenaEternal}
J.~M.~Maldacena,
``Eternal black holes in anti-de Sitter,''
JHEP \textbf{04}, 021 (2003)
doi:10.1088/1126-6708/2003/04/021
[arXiv:hep-th/0106112 [hep-th]].

\bibitem{Bala1}
V.~Balasubramanian, J.~de Boer, V.~Jejjala and J.~Simon,
``The Library of Babel: On the origin of gravitational thermodynamics,''
JHEP \textbf{12}, 006 (2005)
doi:10.1088/1126-6708/2005/12/006
[arXiv:hep-th/0508023 [hep-th]].

\bibitem{Bala2}
V.~Balasubramanian, B.~Czech, K.~Larjo and J.~Simon,
``Integrability versus information loss: A Simple example,''
JHEP \textbf{11}, 001 (2006)
doi:10.1088/1126-6708/2006/11/001
[arXiv:hep-th/0602263 [hep-th]].

\bibitem{Onkar}
V.~Balasubramanian, D.~Berenstein, A.~Lewkowycz, A.~Miller, O.~Parrikar and C.~Rabideau,
``Emergent classical spacetime from microstates of an incipient black hole,''
JHEP \textbf{01}, 197 (2019)
doi:10.1007/JHEP01(2019)197
[arXiv:1810.13440 [hep-th]].

\bibitem{Deutsch}
J.~M.~Deutsch,
``Quantum statistical mechanics in a closed system,''
Phys. Rev. A \textbf{43}, no.4, 2046 (1991)
doi:10.1103/PhysRevA.43.2046

\bibitem{Srednicki}
M.~Srednicki,
``Chaos and Quantum Thermalization,''
Phys. Rev. E \textbf{50}, 888
doi:10.1103/PhysRevE.50.888
[arXiv:cond-mat/9403051 [cond-mat]].

\bibitem{CveticLarsen1}
M.~Cvetic and F.~Larsen,
``General rotating black holes in string theory: Grey body factors and event horizons,''
Phys. Rev. D \textbf{56}, 4994-5007 (1997)
doi:10.1103/PhysRevD.56.4994
[arXiv:hep-th/9705192 [hep-th]].

\bibitem{CveticLarsen2}
M.~Cvetic and F.~Larsen,
``Grey body factors for rotating black holes in four-dimensions,''
Nucl. Phys. B \textbf{506}, 107-120 (1997)
doi:10.1016/S0550-3213(97)00541-5
[arXiv:hep-th/9706071 [hep-th]].


\bibitem{CMS}
A.~Castro, A.~Maloney and A.~Strominger,
``Hidden Conformal Symmetry of the Kerr Black Hole,''
Phys. Rev. D \textbf{82}, 024008 (2010)
doi:10.1103/PhysRevD.82.024008
[arXiv:1004.0996 [hep-th]].

\bibitem{CK-KerrCFT}
C.~Krishnan,
``Hidden Conformal Symmetries of Five-Dimensional Black Holes,''
JHEP \textbf{07}, 039 (2010)
doi:10.1007/JHEP07(2010)039
[arXiv:1004.3537 [hep-th]].

\bibitem{BTZ}
M.~Banados, C.~Teitelboim and J.~Zanelli,
``The Black hole in three-dimensional space-time,''
Phys. Rev. Lett. \textbf{69}, 1849-1851 (1992)
doi:10.1103/PhysRevLett.69.1849
[arXiv:hep-th/9204099 [hep-th]].

\bibitem{BTZH}
M.~Banados, M.~Henneaux, C.~Teitelboim and J.~Zanelli,
``Geometry of the (2+1) black hole,''
Phys. Rev. D \textbf{48}, 1506-1525 (1993)
[erratum: Phys. Rev. D \textbf{88}, 069902 (2013)]
doi:10.1103/PhysRevD.48.1506
[arXiv:gr-qc/9302012 [gr-qc]].

\bibitem{Das}
S.~Das, S.~K.~Garg, C.~Krishnan and A.~Kundu,
``Fuzzballs and random matrices,''
JHEP \textbf{10}, 031 (2023)
doi:10.1007/JHEP10(2023)031
[arXiv:2301.11780 [hep-th]].

\bibitem{BrownHenneaux}
J.~D.~Brown and M.~Henneaux,
``Central Charges in the Canonical Realization of Asymptotic Symmetries: An Example from Three-Dimensional Gravity,''
Commun. Math. Phys. \textbf{104}, 207-226 (1986)
doi:10.1007/BF01211590

\bibitem{StromingerNear}
A.~Strominger,
``Black hole entropy from near horizon microstates,''
JHEP \textbf{02}, 009 (1998)
doi:10.1088/1126-6708/1998/02/009
[arXiv:hep-th/9712251 [hep-th]].


\bibitem{StressNew}
``Vacuum Polarization at the Quantum Horizon,'' To appear.

\bibitem{Festuccia-Liu}
G.~Festuccia and H.~Liu,
``Excursions beyond the horizon: Black hole singularities in Yang-Mills theories. I.,''
JHEP \textbf{04}, 044 (2006)
doi:10.1088/1126-6708/2006/04/044
[arXiv:hep-th/0506202 [hep-th]].


\bibitem{Festuccia}
Festuccia, Guido. (2009). Black hole singularities in the framework of gauge/string duality

\bibitem{Leutheusser}
S.~A.~W.~Leutheusser and H.~Liu,
``Emergent Times in Holographic Duality,''
Phys. Rev. D \textbf{108}, no.8, 086020 (2023)
doi:10.1103/PhysRevD.108.086020
[arXiv:2112.12156 [hep-th]].

\bibitem{ElShowk}
S.~El-Showk and K.~Papadodimas,
``Emergent Spacetime and Holographic CFTs,''
JHEP \textbf{10}, 106 (2012)
doi:10.1007/JHEP10(2012)106
[arXiv:1101.4163 [hep-th]].

\bibitem{Mukund}
V.~Balasubramanian, B.~Czech, V.~E.~Hubeny, K.~Larjo, M.~Rangamani and J.~Simon,
``Typicality versus thermality: An Analytic distinction,''
Gen. Rel. Grav. \textbf{40}, 1863-1890 (2008)
doi:10.1007/s10714-008-0606-8
[arXiv:hep-th/0701122 [hep-th]].

\bibitem{Chapman}
S.~Chapman, M.~P.~Heller, H.~Marrochio and F.~Pastawski,
``Toward a Definition of Complexity for Quantum Field Theory States,''
Phys. Rev. Lett. \textbf{120}, no.12, 121602 (2018)
doi:10.1103/PhysRevLett.120.121602
[arXiv:1707.08582 [hep-th]].

\bibitem{Jefferson}
R.~Jefferson and R.~C.~Myers,
``Circuit complexity in quantum field theory,''
JHEP \textbf{10}, 107 (2017)
doi:10.1007/JHEP10(2017)107
[arXiv:1707.08570 [hep-th]].

\bibitem{Rifath}
R.~Khan, C.~Krishnan and S.~Sharma,
``Circuit Complexity in Fermionic Field Theory,''
Phys. Rev. D \textbf{98}, no.12, 126001 (2018)
doi:10.1103/PhysRevD.98.126001
[arXiv:1801.07620 [hep-th]].

\bibitem{action}
A.~R.~Brown, D.~A.~Roberts, L.~Susskind, B.~Swingle and Y.~Zhao,
``Holographic Complexity Equals Bulk Action?,''
Phys. Rev. Lett. \textbf{116}, no.19, 191301 (2016)
doi:10.1103/PhysRevLett.116.191301
[arXiv:1509.07876 [hep-th]].

\bibitem{Action?}
A.~R.~Brown, D.~A.~Roberts, L.~Susskind, B.~Swingle and Y.~Zhao,
``Complexity, action, and black holes,''
Phys. Rev. D \textbf{93}, no.8, 086006 (2016)
doi:10.1103/PhysRevD.93.086006
[arXiv:1512.04993 [hep-th]].

\bibitem{SusskindMP}
L.~Susskind,
``The Typical-State Paradox: Diagnosing Horizons with Complexity,''
Fortsch. Phys. \textbf{64}, 84-91 (2016)
doi:10.1002/prop.201500091
[arXiv:1507.02287 [hep-th]].


\bibitem{Ishibashi}
N.~Ishibashi and T.~Tada,
``Dipolar quantization and the infinite circumference limit of two-dimensional conformal field theories,''
Int. J. Mod. Phys. A \textbf{31}, no.32, 1650170 (2016)
doi:10.1142/S0217751X16501700
[arXiv:1602.01190 [hep-th]].

\bibitem{Suchetan}
S.~Das,
``Stretched Horizon from Conformal Field Theory,''
[arXiv:2406.10879 [hep-th]].

\bibitem{StanfordShenker}
S.~H.~Shenker and D.~Stanford,
``Stringy effects in scrambling,''
JHEP \textbf{05}, 132 (2015)
doi:10.1007/JHEP05(2015)132
[arXiv:1412.6087 [hep-th]].

\bibitem{synth}
S.~Das, C.~Krishnan, A.~P.~Kumar and A.~Kundu,
``Synthetic fuzzballs: a linear ramp from black hole normal modes,''
JHEP \textbf{01}, 153 (2023)
doi:10.1007/JHEP01(2023)153
[arXiv:2208.14744 [hep-th]].

\bibitem{Riemann}
S.~Das, S.~K.~Garg, C.~Krishnan and A.~Kundu,
``What is the Simplest Linear Ramp?,''
JHEP \textbf{01}, 172 (2024)
doi:10.1007/JHEP01(2024)172
[arXiv:2308.11704 [hep-th]].


\bibitem{ArnabSuman}
S.~Das and A.~Kundu,
``Brickwall in rotating BTZ: a dip-ramp-plateau story,''
JHEP \textbf{02}, 049 (2024)
doi:10.1007/JHEP02(2024)049
[arXiv:2310.06438 [hep-th]].

\bibitem{Murdia}
C.~Murdia, Y.~Nomura and K.~Ritchie,
``Black hole and de Sitter microstructures from a semiclassical perspective,''
Phys. Rev. D \textbf{107}, no.2, 026016 (2023)
doi:10.1103/PhysRevD.107.026016
[arXiv:2207.01625 [hep-th]].

\bibitem{Diptarka}
D.~Das, S.~Mandal and A.~Sarkar,
``Chaotic and thermal aspects in the highly excited string S-matrix,''
JHEP \textbf{24}, 200 (2020)
doi:10.1007/JHEP08(2024)200
[arXiv:2312.02127 [hep-th]].

\bibitem{Souvik1}
S.~Banerjee and G.~Vos,
``Behind-the-horizon excitations from a single 2d CFT,''
JHEP \textbf{05}, 309 (2024)
doi:10.1007/JHEP05(2024)309
[arXiv:2401.00890 [hep-th]].

\bibitem{Souvik2}
S.~Banerjee, S.~Das, M.~Dorband and A.~Kundu,
``Brickwall, normal modes, and emerging thermality,''
Phys. Rev. D \textbf{109}, no.12, 126020 (2024)
doi:10.1103/PhysRevD.109.126020
[arXiv:2401.01417 [hep-th]].

\bibitem{Su1}
P.~Biswas, S.~Das and A.~Dinda,
``Moving interfaces and two-dimensional black holes,''
JHEP \textbf{05}, 329 (2024)
doi:10.1007/JHEP05(2024)329
[arXiv:2401.11451 [hep-th]].

\bibitem{Su2}
P.~Biswas, B.~Ezhuthachan, A.~Kundu and B.~Roy,
``Moving Mirrors, OTOCs and Scrambling,''
[arXiv:2406.05772 [hep-th]].

\bibitem{Cotler}
J.~S.~Cotler, G.~Gur-Ari, M.~Hanada, J.~Polchinski, P.~Saad, S.~H.~Shenker, D.~Stanford, A.~Streicher and M.~Tezuka,
``Black Holes and Random Matrices,''
JHEP \textbf{05}, 118 (2017)
[erratum: JHEP \textbf{09}, 002 (2018)]
doi:10.1007/JHEP05(2017)118
[arXiv:1611.04650 [hep-th]].

\bibitem{superstrata}
I.~Bena, S.~Giusto, E.~J.~Martinec, R.~Russo, M.~Shigemori, D.~Turton and N.~P.~Warner,
``Smooth horizonless geometries deep inside the black-hole regime,''
Phys. Rev. Lett. \textbf{117}, no.20, 201601 (2016)
doi:10.1103/PhysRevLett.117.201601
[arXiv:1607.03908 [hep-th]].

\bibitem{Rychkov}
V.~S.~Rychkov,
``D1-D5 black hole microstate counting from supergravity,''
JHEP \textbf{01}, 063 (2006)
doi:10.1088/1126-6708/2006/01/063
[arXiv:hep-th/0512053 [hep-th]].

\bibitem{Grant}
L.~Grant, L.~Maoz, J.~Marsano, K.~Papadodimas and V.~S.~Rychkov,
``Minisuperspace quantization of 'Bubbling AdS' and free fermion droplets,''
JHEP \textbf{08}, 025 (2005)
doi:10.1088/1126-6708/2005/08/025
[arXiv:hep-th/0505079 [hep-th]].

\bibitem{CK-Avinash}
C.~Krishnan and A.~Raju,
``A Note on D1-D5 Entropy and Geometric Quantization,''
JHEP \textbf{06}, 054 (2015)
doi:10.1007/JHEP06(2015)054
[arXiv:1504.04330 [hep-th]].

\bibitem{FuzzComp}
S.~D.~Mathur,
``A model with no firewall,''
[arXiv:1506.04342 [hep-th]].

\bibitem{Vyshnav-State}
C.~Krishnan and V.~Mohan,
``State-independent black hole interiors from the crossed product,''
JHEP \textbf{05}, 278 (2024)
doi:10.1007/JHEP05(2024)278
[arXiv:2310.05912 [hep-th]].


\bibitem{MarolfPolchinskiSD}
D.~Marolf and J.~Polchinski,
``Violations of the Born rule in cool state-dependent horizons,''
JHEP \textbf{01}, 008 (2016)
doi:10.1007/JHEP01(2016)008
[arXiv:1506.01337 [hep-th]].

\bibitem{HarlowSD}
D.~Harlow,
``Aspects of the Papadodimas-Raju Proposal for the Black Hole Interior,''
JHEP \textbf{11}, 055 (2014)
doi:10.1007/JHEP11(2014)055
[arXiv:1405.1995 [hep-th]].

\bibitem{PR}
K.~Papadodimas and S.~Raju,
``State-Dependent Bulk-Boundary Maps and Black Hole Complementarity,''
Phys. Rev. D \textbf{89}, no.8, 086010 (2014)
doi:10.1103/PhysRevD.89.086010
[arXiv:1310.6335 [hep-th]].

\bibitem{Kaplan}
H.~Chen, C.~Hussong, J.~Kaplan and D.~Li,
``A Numerical Approach to Virasoro Blocks and the Information Paradox,''
JHEP \textbf{09}, 102 (2017)
doi:10.1007/JHEP09(2017)102
[arXiv:1703.09727 [hep-th]].

\bibitem{Witten}
E.~Witten,
``Gravity and the crossed product,''
JHEP \textbf{10}, 008 (2022)
doi:10.1007/JHEP10(2022)008
[arXiv:2112.12828 [hep-th]].

\bibitem{Roji}
K.~Jalan, R.~Pius and M.~Ramchander,
``Half-sided Translations and the Information Recovery from Radiation,''
[arXiv:2404.00773 [hep-th]].

\bibitem{Akers}
C.~Akers, N.~Engelhardt, D.~Harlow, G.~Penington and S.~Vardhan,
``The black hole interior from non-isometric codes and complexity,''
JHEP \textbf{06}, 155 (2024)
doi:10.1007/JHEP06(2024)155
[arXiv:2207.06536 [hep-th]].

\bibitem{Hartman}
T.~Hartman and J.~Maldacena,
``Time Evolution of Entanglement Entropy from Black Hole Interiors,''
JHEP \textbf{05}, 014 (2013)
doi:10.1007/JHEP05(2013)014
[arXiv:1303.1080 [hep-th]].

\bibitem{Penington}
G.~Penington,
``Entanglement Wedge Reconstruction and the Information Paradox,''
JHEP \textbf{09}, 002 (2020)
doi:10.1007/JHEP09(2020)002
[arXiv:1905.08255 [hep-th]].

\bibitem{Almheiri}
A.~Almheiri, N.~Engelhardt, D.~Marolf and H.~Maxfield,
``The entropy of bulk quantum fields and the entanglement wedge of an evaporating black hole,''
JHEP \textbf{12}, 063 (2019)
doi:10.1007/JHEP12(2019)063
[arXiv:1905.08762 [hep-th]].

\bibitem{2105.12313}
S.~Xin, B.~Chen, R.~K.~L.~Lo, L.~Sun, W.~B.~Han, X.~Zhong, M.~Srivastava, S.~Ma, Q.~Wang and Y.~Chen,
``Gravitational-wave echoes from spinning exotic compact objects: Numerical waveforms from the Teukolsky equation,''
Phys. Rev. D \textbf{104}, no.10, 104005 (2021)
doi:10.1103/PhysRevD.104.104005
[arXiv:2105.12313 [gr-qc]].

\bibitem{Mayerson}
D.~R.~Mayerson,
``Fuzzballs and Observations,''
Gen. Rel. Grav. \textbf{52}, no.12, 115 (2020)
doi:10.1007/s10714-020-02769-w
[arXiv:2010.09736 [hep-th]].

\bibitem{Saad}
P.~Saad, S.~H.~Shenker and D.~Stanford,
``JT gravity as a matrix integral,''
[arXiv:1903.11115 [hep-th]].

\bibitem{Liu-Vardhan}
H.~Liu and S.~Vardhan,
``Entanglement Entropies of Equilibrated Pure States in Quantum Many-Body Systems and Gravity,''
PRX Quantum \textbf{2}, no.1, 010344 (2021)
doi:10.1103/PRXQuantum.2.010344
[arXiv:2008.01089 [hep-th]].

\bibitem{Bousso}
R.~Bousso and E.~Wildenhain,
``Gravity/ensemble duality,''
Phys. Rev. D \textbf{102}, no.6, 066005 (2020)
doi:10.1103/PhysRevD.102.066005
[arXiv:2006.16289 [hep-th]].

\bibitem{Vyshnav}
C.~Krishnan and V.~Mohan,
``Hints of gravitational ergodicity: Berry\textquoteright{}s ensemble and the universality of the semi-classical Page curve,''
JHEP \textbf{05}, 126 (2021)
doi:10.1007/JHEP05(2021)126
[arXiv:2102.07703 [hep-th]].

\bibitem{Mehta}
B.~Guo, M.~R.~R.~Hughes, S.~D.~Mathur and M.~Mehta,
``Contrasting the fuzzball and wormhole paradigms for black holes,''
Turk. J. Phys. \textbf{45}, no.6, 281-365 (2021)
doi:10.3906/fiz-2111-13
[arXiv:2111.05295 [hep-th]].

\bibitem{PSSY}
G.~Penington, S.~H.~Shenker, D.~Stanford and Z.~Yang,
``Replica wormholes and the black hole interior,''
JHEP \textbf{03}, 205 (2022)
doi:10.1007/JHEP03(2022)205
[arXiv:1911.11977 [hep-th]].

\bibitem{EngelhardtWall}
N.~Engelhardt and A.~C.~Wall,
``Quantum Extremal Surfaces: Holographic Entanglement Entropy beyond the Classical Regime,''
JHEP \textbf{01}, 073 (2015)
doi:10.1007/JHEP01(2015)073
[arXiv:1408.3203 [hep-th]].

\bibitem{Jude}
C.~Krishnan, V.~Patil and J.~Pereira,
``Page Curve and the Information Paradox in Flat Space,''
[arXiv:2005.02993 [hep-th]].

\bibitem{Matsuo}
K.~Hashimoto, N.~Iizuka and Y.~Matsuo,
``Islands in Schwarzschild black holes,''
JHEP \textbf{06}, 085 (2020)
doi:10.1007/JHEP06(2020)085
[arXiv:2004.05863 [hep-th]].

\bibitem{Thor}
F.~F.~Gautason, L.~Schneiderbauer, W.~Sybesma and L.~Thorlacius,
``Page Curve for an Evaporating Black Hole,''
JHEP \textbf{05}, 091 (2020)
doi:10.1007/JHEP05(2020)091
[arXiv:2004.00598 [hep-th]].

\bibitem{Mahajan}
A.~Almheiri, R.~Mahajan and J.~E.~Santos,
``Entanglement islands in higher dimensions,''
SciPost Phys. \textbf{9}, no.1, 001 (2020)
doi:10.21468/SciPostPhys.9.1.001
[arXiv:1911.09666 [hep-th]].

\bibitem{Critical}
C.~Krishnan,
``Critical Islands,''
JHEP \textbf{01}, 179 (2021)
doi:10.1007/JHEP01(2021)179
[arXiv:2007.06551 [hep-th]].

\bibitem{Kausik}
K.~Ghosh and C.~Krishnan,
``Dirichlet baths and the not-so-fine-grained Page curve,''
JHEP \textbf{08}, 119 (2021)
doi:10.1007/JHEP08(2021)119
[arXiv:2103.17253 [hep-th]].

\bibitem{CKReview}
C.~Krishnan,
``Quantum Field Theory, Black Holes and Holography,''
[arXiv:1011.5875 [hep-th]].

\end{thebibliography}
\end{document}